\documentclass[12pt]{article}  
 \pdfoutput=1

\usepackage{epsfig}
\usepackage{graphics}  

\usepackage{url}  
\usepackage{amsmath,amssymb}  
\usepackage{cite}  
 
\numberwithin{equation}{section}  
\newcommand{\bff}[1]{\mbox{\boldmath ${#1}$}}  
  
\newcommand{\pvec}{{\mathbf p}}  
\newcommand{\qvec}{{\mathbf q}}  
\newcommand{\unit}{{\mathbf 1}}  
\newcommand{\mred}{m_\mathrm{red}}  
\newcommand{\mbar}{M}  
\newcommand{\nus}{\nu_\mathrm{spin}^S}  
\newcommand{\nua}{\nu_\mathrm{ann}^{R_\alpha,S}}  
\newcommand{\ep}{\epsilon}  
\newcommand{\dra}{D_{R_\alpha}}  
  
\newcommand{\logPow}[2]{\if 1#1 \mathrm{L}_{#2} \else \mathrm{L}^#1_{#2} \fi}  
\newcommand{\logh}[1][1]{\logPow{#1}{\mu_h}}  
\newcommand{\logm}[1][1]{\logPow{#1}{2\mbar}}  
\newcommand{\logmr}[1][1]{\logPow{#1}{4\mred}}  
\newcommand{\logE}[2][1]{\if 1#2  
 \logPow{#1}{E_f} \else\logPow{#1}{#2E} \fi}  
\newcommand{\logEs}[2][1]{\if 1#2  
 \logPow{#1}{E_s} \else\logPow{#1}{#2E_s} \fi}  
  
\begin{document}  
  
\allowdisplaybreaks  
\thispagestyle{empty}  
  
\begin{flushright}  
{\small  
SI-HEP-2016-22\\[0.0cm]  
TTP16-031\\[0.0cm]  
TTK-16-31\\[0.0cm]  
TUM-HEP-1057/16 \\
arXiv:1607.07574 [hep-ph]\\[0.1cm]
October 25, 2016
}  
\end{flushright}  
  
\vspace{\baselineskip}  
  
\begin{center}  
\vspace{0.5\baselineskip}  
\textbf{\Large\boldmath  
NNLL soft and Coulomb resummation for squark \\[0.2cm]  
and gluino production at the LHC  
}  
\\  
\vspace{2\baselineskip}  
{\sc M.~Beneke$^a$, J.~Piclum$^{b}$,  C.~Schwinn$^{c}$,   
  C.~Wever$^{d}$}\\  
\vspace{0.7cm}  
{\sl \small    
${}^a$Physik Department T31,  
James-Franck-Stra\ss e~1, Technische Universit\"at M\"unchen,\\  
D--85748 Garching, Germany\\  
\vspace{0.3cm}  
${}^b$  
Theoretische Physik 1, Naturwissenschaftlich-Technische Fakult\"at, Universit\"at Siegen,  
57068 Siegen, Germany\\  
and\\  
Albert Einstein Center for Fundamental Physics, Institute for Theoretical Physics,  
University of Bern, 3012 Bern, Switzerland  
\\  
\vspace{0.3cm}  
${}^c$Institut f\"ur Theoretische Teilchenphysik und   
Kosmologie,  
RWTH Aachen University, D--52056 Aachen, Germany\\
\vspace{0.3cm}  
${}^d$Institute of Nuclear Physics, NCSR "Demokritos", GR 15310 Athens,  
Greece\\  
and\\  
Institute for Theoretical Particle Physics (TTP), Engesserstra\ss e 7, D-76128 Karlsruhe \& Institute  
for Nuclear Physics (IKP), Hermann-von-Helmholtz-Platz 1, D-76344 Eggenstein-Leopoldshafen,  
Karlsruhe Institute of Technology, Germany  
}  
  
\vspace*{.6cm}  
\textbf{Abstract}\\  

\vspace{.5\baselineskip}  
\parbox{0.9\textwidth}{ We present predictions for the total cross sections
  for pair production of squarks and gluinos at the LHC including a combined
  NNLL resummation of soft and Coulomb gluon effects.  We derive all terms in
  the NNLO cross section that are enhanced near the production threshold,
  which include contributions from spin-dependent potentials and so-called
  annihilation corrections.  The NNLL corrections at $\sqrt{s}=13$~TeV 
  range from up to $20\%$ for
  squark-squark production to $90\%$ for gluino pair production relative to
  the NLO results and reduce the theoretical uncertainties of the perturbative
  calculation to the $10\%$ level. Grid files with our  numerical results 
  are publicly available~\cite{SUSYNNLL}.}
\end{center}  
  
\newpage  
\setcounter{page}{1}  
\section{Introduction}  
  
Supersymmetry (SUSY) and its realization in the $R$-parity conserving
Minimally Supersymmetric Standard Model~(MSSM) is a well-studied and motivated
extension of the Standard Model~(SM) of particle physics. It could provide a
solution to shortcomings of the SM such as the absence of a dark matter
candidate and it might stabilize the electroweak scale against quantum
corrections.  The search for SUSY at the TeV scale is therefore a central part
of the physics program of the Large Hadron Collider~(LHC).  The production of
squarks $\tilde q$ and gluinos $\tilde g$, the super-partners of quarks and
gluons, through the strong interaction is expected to be an important
discovery channel of SUSY, provided these particles are kinematically
accessible at the LHC.  The most stringent limits from the $7~\mathrm{TeV}$
and $8~\mathrm{TeV}$ runs of the LHC~\cite{Aad:2014wea,Khachatryan:2015vra}
exclude gluino masses up to $m_{\tilde g}=1.3$~TeV and superpartners of the
quarks of the first two generations below $m_{\tilde q}\lesssim
875$~GeV. Equal squark and gluino masses can be excluded up to $m_{\tilde
  g}\sim 1.7$~TeV.  First results at $\sqrt s=13~\mathrm{TeV}$
raised the mass bounds to $m_{\tilde g}\lesssim 1.75$~TeV and $m_{\tilde
  q}\lesssim 1.26$~TeV~\cite{Khachatryan:2016xvy}.  However, these bounds
depend on assumptions, e.g. on the mass of the lightest supersymmetric
particle and on decay chains, and can be evaded, for instance by compressed
mass spectra or non-degenerate light-flavour squark masses. The search for SUSY
therefore remains a focus of the $13$--$14$~TeV run of the LHC that has the
potential to discover or exclude squarks and gluinos up to the $3$~TeV
range. Turning exclusion limits on production cross sections into bounds on
superparticle masses requires precise predictions for these cross sections,
which motivates the computation of higher-order corrections to squark and
gluino production.  The next-to-leading order (NLO) corrections for production
of the light-flavour squarks and gluinos in the supersymmetric extension of
quantum chromodynamics~(SQCD) have been known for a long
time~\cite{Beenakker:1996ch} and have been implemented in the program
\texttt{PROSPINO} \cite{Beenakker:1996ed}. More recently, additional
higher-order QCD corrections have been added to this result in various
approximations~\cite{Kulesza:2008jb,Kulesza:2009kq,Langenfeld:2009eg,Hagiwara:2009hq,Beenakker:2009ha,Beneke:2010da,Kauth:2011vg,Kauth:2011bz,Beenakker:2011sf,Falgari:2012hx,Langenfeld:2012ti,Pfoh:2013iia,Beneke:2013opa,Beenakker:2014sma}. Corresponding
results for top squarks have been obtained as
well~\cite{Beenakker:1997ut,Younkin:2009zn,Beenakker:2010nq,Langenfeld:2010vu,Falgari:2012hx,Broggio:2013uba,Broggio:2013cia,Kim:2014yaa,Beenakker:2016gmf}.
Complementary work to this improvement of total cross sections by higher-order
QCD corrections is provided by the computation of electroweak
contributions~\cite{Bornhauser:2007bf,Hollik:2007wf,Hollik:2008yi,Hollik:2008vm,Mirabella:2009ap,Germer:2010vn,Germer:2014jpa,Hollik:2015lha},
the automation of NLO calculations in the
MSSM~\cite{GoncalvesNetto:2012yt,Goncalves:2014axa}, the matching of NLO
corrections to a parton
shower~\cite{Gavin:2013kga,Gavin:2014yga,Degrande:2015vaa}, the calculation of
NLO corrections to squark production and
decay~\cite{Hollik:2012rc,Hollik:2013xwa,Gavin:2014yga} and the
estimate of finite-width effects~\cite{Falgari:2012sq}.

The dominant production channels for squark and gluino production at hadron colliders are pair-production processes of the form  
\begin{equation}  
  \label{eq:had-process}  
  N_1 N_2 \to \tilde{s} \tilde{s}'X,  
\end{equation}  
where $N_{1,2}$ denote the incoming hadrons and $\tilde{s}$,  
$\tilde{s}'$ the two sparticles.  In this paper we will consider all  
pair-production processes of gluinos and squarks except top squark  
production.    
The NLO SQCD corrections to squark and gluino production processes can  
become very large for heavy sparticle masses~\cite{Beenakker:1996ch},  
up to $100\%$ of the tree-level result for gluino-pair  
production. This raises the question of the convergence of the  
perturbative series.  A substantial part of the large NLO corrections  
can be attributed to terms that are enhanced in the limit of a small 
relative velocity $\beta$ of the sparticles,   
\begin{equation}  
\label{eq:threshold}  
\beta=\sqrt{1-\frac{(m_{\tilde{s}}+m_{\tilde{s}^\prime})^2}{\hat s}}\to 0 ,   
\end{equation}  
where $\hat s$ is the partonic centre-of-mass energy.  
These corrections arise at each order in perturbation theory through 
threshold  logarithms $\alpha_s\ln^{2,1}\beta$ due to soft-gluon 
corrections and through Coulomb corrections of the form  $\alpha_s/\beta$.    
The large NLO corrections to squark and gluino production and the  
significant contribution of the threshold region motivate the  
resummation of these threshold corrections,  i.e.\  a reorganization of the perturbation theory under the assumption that both types of threshold corrections are of order one,  
\begin{equation}  
\label{eq:count}  
  \alpha_s\ln\beta\sim 1 \,,\quad  
  \frac{\alpha_s}{\beta}\sim 1.  
\end{equation}  
The accuracy of the resummed perturbative series can be defined by  
representing the resummed cross section schematically as  
\begin{equation}  
\label{eq:syst}  
\begin{aligned}  
\hat{\sigma}_{p p'}  
=&\, \hat \sigma^{(0)}_{p p'}\,  
\sum_{k=0}^\infty \left(\frac{\alpha_s}{\beta}\right)^{\!k}   
\Bigl(1+ \alpha_s c_{\text{NNLL}}+\ldots\Bigr)\\   
&\times\, \exp\Big[\underbrace{\ln\beta\,g_0(\alpha_s\ln\beta)}_{\mbox{(LL)}}+   
\underbrace{g_1(\alpha_s\ln\beta)}_{\mbox{(NLL)}}+  
\underbrace{\alpha_s g_2(\alpha_s\ln\beta)}_{\mbox{(NNLL)}}+\ldots\Big]  
\, .   
\end{aligned}  
\end{equation}  
Methods for the separate resummation of the two towers of corrections are well established and have been applied to squark and gluino production.  
The resummation of  threshold logarithms~\cite{Sterman:1986aj,Catani:1989ne,Kidonakis:1997gm,Bonciani:1998vc} with a   
 fixed-order treatment of Coulomb corrections was performed at NLL~\cite{Kulesza:2008jb,Kulesza:2009kq,Beenakker:2009ha,Beenakker:2010nq} and more recently at NNLL accuracy~\cite{Beenakker:2011sf,Pfoh:2013iia,Broggio:2013cia,Beenakker:2014sma}.    
The application of Coulomb-resummation~\cite{Hoang:2000yr} to squark and gluino production  with a fixed-order treatment of threshold logarithms was considered in~\cite{Kulesza:2009kq,Hagiwara:2009hq,Kauth:2011bz,Kauth:2011vg}.   
  
In these approaches, only one of the two variables in~\eqref{eq:count}  
is considered to be of order one in the threshold region, which is not  
justified a priori.  Therefore a combined resummation of soft and  
Coulomb corrections is desirable and was established  
in~\cite{Beneke:2009rj,Beneke:2010da} using effective-theory methods.  
The application of this method to squark and gluino production at NLL  
accuracy~\cite{Falgari:2012hx} has revealed a significant effect of  
Coulomb corrections and soft-Coulomb interference effects that can be  
as large as the soft corrections alone. Since the joint soft and  
Coulomb corrections at NLL can show an enhancement of up to $100\%$  
relative to the NLO cross section for some processes and large  
sparticle masses \cite{Falgari:2012hx}, 
a combined NNLL treatment seems to be required for a  
stabilisation of the perturbative behaviour.  We note that when the 
Coulomb corrections are not summed, some of the sizeable corrections at 
NLL in the combined soft-Coulomb resummation appear only at the 
next order (NNLL) in pure soft-gluon resummation. 
Ref.~\cite{Beenakker:2014sma} indeed confirms the earlier finding of 
a significant soft-Coulomb interference effect. 
In the present paper we perform for the  
first time such a combined soft and Coulomb resummation for squark and  
gluino production at NNLL accuracy. Preliminary results have been  
presented already in~\cite{Beneke:2013opa}. A combination of Coulomb  
corrections and NNLL soft resummation has also been performed for the  
case of top-squark bound states (``stoponium'') 
in~\cite{Kim:2014yaa} using a formalism similar to ours.  
  
With respect to our previous work on NNLL  
resummation for top quark  
production~\cite{Beneke:2011mq,Beneke:2012wb}, this paper contains  
several new theoretical results and features: We derive the extension  
of the spin-dependent non-Coulomb $\alpha_s^2\ln\beta$ terms given  
for top-pair production in~\cite{Beneke:2009ye} to squark and gluino  
production (these results have been quoted already  
in~\cite{Beneke:2013opa}).  We also generalize the additional logarithm  
found in~\cite{Baernreuther:2013caa} for top-pair production to squark  
and gluino production and show how it arises in the effective-theory  
framework.  For the soft-gluon resummation we use the scale choice 
introduced in~\cite{Sterman:2013nya} as a default.  
Our  numerical cross section results are 
publicly available in the form of grids in the squark--gluino 
mass plane~\cite{SUSYNNLL}. 
  
The paper is organized as follows: In Section~\ref{sec:method} we give  
an overview of squark and gluino production, review our resummation  
method and provide the input for NNLL resummation.  We compute the  
single-logarithmic potential corrections and spell out our choice of  
the soft scale in soft-gluon resummation in the momentum-space framework. In  
Section~\ref{sec:numerics} we present our numerical results and  
specify our estimate of the remaining theoretical uncertainties.  Some  
technical details of the NNLL resummation are provided in an Appendix.

%%%%%%%%%%%%%%%%%%%%%%%%%%%%%%%%%%%%%%%%%%%%%%%%%%%%%%%%%%%%%%%  
\section{NNLL soft-Coulomb resummation for squark and gluino production}  
\label{sec:method}  
\subsection{Production processes}  
\label{sec:processes}  
  
The total hadronic cross sections for the processes~\eqref{eq:had-process} 
can be obtained from short-distance production cross sections 
$\hat{\sigma}_{p p'}(\hat s, \mu_f)$ for the partonic processes   
\begin{equation}  
  \label{eq:part-process}  
  pp'\to \tilde{s} \tilde{s}'X\;, \quad p,p'\in\{q,\bar q,g\},  
\end{equation}  
by a convolution with the parton luminosity functions $L_{p p'}(\tau,\mu)$:  
\begin{equation}  
\label{eq:sigma-had}  
\sigma_{N_1 N_2 \to \tilde{s} \tilde{s}'X}(s) = \int_{\tau_0}^1 d \tau \sum_{p, p'=q,\bar{q}, g} L_{p p'}(\tau,\mu_f) \hat{\sigma}_{p p'}(\tau s, \mu_f) \, ,  
\end{equation}   
with $\tau_0=4\mbar^2/s$ and the average sparticle mass  
\begin{equation}  
  \label{eq:mav}  
\mbar=\frac{m_{\tilde{s}}+m_{\tilde{s}'}}{2}.  
\end{equation}  
The parton luminosity functions    
are defined in terms of the parton density functions (PDFs) as  
\begin{equation}  
L_{p p'}(\tau,\mu)=\int_0^1 d x_1 d x_2 \delta(x_1 x_2-\tau) f_{p/N_1}(x_1,\mu) f_{p'/N_2}(x_2,\mu)  \, .  
\end{equation}  
  
At leading order~\cite{Kane:1982hw,Harrison:1982yi,Dawson:1983fw}, the following partonic channels contribute to the  
production of light-flavour squarks   
and gluinos:  
\begin{eqnarray}  
\label{eq:processes}  
 gg, \, q_i \bar{q}_j &\rightarrow& \tilde{q} \bar{\tilde{q}} \,, \nonumber\\  
  q_i q_j&\rightarrow& \tilde{q} \tilde{q}, \qquad \bar{q}_i  
 \bar{q}_j\rightarrow \bar{\tilde{q}} \bar{\tilde{q}} \,, \nonumber\\  
  g q_i &\rightarrow& \tilde{g} \tilde{q},\,\qquad   
  g \bar{q}_i  \rightarrow \tilde{g} \bar{\tilde{q}}  \,, \nonumber\\  
  gg, \, q_i \bar{q}_i  & \rightarrow& \tilde{g} \tilde{g}  \,,   
\end{eqnarray}  
where $i,\, j=u, \, d, \, s, \, c, \, b$.  Flavour indices of squarks have
been suppressed. For the light-flavour squarks a common mass $m_{\tilde{q}}$
will be assumed.  The predictions for the cross sections presented below always
include a sum over the contributions of the ten light-flavour squarks
($\tilde{u}_{L/R}, \, \tilde{d}_{L/R}, \, \tilde{c}_{L/R}, \, \tilde{s}_{L/R},
\, \tilde{b}_{L/R}$).  The partonic cross sections for squark-anti-squark and
squark-squark production differ for equal and unequal initial-state 
(anti-) quarks, but otherwise do not depend on the individual quark flavours. 
Therefore it is possible to express the cross section~\eqref{eq:sigma-had} 
in terms of diagonal and off-diagonal flavour-summed parton luminosities.
  
In this paper we consider higher-order corrections to partonic  
channels where the sparticle pair is dominantly produced with vanishing 
orbital momentum (i.e.\ in an $S$-wave), with a Born cross section 
$\hat \sigma\propto \beta$ in the threshold limit $\beta\to 0$.    
For the purpose of resummation, the partonic cross section  
$\hat{\sigma}_{p p'}$ is decomposed into contributions of definite   
colour and spin of the final-state system. With regard to colour, the product 
of the $SU(3)$ representations $r$ and $r'$ of the initial state particles 
($R$ and $R'$ of the final state particles)  is decomposed into irreducible 
representations
\begin{equation}  
\label{eq:irreducible}  
  r\otimes r' =\sum_{\alpha} r_\alpha\,,\qquad  
R\otimes R'=\sum_{R_\alpha} R_{\alpha}\,.  
\end{equation}  
For squark and gluino production the relevant decompositions are  
\begin{equation}  
\label{eq:susy-reps}  
  \begin{aligned}  
3\otimes \bar 3&=1\oplus 8\,,\\  
 3\otimes 3&= \bar 3 \oplus 6\,,\\   
3\otimes 8&=3\oplus \bar 6 \oplus 15\,,\\  
8 \otimes 8&=1 \oplus 8_s \oplus 8_a \oplus 10  
  \oplus \overline{10} \oplus 27\,.  
  \end{aligned}  
\end{equation}  
The production cross sections can be decomposed into a colour basis characterized by pairs of representations, $P_i=(r_{\alpha}, R_{\beta})$ with equivalent  initial- and final-state representations, $r_\alpha\sim R_\beta$.  
Basis tensors for the pairs $P_i$ can be constructed in terms of Clebsch-Gordan coefficients~\cite{Beneke:2009rj}.  
The colour and spin quantum numbers  
resulting in $S$-wave sparticle production have been classified e.g.\  
in~\cite{Beenakker:2013mva}, see also~\cite{Kauth:2011vg} for gluino  
pair production. The results are collected in  
Table~\ref{tab:s-wave}.   
  
The higher-order corrections are written in terms of scaling 
functions $f^{(n)}_{pp'}$  as  
\begin{equation} \label{eq:partonic}  
\hat{\sigma}_{p p'}= \sum_i\sum_{S=|s-s'|}^{s+s'} 
\hat{\sigma}^{(0),S}_{p p', i} \left[1+  
\sum_{n=1}^\infty  
 \left(\frac{\alpha_s}{4 \pi}\right)^{\!n} f^{(n),S}_{p p', i}  \right]\, .  
\end{equation}  
Here the sum over $i$ runs over the colour basis defined by the pairs 
$P_i$, while $s$ ($s'$) is the spin of the sparticle $\tilde s$ 
($\tilde s'$) and $S$ the total spin of the sparticle pair.

%%%%%%%%%%%%%%%%%%%%%%%%%%%%%%%%%%%%%%%%%%%%%%%%%%%%%%%%%%%%%%%%%%%%%%%%%%%  
\begin{table}[t]  
\centering  
\begin{tabular}{c||c|c|c|c}   
$\tilde s\tilde s'$ &$pp'$  &$(r_\alpha,R_\beta)$ & $S$ & Comments\\ 
\hline\hline   
&&&&\\[-.6cm]  
$\tilde q\bar{\tilde q}$ & $q\bar q$   & $(1,1)$, $(8,8)$ & $0$ \\  
& $gg$  &$(1,1)$, $(8_s,8)$  &$0$ &\\\hline  
$\tilde q_i\tilde q_j $& $qq$  & $(\bar 3,\bar 3)$  &$0$ & $i\neq j$ only\\    
&& $(6,6)$ & $0$\\ \hline  
       $\tilde q\tilde g$& $qg$  & $(3,3)$, $(\bar 6,\bar 6)$, $(15,15)$ & $\frac{1}{2}$ &\\\hline  
         $\tilde g\tilde g$  & $q\bar q$  & $(8,8_a)$ & $1$ & \\  
         & $gg$ &$(1,1)$, $(8_s,8_s)$, $(27,27)$  & $0$&  
\end{tabular}  
\caption{Spin and colour quantum numbers leading to $S$-wave production of 
squarks and gluinos.}  
\label{tab:s-wave}  
\end{table}  
%%%%%%%%%%%%%%%%%%%%%%%%%%%%%%%%%%%%%%%%%%%%%%%%%%%%%%%%%%%%%%%%%%%%%%%%%%%  
  
The colour-separated Born cross sections  $\hat{\sigma}^{(0)}_{p p',i}$   
for squark and gluino production  are available in \cite{Kulesza:2009kq,Beenakker:2009ha,Beenakker:2010nq}.   
The colour-averaged NLO scaling functions were computed in~\cite{Beenakker:1996ch} for degenerate light-flavour squark masses and have been implemented in the computer program \texttt{PROSPINO}~\cite{Beenakker:1996ed}. For general squark spectra, the NLO corrections have been computed recently~\cite{GoncalvesNetto:2012yt,Gavin:2013kga}.  
An approximation of the NNLO scaling functions consisting of all terms that are enhanced in the limit $\beta\to 0$ has been given in~\cite{Beneke:2009ye}, up to an additional $\alpha_s^2\ln\beta$ term that has been calculated for the case of top-quark production in~\cite{Baernreuther:2013caa}. In Section~\ref{sec:potential} we derive the generalization of this contribution for the production of squarks and gluinos.  
  
%%%%%%%%%%%%%%%%%%%%%%%%%%%%%%%%%%%%%%%%%%%%%%%%%%%%%%%%%%%%%%%%%%%%%%%%%%%%%  
\subsection{Resummation formula}  
  
Up to NNLL accuracy,   
the partonic production cross sections for the processes~\eqref{eq:processes} factorize  in the threshold limit $\beta\to 0$   into spin- and colour-dependent  
hard and Coulomb functions $H^{S}_i$ and  $J^S_{R_\alpha}$  
and a soft function $W^{R_\alpha}$ depending only on the total colour charge $R_\alpha$ of the final-state particles~\cite{Beneke:2009rj,Beneke:2010da}:  
\begin{equation}  
\label{eq:fact}  
  \hat\sigma_{pp'}(\hat s,\mu_f)  
= \sum_{i}\sum_{S=|s-s'|}^{s+s'} H^{S}_{i}(m_{\tilde q},m_{\tilde g},\mu_f)  
\;\int d \omega\;  
J^S_{R_\alpha}(E-\frac{\omega}{2})\,  
W_i^{R_\alpha}(\omega,\mu_f)\, .  
\end{equation}  
Here  $E=\sqrt{\hat s}-2 \mbar$ is the partonic centre-of-mass energy measured from threshold.   
The hard function encodes the  
partonic hard-scattering processes and is related to squared on-shell  
scattering amplitudes at threshold. The  
potential function is defined in terms of non-relativistic fields for the sparticles whose interactions are described in potential non-relativistic QCD~(PNRQCD). Solving the Schr\"odinger equation in PNRQCD allows to sum the Coulomb corrections to all orders.   
The soft function is defined in terms of soft Wilson lines and contains the threshold logarithms.  The convolution of the soft- and potential  
functions accounts for the energy loss of the squark/gluino system due  
to soft gluons with energy of the order $\mbar\beta^2$.  
For the colour basis based on the pairs of representations $P_i$  
constructed in~\cite{Beneke:2009rj}, the soft function is diagonal in  
colour space and identical to that of a simpler two-to-one scattering  
process where a single heavy particle with colour charge $R_\alpha$ is  
produced from the two incoming partons. This basis has been assumed in  
writing~\eqref{eq:fact}.  Only production channels with an $S$-wave  
contribution will be taken into account in~\eqref{eq:fact}.   
It can be seen from Table~\ref{tab:s-wave} that only a single spin  
quantum number contributes for the threshold production for a given partonic  
colour channel. In practice the spin sum in~(\ref{eq:partonic})  
therefore collapses to a single term, so the sum over $S$ and the spin label on the hard function will  
be suppressed in the following.  
  
Resummation of threshold logarithms is performed by evolving the soft  
function from a soft scale $\mu_s\sim \mbar\beta^2$ to a hard-scattering  
scale $\mu_f\sim \mbar$ using a renormalization-group equation.  The anomalous 
dimensions required for NNLL resummation are collected 
in~\cite{Beneke:2009rj}. The hard  
function is evolved from a scale $\mu_h\sim 2\mbar$ to $\mu_f$. In the 
momentum-space  formalism~\cite{Becher:2006nr,Becher:2007ty}   
the resummed cross section can be written as \cite{Beneke:2010da}
\begin{equation}  
  \label{eq:resum-sigma}  
\begin{aligned}  
\hat\sigma^{\text{res}}_{pp'}(\hat s,\mu_f)=&  
\sum_{i}  
H_{i}(m_{\tilde q},m_{\tilde g},\mu_h)\,  
U_{R_\alpha}(\mu_h,\mu_s,\mu_f)  
\left(\frac{2\mbar}{\mu_s}\right)^{-2\eta} \\  
&\times  
\tilde{s}_i^{R_\alpha}(\partial_\eta,\mu_s)  
\frac{e^{-2 \gamma_E \eta}}{\Gamma(2 \eta)}\,  
\int_0^\infty \!\!\!\!\! d \omega   
\;\frac{ J^S_{R_{\alpha}}(\mbar\beta^2-\tfrac{\omega}{2})}{\omega}   
\left(\frac{\omega}{\mu_s}\right)^{2 \eta}.  
\end{aligned}  
\end{equation}  
Here the energy variable in the argument of the potential function has been
expanded near threshold which yields the non-relativistic expression $E=
\mbar\beta^2$. This defines our default implementation. The derivation of the
NLO potential function required at NNLL accuracy is the subject of
Section~\ref{sec:potential} and the result is given in~\eqref{JRal} below.
The quantity $\tilde s_i^{R_\alpha}$ is the Laplace transform of the soft
function. For NNLL resummation, the NLO soft function~\cite{Beneke:2009rj} is
required which reads
\begin{align}  
 \tilde  s_i^{R_\alpha}(\rho,\mu)&=\int_{0}^{\infty} d \omega e^{-s \omega}  
\,  W_i^{R_\alpha}(\omega,\mu) \nonumber\\  
&= 1 +\frac{\alpha_s}{4\pi}   
\left[\left(C_r+C_{r'}\right)\left(   \rho^2+\frac{\pi^2}{6}\right)  
- 2C_{R_\alpha}\left( \rho-2\right)  \right]+\mathcal{O}(\alpha_s^2),  
\label{eq:nlo-soft}
\end{align}  
with $s = 1/(e^{\gamma_E} \mu e^{\rho/2})$.  After carrying out the
differentiations with respect to $\eta$ in~\eqref{eq:resum-sigma}, this
variable is identified with a resummation function which contains single
logarithms, $\eta=\frac{2\alpha_s}{\pi}(C_r+C_{r'})\ln(\mu_s/\mu_f)+\dots$,
while the resummation function $U_i$ sums the Sudakov double logarithms
$\alpha_s\ln^2\frac{\mu_h}{\mu_f}$ and
$\alpha_s\ln^2\frac{\mu_s}{\mu_f}$. The precise definitions of these
functions for the case of heavy-particle pair production are given
in~\cite{Beneke:2010da} and the expansions required for NNLL accuracy can be
found in~\cite{Becher:2007ty}. For $\mu_s<\mu_f$ the function $\eta$ is
negative and the factor $\omega^{2\eta-1}$ in the resummed cross
section~\eqref{eq:resum-sigma} has to be understood in the distributional
sense, as discussed in detail in~\cite{Beneke:2011mq}. The 
prescription for the choice of the soft scale is detailed in
Section~\ref{sec:scales}.
  
\subsubsection{Hard functions}  
The perturbative expansion of the hard function in the resummation  
formula~\eqref{eq:resum-sigma} in the $\overline{\rm MS}$ scheme can be 
written as  
\begin{equation}  
\label{eq:hard-def}  
H_{i}(m_{\tilde q},m_{\tilde g},\mu)=
H^{(0)}_{i}(m_{\tilde q},m_{\tilde g},\mu)  
 \left[1+\sum_n   
 \left(\frac{\alpha_s(\mu)}{4\pi}\right)^{\!n} 
 h^{(n)}_{i}(m_{\tilde q},m_{\tilde g},\mu)\right],  
\end{equation}  
where for NNLL resummation the one-loop coefficients $h^{(1)}_{i}$  
are required.  
  
The leading-order hard function $H^{(0)}_{i}$ is related to the threshold limit of the Born cross section for a given colour channel according to~\cite{Beneke:2010da}  
\begin{equation}  
\label{eq:sigma-hard}  
  \hat\sigma_{pp'}^{(0)R_\alpha}(\hat s)  
\underset{\hat s\to 4\mbar^2}{=}  
 \frac{(m_{\tilde s}m_{\tilde s'})^{3/2}}{\mbar}\,\frac{\beta}{2\pi} H^{(0)}_{i} + \mathcal{O}(\beta^3)\, .  
\end{equation}  
In our numerical implementation, we define the leading-order hard functions  
$H^{(0)}_{i}$ in terms of the exact Born-cross sections, instead of  the leading term in the threshold limit, which is seen to improve the accuracy of the threshold approximation in some cases, but not in a systematic fashion.  
However, the hard function for a given production and colour channel  
is set to zero if there is no $S$-wave contribution to the Born cross  
section at threshold, even if the full Born cross section for this  
channel is non-vanishing. This affects the sub-process $q\bar  
q\to\tilde g\tilde g$ in the singlet and symmetric octet channels,  
the sub-processes $gg\to\tilde g\tilde g$ and $gg\to \tilde q\bar{\tilde q} $ in the  
anti-symmetric octet channel, as  well as  $q_iq_i\to \tilde q_i\tilde q_i$ in  
the triplet channel, see Table~\ref{tab:s-wave}.  
  
A prescription to compute the one-loop hard functions from on-shell  
Born and one-loop amplitudes at threshold has been given  
in~\cite{Beneke:2010da}.  Alternatively, the one-loop coefficient can  
be read off from the constant term in the threshold expansion of the  
total NLO cross section given in~\eqref{eq:NLOapprox}.  This allows to  
extract the one-loop hard functions from recent computations of the  
corresponding matching coefficients in the Mellin-space approach to  
threshold resummation~\cite{Beenakker:2011sf,Beenakker:2013mva}, which  
are defined as the constant term in the Mellin-transformed one-loop  
cross section in the threshold limit (for gluino-pair production, see  
also~\cite{Kauth:2011vg,Langenfeld:2012ti}).   
From the Mellin transformation of the NLO threshold cross section in  
momentum space~\eqref{eq:NLOapprox}, we obtain the relation of  the one-loop  
hard coefficients $h_i^{(1)}$ to  the matching coefficients  
$\mathcal{C}^{(1)}_{pp'\to \tilde s\tilde s', I}$ in the notation  
of~\cite{Beenakker:2011sf,Beenakker:2013mva}  
\begin{equation}  
\label{eq:rel-hard-mellin}  
\begin{aligned}  
h_i^{(1)}(m_{\tilde q},m_{\tilde g},\mu)=&   
-4(C_r+C_{r'}) \left (\ln^2\left(\frac{2\mbar}{\mu e^{\gamma_E}}\right)+
\frac{\pi^2}{24}\,\right)\\  
&+  4 C_{R_\alpha}  \left(\ln\left(\frac{2\mbar}{\mu e^{\gamma_E}}\right)
-1\right)  
+4 \,\mathcal{C}^{(1)}_{pp'\to \tilde s\tilde s', I}
(m_{\tilde q},m_{\tilde g},\mu),   
\end{aligned}  
\end{equation}  
where $I$ is the label of the colour basis tensors used in~\cite{Beenakker:2013mva} that correspond to the basis elements $P_i$ in our notation.  
  
In addition to the dependence on the scale $\mu$ and the squark and  
gluino masses as indicated in~\eqref{eq:rel-hard-mellin}, the one-loop  
hard functions in SQCD depend as well on the top-quark mass, with all  
other quarks treated as massless.  Numerical results for the  
coefficients $\mathcal{C}^{(1)}_{pp'\to \tilde s\tilde s', I}$ have  
been plotted in~\cite{Beenakker:2011sf,Beenakker:2013mva}. In the case  
of gluino-pair production and squark-gluino production, the hard  
functions become singular for $m_{\tilde g}=m_{\tilde q}+m_t$ when the  
on-shell decay-channel $\tilde g\to \tilde t t$ opens up.\footnote{Note 
that in the NLO calculations of~\cite{Beenakker:1996ch,Beenakker:2013mva} 
virtual top squarks are treated as mass-degenerate with the light-flavour 
squarks.} This singularity is not physical and arises from neglecting the 
gluino decay width.  In addition, for gluino-pair production from a  
quark-antiquark initial state, the threshold limit of the Born hard  
function goes to zero for $m_{\tilde q}=m_{\tilde g}$, so that the  
$S$-wave contribution to this channel vanishes in this special point  
of parameter space.  As a result, the relative NLO corrections given  
by the one-loop hard coefficient diverge.  Since we only apply  
resummation to the $S$-wave production channel, we set the resummed  
contribution of the quark-antiquark initial state to zero for $  
m_{\tilde q}=m_{\tilde g}$, while it is included in fixed-order at NLO  
through the matching to PROSPINO.  In practice, this prescription is  
implemented by using the threshold limit of the Born hard function  
$H^{(0)}$ for the subprocess $q\bar q\to \tilde g\tilde g$ for $0.9<  
m_{\tilde q}/m_{\tilde g}<1.1$.  The numerical effect of the precise  
choice of this interval is negligible.  
  
\subsection{Potential effects}  
\label{sec:potential} 
 
In the framework of~\cite{Beneke:2009rj,Beneke:2010da}, the  
non-relativistic sparticles are described by the Lagrangian of  
potential non-relativistic SQCD~(PNRSQCD). To the order relevant  
for NNLL resummation, the Lagrangian reads\footnote{Note that the sign  
of the potential term in~\cite{Beneke:2010da} is incorrect, which, however,  
has no consequence for the results presented there. For the case of  
fermions, the sign here is consistent with~\cite{Beneke:2013jia} if  
the different conventions for antiparticles are taken into account.}  
\begin{equation}
\begin{aligned}
\label{eq:NRQCD}
\mathcal{L}_{\mbox{\tiny PNRSQCD}} = \,\,&
\psi^\dagger \left(i D_s^0+\frac{\vec{\partial}^2}{2 m_{\tilde{s}}}
+\frac{\vec{\partial}^4}{8 m_{\tilde{s}}^3}
\right) \psi
+\psi^{\prime\,\dagger} \left(i D_s^0+\frac{\vec{\partial}^2}{2 m_{\tilde{s}'}}
+\frac{\vec{\partial}^4}{8 m_{\tilde{s}'}^3}
\right) \psi'
\\
&-\frac{1}{2^{\delta_{\tilde{s}\tilde{s}'}}}\int d^3 \vec{r}\; 
V_{\{k\}}(\vec{r},\vec\partial)
\psi_{k_4}^{\prime\,\dagger}(x)
\psi_{k_3}^{\dagger}(x+\vec r\,) 
\psi_{k_1}(x+\vec r\,) 
\psi_{k_2}^\prime(x)\,.
\end{aligned}
\end{equation}
Here the fields $\psi_k^\dagger$ and $\psi_k^{\prime \,\dagger}$ are
non-relativistic fields which create the heavy sparticles $\tilde{s}$ and 
$\tilde{s}'$. The label $k$ collectively denotes the flavour, spin and 
colour quantum numbers of the non-relativistic field, 
$\psi_k=\psi_{n,a,\alpha}$, where Latin letters $n$ and $a$ are used 
for flavour and colour indices, respectively, while the 
Greek index $\alpha$ denotes the spin index of the field. For objects 
such as the potential, which depend on the labels of several fields, 
we employ a multi-index convention for the spin indices, 
$\{\alpha\} = \alpha_1\alpha_2\alpha_3\alpha_4$, and analogously for the 
colour ($\{a\}$), flavour ($\{n\}$), and collective ($\{k\}$) index. 
The soft gluon field couples to the non-relativistic
sparticles through the soft covariant derivative $iD_s^0\psi=(i\partial^0 +g_s
{\bf T}^{(R)a}A^{a0})\psi$, where $ {\bf T}^{(R)a}$ are the $SU(3)$ generators
in the representation $R$.  Note that a factor $1/2$ appears in the potential
in the case of identical sparticle species, where we treat particles as
identical that belong to the same spin and $SU(3)$ representation, and 
species (that is, the ten light-flavour squarks are treated as
identical particles with an index $n$ denoting flavour and the helicity
label).
  
For NNLL accuracy, higher-order potential effects beyond the leading  
Coulomb potential have to be taken into account,  
see~\cite{Beneke:2013jia} for a detailed discussion in the PNRQCD  
formalism used here.  The relevant potentials are given by the NLO  
Coulomb potential, the $1/m^2$ corrections to the tree-level  
potential, the one-loop $1/m$ potential, and the so-called annihilation  
contributions,  
\begin{equation}  
\begin{aligned}
V_{\{k\}}&={\bf T}^{(R)a}_{a_3a_1} {\bf T}^{(R')a}_{a_4a_2}  
\left[V_{\mathrm{C}}  \,\delta_{\alpha_3\alpha_1}\delta_{\alpha_4\alpha_2}  
+\delta_{1/m^2} V_{\{\alpha\}}  
\right]\delta_{n_3n_1}\delta_{n_4n_2} \\
&\quad+ \delta_{1/m} V_{\{a\}}  
\delta_{\alpha_3\alpha_1}\delta_{\alpha_4\alpha_2}
\delta_{n_3n_1}\delta_{n_4n_2}+  
\delta_{\mathrm{ann}} V_{\{k\}}\,.  
\end{aligned}
\label{eq::potential}  
\end{equation}  
Note that due to (\ref{eq:count}), ${\cal O}(\beta)$ and 
${\cal O}(\alpha_s\beta, \beta^2)$ 
suppressed potentials appear here on the same footing, if the latter 
generate a logarithm of $\beta$. 
All contributions to the potential apart from the annihilation
contribution are flavour-independent, while only the $1/m^2$ potential
and the annihilation contribution are spin-dependent.  
Following~\cite{Beneke:2009rj,Beneke:2010da}, we perform a projection  
of the potential on states with definite colour charge and spin of the  
heavy particle system by introducing projectors $P^{R_\alpha}_{\{a\}}$  
and $\Pi^S_{\{\alpha\}}$ on colour and spin space, respectively.  The  
colour projectors can be written in terms of Clebsch-Gordan  
coefficients for the combination of the representations $R$ and $R'$  
into the irreducible representation $R_\alpha$,  
\begin{equation}  
\label{eq:project-clebsch}  
P^{R_\alpha}_{a_1a_2a_3a_4}=C^{R_\alpha\ast}_{A a_1a_2}    
C^{R_\alpha}_{A a_3a_4}\,,  
\end{equation}  
where the index $A$ is the colour index for the irreducible  
representation $R_\alpha$.

Following the reasoning of Appendix A of~\cite{Beneke:2009rj}, gauge 
invariance  implies that the  potential can be expanded in terms of the 
colour projectors~\eqref{eq:project-clebsch}\footnote{Strictly speaking these 
arguments imply that the potential can be written in the form 
$ V_{\{a\}}= \sum_I V^{I} C^{R_\alpha *}_{Aa_3a_4}C^{R_\beta}_{Aa_1a_2}$  
where the sum is over pairs $P_I=(R_\alpha,R_\beta)$ of equivalent 
representations $R_\alpha\sim R_\beta$. In squark-gluino production, the only case where equivalent but non-identical representations appear is the production of gluino pairs, that can be in an $8_s$ or $8_a$ state. However, for a given partonic initial state, only one of the two channels appears (see Table~\ref{tab:s-wave}), so in practice it is sufficient to consider the case where the two representations are identical and the decomposition assumes the form~\eqref{eq:Vdecomp}.  
}  
\begin{equation}  
V_{\{k\}}= \sum_{R_\alpha} V_{\{n,\alpha\}}^{R_\alpha} 
P^{R_\alpha}_{a_3a_4a_1a_2}\,.  
\label{eq:Vdecomp}
\end{equation}  
We will also only require potentials which allow for an analogous 
decomposition in spin space. The potential term in the Lagrangian then 
assumes the form  
\begin{eqnarray}  
\label{eq:decompose-v}
V_{\{k\}}\,\psi_{k_4}^{\prime\,\dagger}  
\psi_{k_3}^{\dagger} \psi_{k_1} \psi_{k_2}^\prime  
=\sum_{R_\alpha,S} V^{R_\alpha,S}_{\{n\}}  
\left[(\psi\otimes\psi')^{R_\alpha,S}_{n_3n_4}\right]^\dagger  
(\psi\otimes\psi')^{R_\alpha,S}_{n_1n_2}  
\\[-0.7cm]
\nonumber
\end{eqnarray}  
in tensor-product notation  
$(\psi\otimes\psi')_{12}(t,\vec r,\vec R)=
\psi^{(0)}_{1}(\vec  R+\tfrac{\vec r}{2}) 
\psi^{\prime \,(0)}_{2}(\vec R-\tfrac{\vec r}{2}) $.  
  
As shown in~\cite{Beneke:2010da} for the case of the Coulomb potential, the 
interaction of the non-relativistic particles with soft gluons can be 
eliminated from the PNR(S)QCD Lagrangian through a field redefinition.  
It can be seen that the same transformation also decouples soft gluons from a  
general gauge invariant potential.   
Therefore the fields in the Lagrangian~\eqref{eq:NRQCD} can be replaced by the
decoupled fields $\psi^{(0)}$ and the covariant derivatives can be replaced by
ordinary derivatives.  This decoupling holds to all orders in the strong
coupling but at leading power in the non-relativistic expansion in
$\beta$. Non-decoupling effects appear at $\mathcal{O}(\beta)$ through the
chromo-electric interaction, but do not contribute NNLL corrections to
the total cross section~\cite{Beneke:2010da}.

The potential function is defined as the correlation function of the decoupled
potential fields, 
\begin{equation}
  J_{\{k\}}(q)=\sum_{R_\alpha,S }P^{R_{\alpha}}_{a_3a_4a_1a_2}\;
\Pi^{S}_{\alpha_3\alpha_4\alpha_1\alpha_2}
\; J^{S}_{R_{\alpha},\{n\}}(q)\,.
\end{equation}
For identical bosonic (fermionic) sparticles, the potential function satisfies
the symmetry (antisymmetry) property
\begin{equation}
  J_{1234}=\pm J_{2134}=\pm J_{1243} 
\end{equation}
that implies the symmetry properties of the colour and spin-projected
potential function $J^{S}_{R_{\alpha},\{n\}}$ together with the symmetry or
antisymmetry of the colour and spin representations.
Since the Coulomb potential, the one-loop $1/m$ and the tree-level $1/m^2$
potentials are flavour-independent, the flavour structure can
be neglected in all contributions apart from the annihilation contribution,
which will be discussed in Section~\ref{sec:annihilate}.

The LO potential function, which resums all corrections of the form
$(\alpha_s/\beta)^n$, is given by the imaginary part of the zero-distance
Green function of the Schr\"odinger equation with the leading Coulomb potential,
i.e.\ the $\mathcal{O}(\alpha_s/\beta)$ contribution to~\eqref{delV}, 
\begin{equation}  
J^{S,(0)}_{R_\alpha}(E)=2 \,\mbox{Im} \left[\,  
G^{(0)}_{R_\alpha}(0,0;E)  \right]\, .  
\label{JRal0}  
\end{equation}  
The explicit  expression can  
be obtained by the simple replacement $m_t\to 2\mred$ from the  
corresponding result for top-pair production quoted e.g.\ in  
Eq.~(A.1) of~\cite{Beneke:2011mq}.
For the NNLL prediction, the NLO contributions to the
potential~\eqref{eq::potential}  are taken into account
perturbatively, 
\begin{equation}
\label{eq:dG-nlo}
 \delta  G^{(1), S }_{R_\alpha,\{n\}}(0,0,E)=
\int d^3 z\,  G^{(0)}_{R_\alpha }(0,\vec z,E)
\,(i\delta V^{R_\alpha,S}_{\{n\}}(\vec z\,))\,
i G^{(0)}_{R_\alpha}(\vec z,0,E)\,,
\end{equation}
where it was used that all NLO potentials are diagonal with respect to the
colour representations and that the leading Coulomb Green function is
spin-independent.  It is not necessary to (anti-)symmetrize the Green 
function with respect to the flavour indices. The contribution to the 
cross section automatically inherits the  correct symmetry properties 
from those of the potential and the hard function. This allows us to omit 
the flavour indices of $G^{(0)}_{R_\alpha}$.
The solution to~\eqref{eq:dG-nlo} for the potential~\eqref{eq::potential} is
given in Section~\ref{sec:nlo-pot}.

%%%%%%%%%%%%%%%%%%%%%  
\subsubsection{Coulomb and  non-Coulomb potential terms}  
\label{sec:non-C}  
  
In momentum space, the colour-projected Coulomb potential up to NLO reads  
\begin{equation}  
\tilde{V}^{R_\alpha}_{\mathrm{C}}(\bff{p},\bff{q}) =   
\frac{4\pi D_{R_\alpha}\alpha_s(\mu)}{\bff{q}^2} 
\left[1+ \frac{\alpha_s(\mu)}{4\pi}   
\left(a_1-\beta_0\ln\frac{\bff{q}^2}{\mu^2}\right)+\dots\right],  
\label{delV}  
\end{equation}  
where $\beta_0= \frac{11}{3} C_A-\frac{4}{3} n_l T_f$ is the one-loop 
beta-function coefficient, and $a_1  
=\frac{31}{9} C_A-\frac{20}{9} n_l T_f$.  The coefficient $D_{R_\alpha}$ of 
the Coulomb potential for a pair of heavy particles in the  
$SU(3)$ representations $R$, $R'$ in the irreducible product representation 
$R_\alpha$ is given in terms of the quadratic Casimir operators for the 
various representations by
\begin{equation}  
\label{eq:dralpha}  
  D_{R_\alpha}=\frac{1}{2}(C_{R_\alpha}-C_R-C_{R'})\,,  
\end{equation}  
where negative values correspond to an attractive Coulomb potential,  
positive values to a repulsive one.  The numerical values for the  
representations relevant for squark and gluino production can be found  
in~\cite{Kats:2009bv,Beneke:2010da}.  
 
The following potentials are all suppressed by two powers of velocity 
in the non-relativistic expansion, but have to be considered, since, 
contrary to the Coulomb potential, they generate logarithms of $\beta$ 
not related to the running coupling. At the next order in $1/m$,  
the colour-projected one-loop potential of order $m^{-1}$   
in $D=4-2\epsilon$ dimensions is given by  
\begin{eqnarray}  
\delta_{1/m} \tilde{V}^{R_\alpha}(\qvec) &=& 
\frac{\pi^2\alpha_s^2 D_{R_\alpha}}
{2\mred}  \frac{\mu^{2\epsilon}}{|\qvec|^{1+2\ep}}
  \frac{e^{\epsilon\gamma_E}\Gamma^2(\frac{1}{2}-\ep)\, 
    \Gamma(\frac{1}{2}+\ep)}{\pi^{3/2}\, \Gamma(1-2\ep)} 
\nonumber\\
&&\times \left(  
    \frac{D_{R_\alpha}}{2} (1-2\ep) \frac{2\mred}{\mbar} + C_A (1-\ep)  
    \right).  
\label{eq:V1overm}
\end{eqnarray}  
  
We obtain the $1/m^2$ potential at tree-level for squark-and gluino 
production from the generalization of the spin-dependent non-Coulomb  
terms for top-quark production~\cite{Beneke:2009ye,Beneke:2011mq} to  
squarks and gluinos (the result has been quoted already  
in~\cite{Beneke:2013opa}). This derivation is analogous to the one for
the $1/m^2$ potential for threshold production of top-quark
pairs. Details for the latter can be found in~\cite{Beneke:2013jia}.
The general expression is   
\begin{eqnarray}  
\delta_{1/m^2} \tilde{V}^{R_\alpha}(\bff{p},\bff{q})&& =  
\frac{4\pi D_{R}\alpha_s(\mu^2)}{\bff{q}^2}   
\left[  
\frac{\pvec^2}{m_{\tilde s}m_{\tilde{s}'}}   
 - \frac{\qvec^2}{8m_{\tilde s}^2m_{\tilde{s}'}^2} 
(2m_{\tilde s}m_{\tilde{s}'} +  
m_{\tilde s}^2\, c_2^{\tilde{s}'} + m_{\tilde{s}'}^2\, c_2^{\tilde s} )  
\right. \nonumber \\  
&&   
+ \,\frac{c_2^{\tilde s}c_2^{\tilde{s}'}}{16m_{\tilde s}m_{\tilde{s}'}} 
[\sigma^i,\sigma^j] q^j \otimes  
  [\sigma^i,\sigma^k] q^k +  
c_2^{\tilde s}\left( \frac{1}{8m_{\tilde s}^2} + 
\frac{1}{4m_{\tilde s}m_{\tilde{s}'}} \right)  
    [\sigma^i,\sigma^j] q^i p^j \otimes \unit 
\nonumber \\  
  &&  \left.  
 +\,c_2^{\tilde{s}'}   \left( \frac{1}{8m_{\tilde{s}'}^2} + 
\frac{1}{4m_{\tilde s}m_{\tilde{s}'}} \right)  
    \unit \otimes [\sigma^i,\sigma^j] q^i p^j  
  \right],  
\label{delV2}  
\end{eqnarray}  
with $\unit$ the $2\times2$ unit matrix in spin space, $\bff{q}=
\bff{p}^\prime-\bff{p}$, and $\bff{p}$ ($\bff{p}^\prime$) the in-going 
(out-going) three-momentum of the heavy particle in the scattering 
amplitude. The coefficient $c_2$ has the tree-level value zero (one) 
for scalar (fermionic) sparticles.  For scalars it also sets the corresponding 
spin-dependent terms to zero.
  
Projecting on the relevant spin states (see Section~4.5 
of~\cite{Beneke:2013jia}) and setting $D\to 4$, which is justified when one 
is only interested in the logarithmically enhanced term generated by the 
potential insertion, the non-Coulomb corrections can be cast in the form  
\begin{equation}  
   \delta_{1/m^2} \tilde{V}^{R_\alpha,S}(\bff{p},\bff{q})=  
  \frac{4\pi D_{R_\alpha}\alpha_s}{\qvec^2}  
  \Biggl[  \frac{\pvec^2}{m_{\tilde s}m_{\tilde{s}'}} + 
\frac{\qvec^2}{4\mred^2}\, \nus \Biggr],  
\end{equation}  
where the spin-dependent coefficient for the squark and gluino  
production processes is given by  
\begin{equation}  
\label{eq:nus}
  \begin{aligned}  
\nu_\mathrm{spin}(\tilde q\bar{\tilde{q}})=  
\nu_\mathrm{spin}(\tilde q\tilde q) &= -\frac{2\mred}{4\mbar} ,&  
\quad \nu_\mathrm{spin}^{{\displaystyle s}=\frac{1}{2}}(\tilde q\tilde g)  &=  
 \frac{1}{2}\left(\frac{m_{\tilde g}^2}{(m_{\tilde q}+m_{\tilde g})^2}-1\right),\\  
\nu_\mathrm{spin}^{S=0}(\tilde g\tilde g)&= 0  ,&  
\nu_\mathrm{spin}^{S=1}(\tilde g\tilde g)&=  -\frac{2}{3}.      
  \end{aligned}  
\end{equation}  
Together with the Coulomb potential~\eqref{delV} and the $1/m$
potential in~\eqref{eq:V1overm}, this result is the needed
generalization of Eq.~(4.95) of~\cite{Beneke:2013jia}, up to the
so-called annihilation contribution, which is derived in the next
section.
  
Using these results we can now determine the corresponding logarithm of 
$\beta$ in the NNLO cross section. For this purpose we use the known 
results for the NNLO Green function of a system of two particles with 
equal masses from\cite{Beneke:1999qg} (given explicitly 
in~\cite{Pineda:2006gx}) and generalize them to the case of unequal masses. 
In the following, we briefly outline this derivation, leaving a more
detailed description for a momentum independent potential 
to~Section~\ref{sec:annihilate}. It is straightforward to adapt the 
coefficients of the potentials in the expressions for the Green function to 
the more general case. Afterwards, the remaining mass dependence is due to 
the equation of motion and thus has to be identified with the reduced
mass. We then expand the expressions to order $\alpha_s^2$,
keeping only logarithms of $\beta$. Note that in addition to the $1/m$
and $1/m^2$ potentials discussed above, we also have to include the
kinetic energy correction $\pvec^4/(8m_{\tilde{s}}^3)$ from the terms
with a fourth power of the spatial derivative in~\eqref{eq:NRQCD}. The 
final result reads
\begin{eqnarray}  
  \Delta\hat\sigma_{pp',\mathrm{nC}}^{(2)R_\alpha,S}(\hat s,\mu_f) &=&
  \hat\sigma_{pp'}^{(0)R_\alpha,S}(\hat s) \, \alpha_s^2 \ln\beta \, 
  \left[ -D_{R_\alpha}\, b_1 - 2\, D_{R_\alpha}^2 \left( 1 + \nus +
    \frac{\mred}{2\mbar} \right) \right] 
\nonumber \\[0.1cm]
   &=&
  \hat\sigma_{pp'}^{(0)R_\alpha,S}(\hat s) \, \alpha_s^2 \ln\beta \, 
  \Big[ C_A D_{R_\alpha} - 2\, D_{R_\alpha}^2 \left( 1 + \nus \right)  
  \Big], 
\label{eq:nCcomplete}
\end{eqnarray}
where $b_1 = -C_A - D_{R_\alpha}\mred/\mbar$ is the 1-loop
coefficient of the $1/m$ potential, cf. (\ref{eq:V1overm}). Combining the contributions of the
$1/m$ and $1/m^2$ potential, we obtain the same expression as in
Eq.~(3) of~\cite{Beneke:2009ye}, which was derived for the equal
mass case. Remarkably, even for the case of unequal masses, the
process dependence is completely contained in the coefficient $\nus$
and the leading order cross section.

%%%%%%%%%%%%%%%%%%%%%%%%%%%%%%  
\subsubsection{Annihilation contributions}  
\label{sec:annihilate}  

We next derive the annihilation contribution
$\delta_{\mathrm{ann}}\tilde{V}_{\{k\}}$ to the
potential~\eqref{eq::potential} for the squark and gluino
pair-production processes as well as the resulting corrections to the
potential function~\eqref{JRal} and the threshold expansion of the
NNLO cross section. For the case of top-quark pair production this 
single-logarithmic correction of order $\alpha_s^2\ln\beta$ 
appears only in the $q\bar q$ partonic channel. It has been 
identified in~\cite{Baernreuther:2013caa} and was not included in 
the approximate NNLO cross section of~\cite{Beneke:2009ye}. 

\begin{figure}[t]  
\begin{center}  
\includegraphics[width=0.95\textwidth]{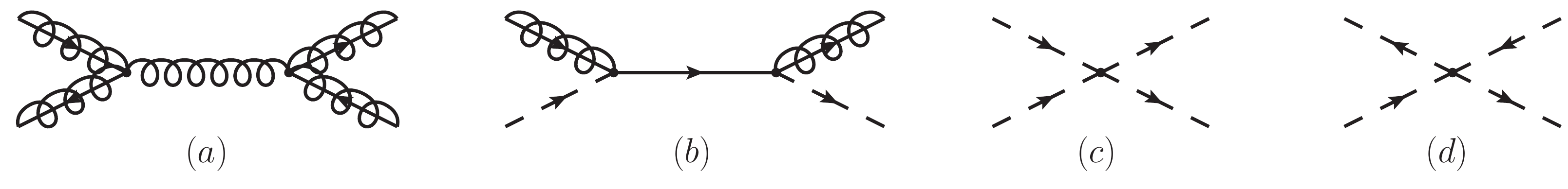}  
\end{center}  
\caption{\label{fig::ann} Tree-level contributions to the matching of  
four-field annihilation operators in NR(S)QCD. Dashed lines represent 
scalars, solid lines quarks and the solid-curly lines the gluino.}  
\end{figure}

The annihilation corrections arise from four-field operators in NRSQCD that
match onto a local contribution to the potential~\eqref{eq::potential}
in PNRSQCD and  contribute to the cross section at NNLO provided their
matching coefficients are generated at tree level.
The matching coefficients are obtained by equating
the EFT matrix element with an insertion of the potential to the
non-relativistic expansion of the matrix element of the two-to-two
sparticle scattering process
$\tilde s\tilde s'\to \tilde s\tilde s'$,
\begin{eqnarray}
\label{eq:match-v}
&& \frac{1}{2^{\delta_{\tilde{s}\tilde{s}'}}}
(-i) \delta_{\mathrm{ann}}\tilde{V}_{\{k\}}
\langle\tilde s_3\tilde s'_4|\psi_{k_4}^{\prime\,\dagger}\psi_{k_3}^{\dagger}
\psi_{k_1}\psi_{k_2}^\prime |\tilde s_1\tilde s'_2\rangle_{\mathrm{EFT}}
\nonumber\\
&&\hspace*{3cm}
=\,\frac{1}{4\sqrt{m_{\tilde s_1}m_{\tilde s_2'}
m_{\tilde s_3}m_{\tilde s_4'}}}\, 
 i \mathcal{M}(\tilde s_1\tilde s'_2\to\tilde s_3\tilde
 s'_4)\big|_{\hat s=4M^2}^{\mathrm{ann}}\,,\qquad
\end{eqnarray}
where the pre-factor on the right-hand side arises from the non-relativistic
normalization of the one-particle states.  
 As indicated by the superscript ``ann'', only the
contributions to the matrix element matching to a local four-fermion operator
must be taken into account. Also $t$-channel gluon exchange  
contributions are excluded since they are  assigned to the 
non-Coulomb scattering potential. The relevant tree-level diagrams for the 
various squark and gluino pair-production processes are shown in
Fig.~\ref{fig::ann}. 
Diagram $(a)$ is the typical diagram for fermion-antifermion annihilation 
through a gluon, which arises for the gluino-gluino process. Diagram
$(b)$ is the corresponding diagram for the squark-gluino process with an
s-channel quark and diagrams $(c)$ and $(d)$ are the scalar four-point 
interactions for the squark-squark and squark-antisquark processes, 
respectively. Note that the $s$-channel gluon annihilation diagram is 
$P$-wave suppressed for the squark-antisquark case so the only contribution 
comes from the four-squark vertex~$(d)$. Note that we shall assume that the 
difference of squark and gluino masses is sufficiently large,
$|m_{\tilde g}-m_{\tilde q}|> \mbar \beta^2$,  so that
annihilation contributions that change the sparticle species (e.g. $\tilde q
\bar{\tilde{q}}\to \tilde g\tilde g$ through $t$-channel quark 
exchange) do not lead to threshold-enhanced
contributions to the cross section.

%%%%%%%%%%%%%%%%%%%%%%%%%%%%%%%%%%%%%%%%%%%%%%%%%%%%%%%%%%%%%%%%%%%%%%%%%%
%%%%%%%%%%%%%%%%%%%%%%%%%%%%%%%%%%%%%%%%%%%%%%%%%%%%%%%%%%%%%%%%%%%%%%%%%%
 
As an example, we discuss the case of gluino-pair production in detail.  In
this case it is convenient to identify the operator $\psi'$ in the PNRSQCD
Lagrangian~\eqref{eq:NRQCD} and the EFT matrix element in the matching
condition~\eqref{eq:match-v} with the creation operator of the
charge-conjugate gluino field $\psi^{c\dagger}$.  The non-relativistic limit
of the annihilation contribution to the matrix element corresponding to
Fig.~\ref{fig::ann}$(a)$ is given by
\begin{equation}
\label{eq:m-gluino}
\begin{aligned}
\frac{1}{4m_{\tilde g}^2}
 i \mathcal{M}(\tilde g_1\tilde g_2\to\tilde g_3\tilde g_4)
\big|_{\hat s=4M^2}^{\text{ann}}
&=(-i) \frac{g_s^2}{4m_{\tilde g}^2}
F^{a}_{a_3a_4}F^{a}_{a_2a_1} (\eta^\dagger_2\sigma^i\xi_1) 
(\xi_3^\dagger\sigma^i\eta_4) \\
&= (-i) \frac{g_s^2}{4m_{\tilde g}^2}\, 2 N_c \, 
P^{(8_a)}_{\{a\}}\Pi^{S=1}_{\{\alpha\}}\,
\eta^\dagger_{\alpha_2}\xi_{\alpha_1} 
\xi_{\alpha_3}^\dagger \eta_{\alpha_4}
\end{aligned}
\end{equation}
with the generators of the adjoint representation $F^{a}_{a_1a_2}=i f^{a_1a
  a_2}$ and the two-component particle (antiparticle) spinors $\xi$ ($\eta)$.
In the second line, we have introduced the projection
operators on the   $8_a$ colour
representation and the spin-triplet, 
\begin{equation}
P^{(8_a)}_{\{a\}}=\frac{1}{N_c} F^{a}_{a_3a_4}F^a_{a_2a_1},
\qquad\qquad
\Pi^{S=1}_{\{\alpha\}}= \frac{1}{2}\sigma^i_{\alpha_3\alpha_4}
\sigma^i_{\alpha_2\alpha_1}\,.
\end{equation}

To evaluate the matrix element on the left-hand side
of~\eqref{eq:match-v}, note that the property of Majorana fermions in an S-wave
state (see e.g.~\cite{Beneke:2012tg})
\begin{equation}
\label{eq:majorana-symm}
  (\psi_i^\dagger \sigma^i \psi_j^c)=-
  (\psi_j^\dagger \sigma^i \psi_i^c)
\end{equation}
and the anti-symmetry of the $F^a_{a_ia_j}$ imply that all possible
four contractions of the external states with the field operators give an
identical contribution to the projection of the matrix element on the $8_a$,
$S=1$ state.  Taking the factor $1/2$ for identical particles
in~\eqref{eq:match-v} into account, the potential is therefore obtained by
multiplying the matrix element~\eqref{eq:m-gluino} by a symmetry factor of
$1/2$.  The final result for the  coefficients in the
decomposition~\eqref{eq:decompose-v}  of the annihilation potential reads
\begin{equation}  
  \delta_{\mathrm{ann}} \tilde{V}^{R_\alpha,S} =  
  \frac{\pi \alpha_s}{m_{\tilde g}^2} \left[  
    N_c\,  
    \delta_{R_\alpha,8}\, \delta_{S,1} \right] \,. \label{eq::Vqqbar}  
\end{equation}  
The corresponding result for quark-antiquark annihilation~\cite{Pineda:1998kj}
is obtained by changing the colour factor $N_c$ to $T_F=\frac{1}{2}$ due to the
normalization of the colour projector $P^{(8)}$ in the fundamental
representation and multiplying by a factor of two due to the absence of the
symmetry factor for Dirac fermions.

%%%%%%%%%%%%%%%%%%%%%%%%%%%%%%%%%%%%%%%%%%%%%%%%%%%%%%%%%%%%%%%%%%%%%%%%%%%%
\begin{table}[t]
  \begin{center}
    \begin{tabular}{l|c}
      $(\tilde s\tilde s')^{R_\alpha}_S$ &
       $A_{\{n\}}^{R_\alpha,S}$ \\[1ex] \hline\hline & \\[-2.8ex]
%%%%%%%%%%%%%%%%%%%%%%%%%%%%%%%%%%%%%%%
      $(\tilde q \bar{\tilde{q}}\,)^1$ & 
      $\displaystyle - D_1 \frac{\mbar}{2\mred}\, 
      X^{i_1}_{\lambda_3\lambda_1}X^{i_2}_{\lambda_2\lambda_4}$
      \\[2ex]
%%%%%%%%%%%%%%%%%%%%%%%%%%%%%%%%%%%%%%%%%
       $(\tilde q \bar{\tilde{q}}\,)^8$  & 
       $\displaystyle  \frac{\mbar}{2\mred}\left(T_F   
         X^{i_1}_{\lambda_1\lambda_2} X^{i_3}_{\lambda_3\lambda_4}
         - D_8 X^{i_1}_{\lambda_3\lambda_1}X^{i_2}_{\lambda_2\lambda_4}\right)$
       \\[2ex] \hline & \\[-2.5ex]
%%%%%%%%%%%%%%%%%%%%%%%%%%%%%%%%%%%%%%%

      $(\tilde q\tilde q)^{\bar 3}$  &
      $ \displaystyle D_{\bar 3}T_F\frac{\mbar}{2\mred}
      \left(  X^{i_1}_{\lambda_1\lambda_3} X^{i_2}_{\lambda_2\lambda_4}
       - X^{i_1}_{\lambda_1\lambda_4} X^{i_2}_{\lambda_2\lambda_3}
      \right) $  \\[2ex]
%%%%%%%%%%%%%%%%%%%%%%%%%%%%%%%%%%%%%%% 
      $(\tilde q\tilde q)^{6}$  &  
      $\displaystyle D_{6}T_F \frac{\mbar}{2\mred}
      \left(  X^{i_1}_{\lambda_1\lambda_3} X^{i_2}_{\lambda_2\lambda_4}
        +X^{i_1}_{\lambda_1\lambda_4} X^{i_2}_{\lambda_2\lambda_3}
      \right)$ \\[2ex] \hline & \\[-2.5ex]
%%%%%%%%%%%%%%%%%%%%%%%%%%%%%%%%%%%%%%%
      $(\tilde q \tilde g )^3_{\frac{1}{2}}$ & 
      $\displaystyle C_F \frac{m_{\tilde q}+m_{\tilde g}}{2m_{\tilde
          q}}$\\[0.7ex]
      \hline & \\[-2.5ex]
%%%%%%%%%%%%%%%%%%%%%%%%%%%%%%%%%%%%%%%
  $(\tilde g\tilde g)^{8_a}_1$ & 
      $\displaystyle N_c\,$
%%%%%%%%%%%%%%%%%%%%%%%%%%%%%%%%%%%%%%%
    \end{tabular}
  \end{center}
  \caption{\label{tab:A-ann}
    Non-vanishing values for the coefficients $A_{\{n\}}^{R_\alpha,S}$ of the
    annihilation potential~\eqref{eq:Vann} for the different squark and gluino
    production  processes for the colour state $R_\alpha$ and spin state
    $S$ (if applicable). 
    The helicity labels for the squark $i$ are denoted  by $\lambda_i$.
    The matrix appearing in the squark potentials is defined in
    Eq.~\eqref{eq:def-X}.
}
\end{table}
%%%%%%%%%%%%%%%%%%%%%%%%%%%%%%%%%%%%%%%%%%%%%%%%%%%%%%%%%%%%%%%%%%%%%%%%%%%%

The results for the remaining squark and gluino pair-production processes are
obtained in a similar way and collected in Table~\ref{tab:A-ann} in the form
of coefficients $A_{\{n\}}^{R_\alpha,S}$ defined by
\begin{equation}
\label{eq:Vann}
  \delta_{\mathrm{ann}} \tilde{V}_{\{n\}}^{R_\alpha, S}  =
\frac{\pi\alpha_s}{\mbar^2}\,
 A_{\{n\}}^{R_\alpha,S}\,.
\end{equation} 
The flavour labels are only required for the cases of squark-squark
and squark-antisquark scattering where the label $n_i$
denoting the ten light-flavour squark states is considered as a pair
$(i, \lambda_i)$ of a flavour label $i$ and a helicity label
 $\lambda_i=1$ ($2$) for left (right) squarks.  The
results are given with a general squark-mass dependence, although we
only require the equal-mass case for our numerical results.  We have
assumed vanishing squark mixing which implies that the matrix
$ X^{i}_{\lambda_i \lambda_j}$ appearing in the four-squark
vertex~\cite{Rosiek:1989rs} has the form
\begin{equation}
\label{eq:def-X}
 X^{i}_{\lambda_i \lambda_j}=(-1)^{\frac{\lambda_i+\lambda_j}{2}}\,
    \delta_{\lambda_i\lambda_j}\delta_{ij}.
\end{equation}

Since the annihilation potential~\eqref{eq:Vann} is momentum independent 
(proportional to $\delta^{(3)}(z)$ in coordinate space), 
the resulting Green function correction~\eqref{eq:dG-nlo} is
proportional to the square of the Coulomb Green function at the origin, 
\begin{equation}
  \begin{aligned}
  \delta_{\mathrm{ann}} G_{R_\alpha,\{n\}}^{(1)S}   
  & = -\frac{\pi \alpha_s}{\mbar^2} A_{\{n\}}^{R_\alpha,S}
   \left[ G_{R_\alpha}^{(0)}(0,0;E) \right]^2 \,.  
  \end{aligned}
\label{eq:dG-ann}
\end{equation}
Note that the squark-squark annihilation potential in
Table~\ref{tab:A-ann}  shares the (anti-)symmetry with respect to the
exchange of initial- or final-state flavours of the potential function
in the $\bar 3$ ($6$)
colour channel,  as assumed in~\eqref{eq:dG-nlo}.

The correction to the potential function $J_{R_\alpha}^S$ is given by  
twice the imaginary part of $\delta G_{R_\alpha}^S$,
\begin{equation}
\label{eq:dJ-ann}  
  \delta_{\mathrm{ann}} J_{R_\alpha,\{n\}}^S =  
  -\frac{4\pi\alpha_s}{\mbar^2} \, 
    A_{\{n\}}^{R_\alpha, S}\,
  \mathrm{Re} \left[ G_{R_\alpha}^{(0)}(0,0;E) \right]  
  \mathrm{Im} \left[ G_{R_\alpha}^{(0)}(0,0;E) \right] \,, 
\end{equation}  
with 
\begin{eqnarray}  
  \mathrm{Re} \left[ G_{R_\alpha}^{(0)}(0,0;E) \right] &=&  
  \frac{\mred^2\dra\alpha_s}{\pi} \ln\beta + \dots \,, \\  
  \mathrm{Im} \left[ G_{R_\alpha}^{(0)}(0,0;E) \right] &=&  
  \frac{\beta}{4\pi} \frac{(m_{\tilde s}m_{\tilde s'})^{3/2}}{\mbar} + {\cal O}(\alpha_s) \,,  
  \label{eq::G0im}  
\end{eqnarray}  
where the
ellipsis denotes non-logarithmic terms or terms of higher order in $\alpha_s$.

To obtain the annihilation correction to the cross section from the
factorization formula~\eqref{eq:fact}, the non-trivial flavour structure of
the hard production process must be taken into account by introducing a
flavour dependent hard function, $H_{i,\{n\}}^S$. This is schematically of
the form $H_{i,\{n\}}^S\sim C^S_{i,n_1n_2}C^{S*}_{i,n_3n_4}$, where the
matching coefficient $C^S_{i,n_in_j}$ is related to the amputated production
amplitude of the sparticle state $(\tilde s_{n_i}\tilde s_{n_j})^{R_\alpha}_S$
at threshold (see Eqs.~(2.60) and~(3.10) in~\cite{Beneke:2010da}).  The
product of the flavour-dependent hard function and the annihilation
contribution to the potential function therefore takes the interference of
the different production channels $\tilde s_{n_1}\tilde s_{n_2}$
and $\tilde s_{n_3}\tilde s_{n_4}$ into account, which are connected by a
rescattering through the annihilation potential $\delta_{\mathrm{ann}}
\tilde{V}_{\{n\}}$.  
The potential function  $J_{R_\alpha}^S(E)$ appearing in~\eqref{eq:fact}
is the flavour-averaged potential function defined by 
\begin{equation}
J_{R_\alpha}^S(E)= \frac{J_{R_\alpha,\{n\}}^S(E)H^{S(0)}_{i,\{n\}}}
{H^{S(0)}_i}\,,
\end{equation}
where $H^{(0)S}_i$ is the flavour-summed LO hard function that appears in the
LO cross section~\eqref{eq:sigma-hard}.

The NNLO annihilation correction to the cross
section in the colour and spin state $R_\alpha$ and $S$ is then given by
\begin{align}
  \Delta\hat\sigma_{pp',\mathrm{ann}}^{(2)R_\alpha,S}(\hat s,\mu_f)
&=  H^{S(0)}_{i} \, 
 \delta_{\mathrm{ann}} J_{R_\alpha}^S(E)|_{\mathcal{O}(\alpha_s^2)}\nonumber\\
  &=\hat\sigma_{pp'}^{(0)R_\alpha,S}(\hat s) \,
\alpha_s^2\ln\beta \,
\left(-  \frac{\dra}{2}\frac{4\mred^2}{\mbar^2}\nua\right),
\label{eq:dnnlo-ann}
\end{align}
 where it was used that only the leading-order soft function contributes to
 the $\mathcal{O}(\alpha_s^2\ln\beta)$ correction, which renders the
 convolution in~\eqref{eq:fact} trivial. 
The annihilation correction relative to the LO cross section 
has been defined in terms of the coefficient
\begin{equation}
\label{eq:nua}
  \nua= \frac{A_{\{n\}}^{R_\alpha,S} H^{S(0)}_{i,\{n\}}}{H^{S(0)}_i}\,,
\end{equation}
whose values  for all squark and gluino
pair-production processes are collected in Table~\ref{tab:nua}.

%%%%%%%%%%%%%%%%%%%%%%%%%%%%%%%%%%%%%%%%%%%%%%%%%%%%%%%%%%%%%%%%%%%%%%%%%%%%
\begin{table}[t]
  \begin{center}
    \begin{tabular}{l|c}
        $pp'\to (\tilde s\tilde s')^{R_\alpha}_S$ &  $\nua$ 
        \\[0.5ex] \hline\hline & \\[-2.5ex]
        %%%%%%%%%%%%%%%%%%%%%%%%%%%%%%%%%%%%%%% 
       $gg\to(\tilde{q}\bar{\tilde{q}}\,)^{R_\alpha}$ &
      $-\displaystyle  \dra \frac{\mbar}{2\mred}$
      \\[2ex]
      $q\bar{q}\to(\tilde{q}\bar{\tilde{q}}\,)^{R_\alpha}$ &
      $\phantom{+}\displaystyle \dra \frac{\mbar}{2\mred}$
  \\[2ex]    \hline & \\[-2.5ex]
%%%%%%%%%%%%%%%%%%%%%%%%%%%%%%%%%%%%%%%
      $q_i q_j\to (\tilde q_i\tilde q_j)^{R_\alpha}$ &
      $\displaystyle
      2\,T_F \dra\frac{\mbar}{2\mred}$   \\[2ex]    \hline & \\[-2.5ex]
%%%%%%%%%%%%%%%%%%%%%%%%%%%%%%%%%%%%%%%  
      $qg\to (\tilde q \tilde g )^3_{\frac{1}{2}}$ &
      $C_F \displaystyle  \frac{m_{\tilde q}+m_{\tilde g}}{2m_{\tilde q}}$
  \\[2ex]    \hline & \\[-2.5ex]
%%%%%%%%%%%%%%%%%%%%%%%%%%%%%%%%%%%%%%%  
 $q\bar q\to (\tilde g\tilde g)^{8_a}_1$ &
      $\displaystyle N_c  $ 
   \end{tabular}
  \end{center}
  \caption{\label{tab:nua}
    Non-vanishing values of the coefficient $\nua$ of the
    annihilation corrections for the different squark and gluino production 
    processes for the colour state $R_\alpha$ and spin state
    $S$ (if applicable).}
\end{table}
%%%%%%%%%%%%%%%%%%%%%%%%%%%%%%%%%%%%%%%%%%%%%%%%%%%%%%%%%%%%%%%%%%%%%%%%%%%%

The results for the squark-antisquark production process can be derived
from the potential coefficients given in Table~\ref{tab:A-ann} and the
definition~\eqref{eq:nua} using the fact that, in the absence of
flavour violation, the matching coefficients for the gluon initial
state are non-vanishing only for equal squark flavours and helicity
labels,
$gg\to \tilde q_{i\lambda_i}\bar{\tilde q}_{i\lambda_i}$. For the
quark-antiquark initial state only opposite helicity labels
contribute, while the flavours are fixed by the initial-state quarks,
$q_i\bar q_j\to \tilde q_{iL}\bar{\tilde q}_{jR}$ and
$q_i\bar q_j\to \tilde q_{iR}\bar{\tilde q}_{jL}$ (see
e.g. Ref.~\cite{Beneke:2010da}).  For squark-squark production, the
helicity labels of the two squarks agree,
$q_i q_j\to \tilde q_{iL}\tilde q_{jL}$ and
$q_i q_j\to \tilde q_{iR}\tilde q_{jR}$, while the symmetry properties
of the matching coefficients under exchange of the squarks follow from
the respective colour representation, i.e. the coefficients for the
$\bar 3$ (6) state are anti-symmetric (symmetric).

For the example of gluino pair-production we obtain the correction
\begin{equation}  
  \Delta \hat\sigma_{q\bar q,\mathrm{ann}}^{(2)R_\alpha=8,S=1}(\hat s)
  = \hat\sigma_{q\bar q}^{(0)R_\alpha=8}(\hat s) \, \alpha_s^2 \ln\beta  
  \left( -\frac{D_8 N_c}{2} \right) \,,  
\end{equation}  
which agrees with the one for top-quark pair production in Eq.~(4.15) 
of~\cite{Baernreuther:2013caa} after the appropriate changes of the
group-theory factors, $N_c\to T_F$ and $D_8\to \frac{1}{2}(C_A-2C_F)$, and 
multiplication by a factor of $2$ due to the Dirac nature of the top quark.
  
\subsubsection{NLO potential function}  
\label{sec:nlo-pot}
For resummation at NNLL accuracy, the NLO potential function has to be
inserted into the resummation formula~\eqref{eq:resum-sigma}.
It is given by the perturbative
solution~\eqref{eq:dG-nlo} of the Schr\"odinger equation with the
potential~\eqref{eq::potential} and can be written in the form
\begin{equation}  
J^S_{R_\alpha}(E)=2 \,\mbox{Im} \left[\,  
G^{(0)}_{R_\alpha}(0,0;E) \Delta^{R_\alpha, S}_{\rm nC}(E)+ 
G^{(1)}_{R_\alpha}(0,0;E) \right]\, .  
\label{JRal}  
\end{equation}  
The function $G^{(1)}_{R_\alpha}$ is obtained from one insertion of the NLO
Coulomb potential and includes all terms of the form $\alpha_s \times
(\alpha_s/\beta)^n$. Its explicit expression can be obtained by the simple
replacement $m_t\to 2\mred$ from the corresponding result for top-pair
production quoted in Eq.~(A.1) of~\cite{Beneke:2011mq}.  The factor
$\Delta^{R_\alpha,S}_{\rm nC}$ arises from an insertion of the 
one-loop $1/m$, the tree-level spin-dependent non-Coulomb, and the 
annihilation potentials. It is given in terms of the results of 
Sections~\ref{sec:non-C} and~\ref{sec:annihilate} as
\begin{equation}  
\label{eq:non-C}  
\Delta_{\rm nC}^{R_\alpha,S}(E) = 1+   
\alpha_s^2(\mu_C)\ln\beta \left[ 
 C_A\dra -2\dra^2(1+\nus) - \frac{\dra}{2} \frac{4 \mred^2}{\mbar^2}\nua   
\right] \theta(E)\,.
\end{equation}  
Eq.~\eqref{JRal} combines the non-Coulomb 
correction~\eqref{eq:non-C} with all-order Coulomb resummation and therefore
includes corrections of the form $\alpha_s^2\ln\beta \times 
(\alpha_s/\beta)^n$. As in \cite{Beneke:2011mq} we do not resum the 
logarithms arising from the non-Coulomb corrections, which formally are 
also an NNLL contribution. Such a resummation can in principle be
performed using renormalization group methods in PNR(S)QCD, but is left for 
future work.
  
%%%%%%%%%%%%%%%%%%%%%%%%%%%%%%%%%%  
\subsubsection{Bound-state effects}  
\label{sec:bound}
In the colour channels with an attractive Coulomb potential 
($D_{R_\alpha}<0$), the Coulomb Green function develops bound-state poles 
below threshold,   
\begin{equation}   
\label{eq:J_bound}  
 J_{R_\alpha}(E)=2  
\sum_{n=1}^\infty \delta(E-E_n)\,R_n \; \theta(-D_{R_\alpha})  
\,,\qquad E<0  
\end{equation}  
with binding energies   
\begin{equation}  
E_n  
= -\frac{2\mred \alpha_s^2\dra^2}{4n^2}\,(1+\frac{\alpha_s}{4\pi}\,e_1)    
\label{eq:E-bound}
\end{equation}  
and residues  
\begin{equation}  
  R_n =\left(\frac{2\mred (-D_{R_\alpha})\alpha_s}{2 n}\right)^{\!3} \,  
  (1+\frac{\alpha_s}{4\pi}\,\delta r_1).  
\end{equation}  
The values of the NLO corrections $e_1$ and 
$\delta r_1$~\cite{Pineda:1997hz,Beneke:2005hg} are quoted 
in~\cite{Beneke:2011mq}, where again the replacement $m_t\to 2\mred$ is 
implied.  

For long-lived squarks or gluinos, the poles correspond to 
physical gluinonium or squarkonium bound states,
which subsequently decay to di-photon or di-jet final states. In this paper we
do not consider this case with the resulting very different collider signals
compared to the usual missing-energy signatures.  For squarks and gluinos that
decay within the LHC detectors, the bound-state poles are smeared out by the
finite decay widths.  The resulting contribution to the total cross section
from partonic centre-of-mass energies below the nominal production threshold
can be included in the resummation formula~\eqref{eq:resum-sigma} by using the
bound-state contributions~\eqref{eq:J_bound} for vanishing decay widths. The
convolution of these corrections with the soft corrections is performed as
described in~\cite{Beneke:2011mq}.  In~\cite{Falgari:2012sq} this
procedure has been compared at NLL accuracy to the description of finite-width
effects through a complex energy $E\to E+ i(\Gamma_{\tilde s}+\Gamma_{\tilde
  s'})/2$ in the potential function.  It was found that finite width effects
on the NLL $K$-factors for squark-squark and squark-antisquark production
processes are well below the $5\%$ level while they can become of the order of
$10\%$ or even larger for gluino production processes with a gluino decay
width above $\Gamma_{\tilde g}/m_{\tilde g}\gtrsim 5\%$.  However, this case
only occurs for SQCD two-body decays $\tilde g\to \tilde q q$ in the region
$m_{\tilde g}\gtrsim 1.3 m_{\tilde q}$, where gluino production is
kinematically suppressed and gives a small contribution to the total SUSY
production rate.
In phenomenologically relevant parameter-space regions of the MSSM, the
finite width effects are therefore smaller than the remaining perturbative
uncertainty of the NNLL calculation, which justifies
 the use of the narrow-width approximation in this paper. 

\subsubsection{Fixed-order treatment of Coulomb corrections }  
 In order to assess the impact  
of Coulomb resummation  
 and to compare to NNLL predictions treating Coulomb corrections at  
 fixed order,  
 we also consider an approximation  
$\text{NNLL}_{\text{fixed-C}}$ where the product of hard and Coulomb  
corrections in the resummation formula~\eqref{eq:resum-sigma} is replaced by its fixed-order expansion up to  
$\mathcal{O}(\alpha_s^2)$,   
\begin{equation}  
 H_{i}(\mu_h)  
J^S_{R_\alpha}(E)\;\;\Rightarrow\;\;   
H^{(0)}_{i}(\mu_h) 
\frac{\beta}{2\pi} \frac{(m_{\tilde s}m_{\tilde s'})^{3/2}}{\mbar} 
\Delta_{\text{hC}}^{\text{NNLO}}(\hat s, \mu_h,\mu_f).  
\end{equation}  
The correction factor is given by 
\begin{eqnarray}  
\Delta_{\text{hC}}^{\text{NNLO}}(\hat s,\mu_h,\mu_f) &= & 
\Biggl\{  
\bigg(1-\alpha_s(\mu_f)\frac{\pi D_{R_\alpha}}{2} \sqrt{\frac{2 \mred}{E}}  
\bigg)  
\left(1+\frac{\alpha_s(\mu_h)}{4\pi} h^{(1)}_{i}(\mu_h)\right)
\nonumber\\  
&&\hspace*{-2cm} +\,\alpha_s^2(\mu_f)\Biggl[  
\frac{\pi^2 D_{R_\alpha}^2}{12} \left(\frac{2 \mred}{E}\right)   
+\frac{D_{R_\alpha}}{8}\sqrt{\frac{2 \mred}{E}}  
\bigg(\beta_0\ln\bigg(\frac{8\mred E}{\mu_f^2}\bigg)  
-a_1 \bigg)  
\nonumber\\  
&& \hspace*{-2cm} +\,  
\frac{1}{2}\ln\frac{E}{\mbar}\left(  
C_A\dra -2\dra^2(1+\nus) -\frac{\dra}{2}\frac{4 \mred^2}{\mbar^2}\nua   
 \right)  
\Biggr]\,  
\Biggr\}.  
\label{eq:fixed-C}  
\end{eqnarray}  
In this approximation one can derive an analytic formula for the NNLL
cross-section that is given in Appendix~\ref{app:fixC}. 

\subsubsection{Numerical size of non-Coulomb potential, annihilation 
and bound state contributions} 

In Table~\ref{tab:potential-num} we illustrate the numerical impact of the
non-Coulomb and annihilation potentials computed in Sections~\ref{sec:non-C}
and~\ref{sec:annihilate}, respectively, as well as the bound-state corrections
discussed in Section~\ref{sec:bound}.  The correction to the full NNLL 
cross section is obtained by removing the respective contribution to the 
NLO potential
function~(\ref{JRal}) from the resummation formula~\eqref{eq:resum-sigma}.  In
Table~\ref{tab:potential-num} the results for a given partonic initial state
are normalized to the total NLO cross section, so that they specify the
contribution to the NNLL K-factor defined in~\eqref{eq:Kfact} below. 
The choices of the input parameters, PDFs and the
various scales are discussed in Sections~\ref{sec:scales} and~\ref{sec:setup}.

The corrections from the bound-state contributions and the non-Coulomb
potential are of comparable size and generally in the per-cent range.
The size of the potential corrections for the different processes follows a
similar pattern as the full NNLL corrections discussed in
Section~\ref{sec:numerics}. The largest values are observed for 
gluino-pair production, where they grow above $10\%$ at high masses. In
contrast, for squark-squark production all types of corrections stay below a
percent. The annihilation 
correction is typically an order of
magnitude smaller than the two other types of corrections and therefore
phenomenologically negligible.  For the case of squark-antisquark and
gluino-pair production, the size of the corrections from the quark-antiquark
and gluon initial states reflects the relative contribution of these partonic
channels to the total cross section. For gluino-pair production, the
quark-antiquark channel contribution to the numbers in 
Table~\ref{tab:potential-num} is further suppressed, since only a single colour
channel contributes to S-wave production at threshold,
c.f. Table~\ref{tab:s-wave}.

\begin{table}[t]  
\centering  
\begin{tabular}{c|c||p{3em}|p{3em}|p{3em}||p{3em}|p{3em}|p{3em}}
  \multicolumn{2}{c||}{}& \multicolumn{3}{c||}{$m_{\tilde q}=1\,\mathrm{TeV}, m_{\tilde
 g}=1.5\,\mathrm{TeV}$ }& 
\multicolumn{3}{c}{$m_{\tilde q}=2.5\,\mathrm{TeV}, m_{\tilde
 g}=3\,\mathrm{TeV}$ }
\\\hline
$\tilde s\tilde s'$ &$pp'$  & $\delta_{\text{BS}}$& 
$\delta_{\text{nC}}$ &  $\delta_{\text{ann}}$
& $\delta_{\text{BS}}$& 
$\delta_{\text{nC}}$ &  $\delta_{\text{ann}}$
 \\ 
\hline\hline   
&&&\\[-.6cm]  
%%%%%%%%%%%%%%%%%%%%%%%%%%%%%%%%%%%%%%%%%%
$\tilde q\bar{\tilde q}$ &
$q\bar q$   & $1.5\%$ &$3.2\%$ &$\phantom{-}0.4\%$     & $\phantom{-}4.2\%$ &$\phantom{-}5.7\%$  &$\phantom{-}0.8\%$  
\\  
  & $gg$  &  $0.3\%$ & $0.8\%$ &$-0.1\%$ &  $\phantom{-}0.4\%$ &  $\phantom{-}0.6\%$& $-0.1\%$\\\hline  
%%%%%%%%%%%%%%%%%%%%%%%%%%%%%%%%%%%%%%%%%%
  $\tilde q_i\tilde q_j $& $qq$  & $0.1\%$& $0.5\%$ &$\phantom{-}0.1\%$ &$\phantom{-}0.2\%$ &
  $\phantom{-}0.4\%$ & $\phantom{-}0.1\%$ \\\hline
%%%%%%%%%%%%%%%%%%%%%%%%%%%%%%%%%%%%%%%%%%    
 $\tilde q\tilde g$& $qg$  &  $0.8\%$ & $1.7\%$&$-0.3\%$ & $\phantom{-}2.4\%$ & $\phantom{-}3.3\%$  
  & $-0.5\%$\\\hline  
%%%%%%%%%%%%%%%%%%%%%%%%%%%%%%%%%%%%%%%%%%
  $\tilde g\tilde g$  & $q\bar q$  & $0.2\%$ &  $0.3\%$ &$-0.1\%$ &
  $-0.1\%$ &$-0.1\%$  & $\phantom{-}0.03\%$\\  
         & $gg$ &  $5.1\%$  & $8.5\%$& \hskip0.4cm --- & $14.9\%$ & $16.2\%$ & \hskip0.4cm ---
%\\\hline
\end{tabular}  
\vskip0.3cm
\begin{tabular}{c|c||p{3em}|p{3em}|p{3em}||p{3em}|p{3em}|p{3em}}
  % \hline
\multicolumn{2}{c||}{}& \multicolumn{3}{|c||}{$m_{\tilde q}=1.5\,\mathrm{TeV}, m_{\tilde
 g}=1\,\mathrm{TeV}$ }& 
\multicolumn{3}{c}{$m_{\tilde q}=3\,\mathrm{TeV}, m_{\tilde
 g}=2.5\,\mathrm{TeV}$ }
\\\hline
$\tilde s\tilde s'$ &$pp'$  & $\delta_{\text{BS}}$& 
$\delta_{\text{nC}}$ &  $\delta_{\text{ann}}$
& $\delta_{\text{BS}}$& 
$\delta_{\text{nC}}$ &  $\delta_{\text{ann}}$
 \\ 
\hline\hline   
&&&\\[-.6cm]  
%%%%%%%%%%%%%%%%%%%%%%%%%%%%%%%%%%%%%%%%%%
$\tilde q\bar{\tilde q}$ &
$q\bar q$   & $1.8\%$ &$3.6\%$ &$\phantom{-}0.5\%$     & $\phantom{-}5.1\%$ &$\phantom{-}6.2\%$  &$\phantom{-}0.8\%$  
\\  
  & $gg$  &  $0.2\%$ & $0.4\%$ &$-0.1\%$ &  $\phantom{-}0.4\%$ &  $\phantom{-}0.5\%$& $-0.1\%$\\\hline  
%%%%%%%%%%%%%%%%%%%%%%%%%%%%%%%%%%%%%%%%%%
  $\tilde q_i\tilde q_j $& $qq$  & $0.1\%$& $0.3\%$ &$\phantom{-}0.1\%$ &$\phantom{-}0.3\%$ &
  $\phantom{-}0.4\%$ & $\phantom{-}0.1\%$ \\\hline
%%%%%%%%%%%%%%%%%%%%%%%%%%%%%%%%%%%%%%%%%%    
 $\tilde q\tilde g$& $qg$  &  $1.5\%$ & $2.8\%$&$-0.3\%$ & $\phantom{-}3.2\%$ & $\phantom{-}4.0\%$  
  & $-0.5\%$\\\hline  
%%%%%%%%%%%%%%%%%%%%%%%%%%%%%%%%%%%%%%%%%%
  $\tilde g\tilde g$  & $q\bar q$  & $0.3\%$ &  $0.4\%$ &$-0.1\%$ &
  $\phantom{-}0.6\%$ &$\phantom{-}0.6\%$  & $-0.2\%$\\  
         & $gg$ &  $4.0\%$  & $7.8\%$& \hskip0.4cm --- & $10.5\%$ & $13.3\%$ &\hskip0.4cm ---
%\\\hline
\end{tabular}  
\caption{
Contribution of the bound-state, non-Coulomb and
  annihilation corrections to the NNLL result for
the LHC  with $\sqrt{s}=13\,\mathrm{TeV}$. The corrections are normalized to
the total NLO cross sections.}  
\label{tab:potential-num}  
\end{table}  

%%%%%%%%%%%%%%%%%%%%%%%%%%%%%%%%%%%%%%%%%%  
\subsection{Scale choices}  
\label{sec:scales}  
  
The resummed cross section~\eqref{eq:resum-sigma} depends on the factorization
scale $\mu_f$, which we set to $\mu_f=\mbar$ as a default, the soft scale $\mu_s$, the hard scale $\mu_h$, as well as on the scale $\mu_C$ used in the Coulomb function. While the solution to the renormalization group equations for the hard and soft functions is formally independent on $\mu_h$ and $\mu_s$, due to the truncation of the perturbative series the NNLL resummed cross section contains a residual dependence on these scales at $\mathcal{O}(\alpha_s^2)$. We specify our scale choices here which, with the exception of the soft scale, follow the treatment in~\cite{Beneke:2011mq,Falgari:2012hx}.

\paragraph{Hard scale}  
Our default value  for the hard scale is $\mu_h=2\mbar$ which can be motivated by  
the logarithmic structure of the renormalization group equation of the  
hard function~\cite{Beneke:2009rj}. This choice is also seen to  
eliminate the logarithms in the expression of the hard  
function~\eqref{eq:rel-hard-mellin}. To estimate the theoretical  
uncertainty from the choice of the hard scale, we include a variation  
in the interval $\mbar\leq \mu_h\leq 4\mbar$ in our estimate of the  
theoretical uncertainties.  
  
\paragraph{Soft scale}  
  
A resummation of all logarithms of $\beta$ in the partonic cross section could
be achieved by the running soft scale $\mu_s\sim \mbar\beta^2$ which, however,
renders the convolution of the partonic cross-section~\eqref{eq:resum-sigma}
with the PDFs unintegrable and leads to a 
Landau pole in $\alpha_s(\mu_s)$.  Instead, Ref.~\cite{Becher:2006nr} proposed
a fixed soft scale that minimizes the one-loop soft corrections to the
hadronic cross section.  Alternatively, a running soft scale
$\mu_s=k_s\,\mbar\,\text{Max}\lbrack \beta^2 ,\beta_{\text{cut}}^2\rbrack$ was
introduced in~\cite{Beneke:2011mq}. This method was applied to squark and
gluino production at NLL accuracy in~\cite{Falgari:2012hx} with $k_s=1$
and choosing the parameter $\beta_{\text{cut}}$ 
small enough to justify the resummation of logarithms
$\ln\beta_{\text{cut}}$ in the lower interval, and large enough so that
threshold logarithms can be treated perturbatively in the upper
interval, following the procedure of~\cite{Beneke:2011mq}.

Recently, Sterman and Zeng~\cite{Sterman:2013nya} considered an   
expansion of the logarithm of the  parton luminosity function,   
\begin{equation}  
\label{eq:sz-lumi}  
  \ln L_{p p^\prime}\left(\tau_0/z,\mu\right)=   
s^{(0)}_{pp'}(\tau_0,\mu)+ s^{(1)}_{pp'}(\tau_0,\mu)  \ln z+\dots  
\end{equation}  
with  
\begin{equation}  
\label{eq:def-s1}  
  s^{(1)}_{pp'}(\tau_0,\mu)  
  =-\frac{d\ln L_{p p^\prime}(\tau,\mu)}{d\ln\tau}|_{\tau=\tau_0}\,.  
\end{equation}  
To the extent that the convolution of the parton luminosity with the  
partonic cross section is dominated by the $\ln z$ term, it was shown  
in~\cite{Sterman:2013nya} that the momentum-space resummation method  
is equivalent to the traditional resummation in Mellin-moment space if  
the soft scale is chosen as  
\begin{equation}  
\label{eq:sz-scale}  
  \mu_s= \frac{2\mbar e^{-\gamma_E}}{ s^{(1)}_{pp'}}\,.  
\end{equation}  
Note that the scale choice~\eqref{eq:sz-scale} amounts to the use of a  
different soft scale in every partonic channel.  As demonstrated  
in~\cite{Sterman:2013nya} for the case of Higgs production, the  
single-power approximation of the parton luminosity provides a  
dominant contribution to the convolution with the partonic cross  
section, so the scale choice~\eqref{eq:sz-scale} is motivated also for  
the use of the exact PDFs.   
This conclusion is also supported by an analysis using the saddle-point approximation~\cite{Bonvini:2014qga}.  
 We therefore adopt~\eqref{eq:sz-scale} as our default choice for the soft scale.  
This choice is also convenient for the numerical implementation, since it  
can be determined during the evaluation of the cross section at very  
small computational cost without a prior minimization procedure as for  
the other scale-setting procedures.  In our implementation, the  
flavour-summed parton luminosities mentioned  
below~\eqref{eq:processes} are used for the determination of the soft  
scale in the case of initial-state quarks.  The theoretical  
uncertainty due to the scale choice is estimated by varying $\mu_s$  
from one-half to twice the default scale.  Note that we keep the 
factorization scale fixed in the determination of the soft scale, i.e.\ we  
always use the default value $\mu_f=\mbar$ in~\eqref{eq:def-s1}.

\paragraph{Coulomb scale}  
At NNLL accuracy the scale $\mu$ in the potential function~\eqref{JRal} can be chosen independently of the other scales.  
NLL effects related to Coulomb exchange  
can be resummed by choosing the scale of the order of $\sqrt{2 \mred \mbar} \beta$, which is the typical virtuality of  
Coulomb gluons. For small $\beta$, in production channels with an   
attractive Coulomb potential the relevant scale is instead given by the Bohr scale $2\mred |D_{R_\alpha}| \alpha_s$ set by the first bound state, as can be seen from the $\beta\to 0$ limit of the NLO potential function quoted in~\cite{Beneke:2011mq}.  
We thus  
choose the scale in $J_{R_\alpha}$ to be  
\begin{equation} \label{eq:muC}  
\mu_C = \text{Max} \left\{2 \alpha_s(\mu_C) \mred |D_{R_\alpha}|,2 \sqrt{2 \mred \mbar} \beta\right\} \, .   
\end{equation}   
Note that no bound states arise for a repulsive potential,  
$D_{R_\alpha}>0$, in which case $J_{R_\alpha}$ vanishes for small  
$\beta$. Therefore the above argument does not determine the Coulomb scale for $\beta\to 0$ in this  case. We nevertheless use the prescription~\eqref{eq:muC} also for a repulsive potential where  resummation of Coulomb corrections leads to small effects, so that the  
precise choice of $\mu_C$ has a negligible numerical  
impact on predictions of the cross section.  We vary $\mu_C$  
from one-half to twice the default value~\eqref{eq:muC} to estimate the
theoretical uncertainty.  

\section{Numerical results}  
\label{sec:numerics}  
  
\subsection{Setup}  
\label{sec:setup}  
  
In order to include the known fixed-order NLO corrections without kinematic
approximation, we match the NNLL-resummed result to the NLO result from
\texttt{PROSPINO}, supplemented with the threshold approximation of the NNLO
cross sections, where all constant contributions at
$\mathcal{O}(\alpha_s^2)$ are set to zero.  In order to avoid double counting,
the fixed-order expansion of the NNLL corrections up to
$\mathcal{O}(\alpha_s^2)$ is subtracted from the cross section. Our matched
predictions are therefore given by
\begin{equation}  
\label{eq:match2}  
  \hat\sigma^{\text{NNLL}}_{pp'\text{matched}}(\hat s)  
  =\left[\hat\sigma^{\text{NNLL}}_{pp'}(\hat s)-  
    \hat\sigma^{\text{NNLL}(2)}_{pp'}(\hat s)\right]  
+ \hat\sigma^{\text{NLO}}_{pp'}(\hat s)  
+ \hat\sigma^{\text{NNLO}}_{\text{app},pp'}(\hat s) \, .  
\end{equation}  
The approximate NNLO cross section is obtained from Eq.~(A.1) in~\cite{Beneke:2009ye} by inserting the one-loop hard functions
$2\mathrm{Re}(C_X)=h_1^{(i)}$, the results for $\nus$ given in~\eqref{eq:nus}
and adding the annihilation contribution~\eqref{eq:dnnlo-ann}.  The expansion
of the NNLL correction to ${\cal O}(\alpha_s^2)$,
$\hat\sigma^{\text{NNLL}(2)}_{pp'}$, is given in Appendix~\ref{app:expansion}.
The partonic cross sections in~\eqref{eq:match2} are convoluted with the
parton luminosities determined using the \texttt{PDF4LHC15\_nnlo\_30}
PDFs~\cite{Butterworth:2015oua},\footnote{The \texttt{PDF4LHC15\_nnlo\_mc}
  set recommended for predictions for the search for new
  physics~\cite{Butterworth:2015oua} yields unphysical negative cross sections
  at large sparticle masses for some member PDFs. We have checked that the
  central value and the PDF uncertainty of the \texttt{PDF4LHC15\_nnlo\_30}
  set are in good agreement with the $68\%$ confidence level predictions of
  the MC set, with differences of the central prediction in general below
  $1\%$ and staying below $10\%$ for gluino pair production at the highest
  considered masses.}  which combine the MMHT14~\cite{Harland-Lang:2014zoa},
CT14~\cite{Dulat:2015mca} and NNPDF3.0~\cite{Ball:2014uwa} sets according
to~\cite{Watt:2012tq,Gao:2013bia,Carrazza:2015aoa}. We have used the \texttt{ManeParse}~\cite{Clark:2016jgm}
interface for some calculations. In the hard functions we use $m_t=173.2$~GeV.
  
In our results we include an estimate of the theoretical uncertainties of our default implementation from various sources.\footnote{      
The terminology used here follows~\cite{Falgari:2012hx} and differs slightly from the one for $t \bar{t}$ production in  \cite{Beneke:2011mq} where the errors from variation of the hard and Coulomb scales, and of the soft scale for the fixed-scale  
implementation, were included in the scale uncertainty instead of the resummation uncertainty.   
}  
\begin{description}  
\item[Factorization scale uncertainty:]  The factorization scale $\mu_f$ is varied between half and twice the   
default value, i.e. $\mbar/2<\mu_f<2 \mbar$. For the resummed result, this is done keeping the other scales $\mu_h$, $\mu_C$ and $\mu_s$ fixed.  
Note that in the fixed-order results we identify the factorization and renormalization scales, which is the default   
procedure implemented in the numerical code \texttt{PROSPINO} used for the computation of the fixed-order NLO result \cite{Beenakker:1996ed}.  
  
\item[Resummation uncertainty:] The soft, hard and Coulomb scale are  
  separately varied around their default values as discussed in  
  Section~\ref{sec:scales}.  Power-suppressed terms are estimated by  
  using the expression $E=\sqrt{\hat s}-2\mbar$ in the argument of the  
  potential function instead of the non-relativistic limit  
  $E=\mbar\beta^2$.  The resulting uncertainties  
  from all sources are added in quadrature.  
\item[Missing higher-order corrections]  
  In order to estimate the uncertainty from  
  uncalculated NNLO corrections we follow~\cite{Beneke:2011mq} and  
  vary the unknown two-loop constant term $C_{pp',i}^{(2)}$ in the  
  threshold expansion of the NNLO scaling function $f^{(2)}_{pp',i}$  
  in~\eqref{eq:partonic} in the interval $-(C^{(1)}_{pp',i})^2 \leq  
  C_{pp',i}^{(2)}\leq +(C^{(1)}_{pp',i})^2$, where the one-loop  
  constant $C^{(1)}_{pp',i}$ is defined in~\eqref{eq:C1}.  
\item[PDF$+\alpha_s$ uncertainty:] we estimate the error due to uncertainties
  in the PDFs and the strong coupling using the
  \texttt{PDF4LHC15\_nnlo\_30\_pdfas} set~\cite{Butterworth:2015oua},
adding the PDF uncertainty at $68\%$ confidence
  level in quadrature to the uncertainty from the variation of the strong
  coupling constant $\alpha_s(M_Z) = 0.118 \pm 0.0015$.
\end{description}  
In the following we will often refer to the sum in quadrature of scale and resummation uncertainty and the variation of the two-loop constant as ``total theoretical uncertainty".     
  
%%%%%%%  
\begin{figure}[t]  
\begin{center}  
\includegraphics[width=0.49 \linewidth]{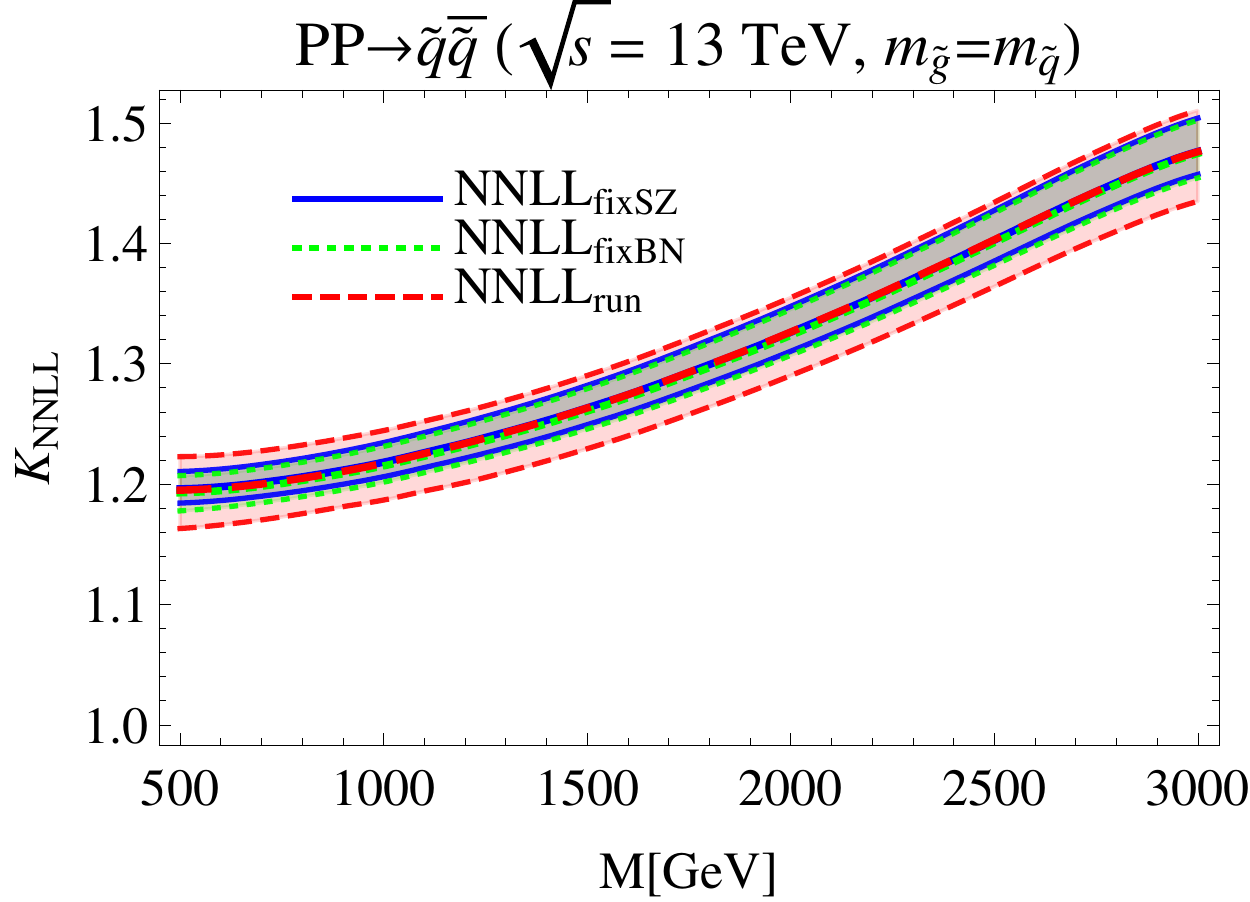}  
\includegraphics[width=0.49 \linewidth]{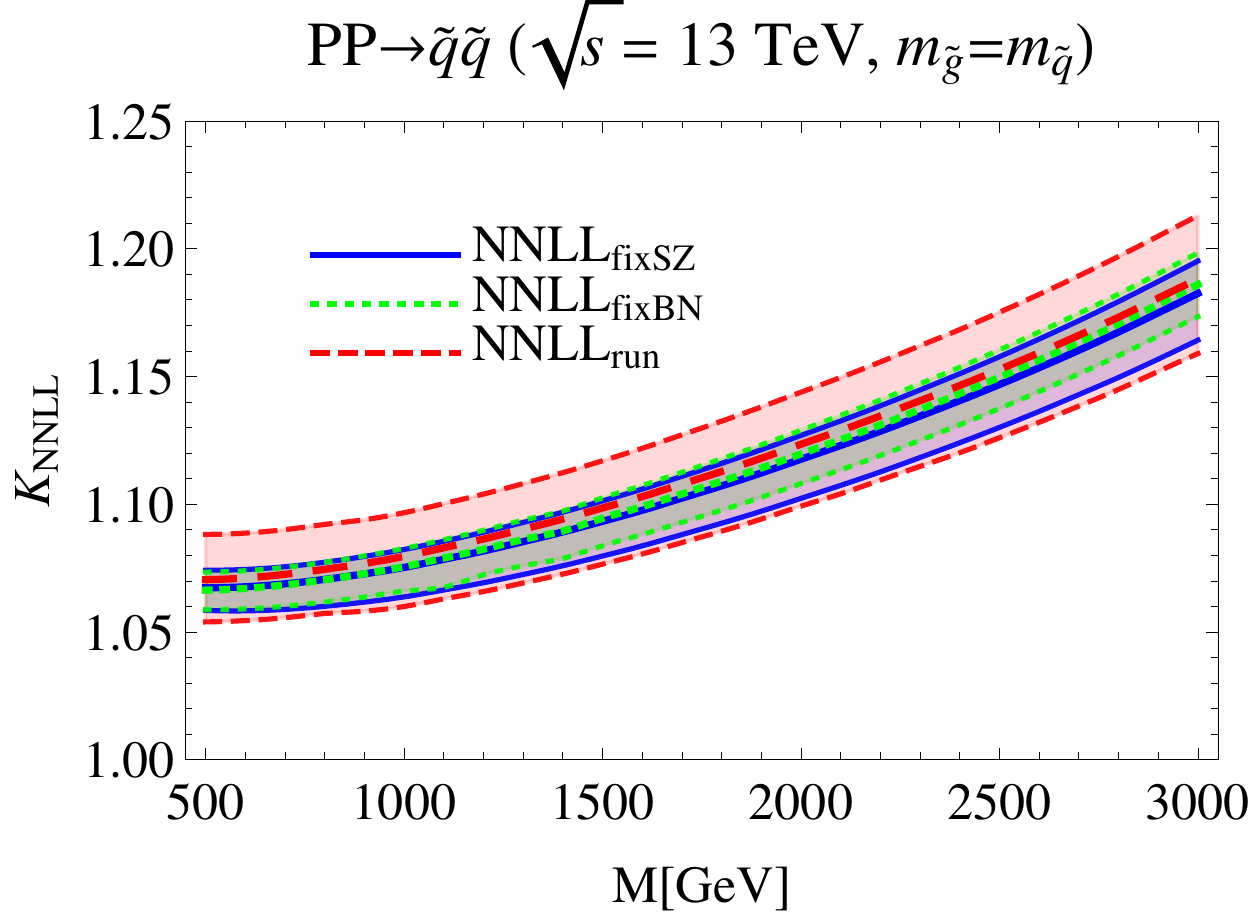}\\[.5cm]   
\includegraphics[width=0.49 \linewidth]{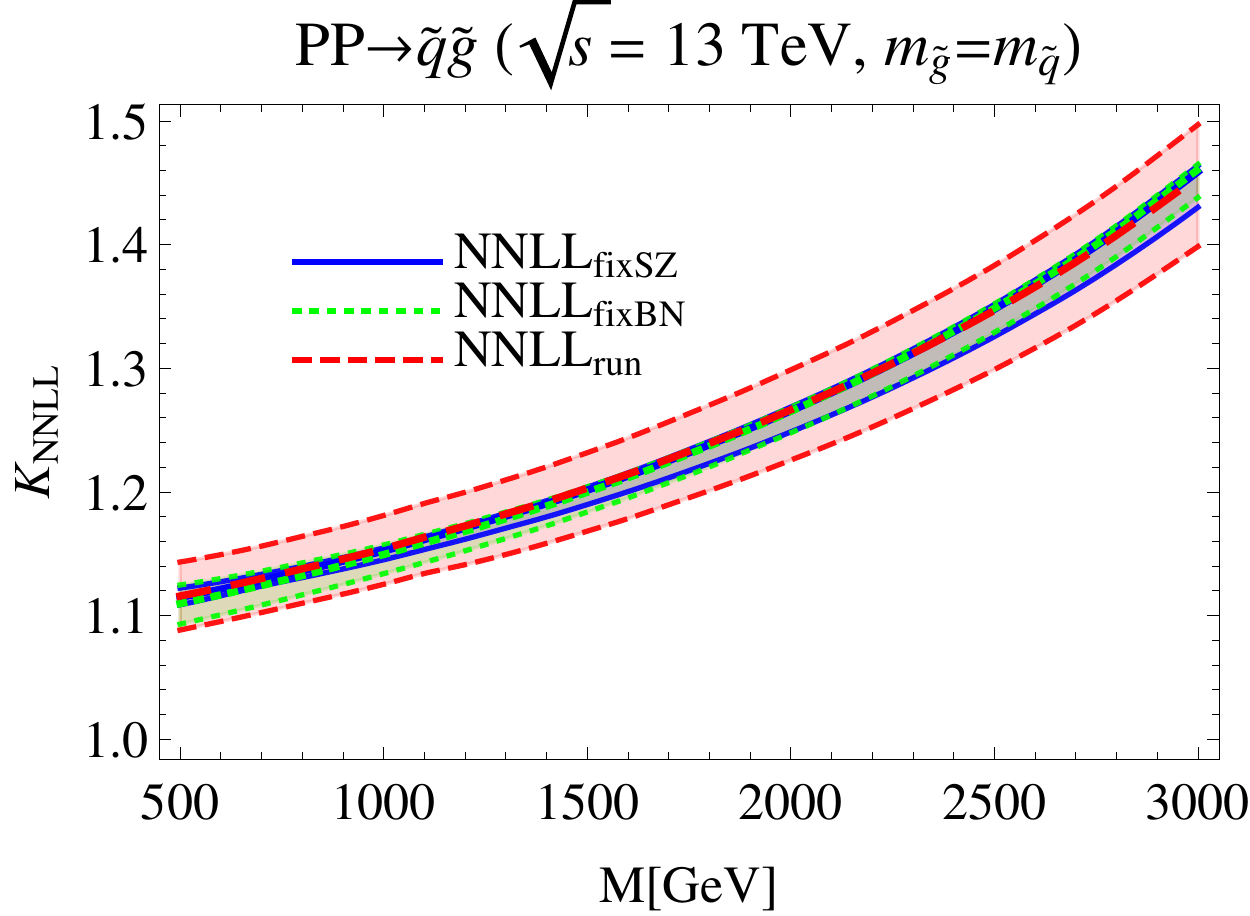}  
\includegraphics[width=0.49 \linewidth]{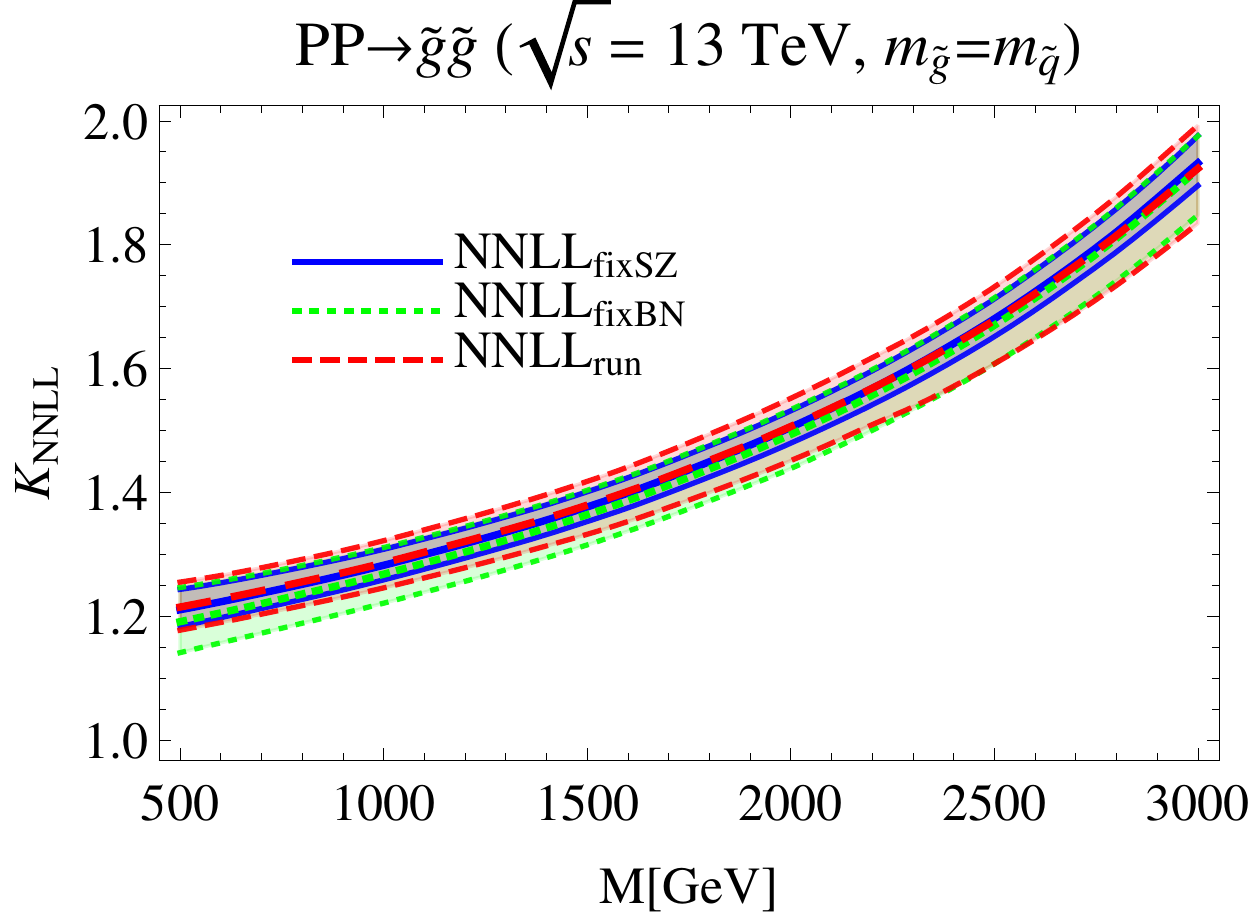}  
\end{center}  
\caption{ Resummation uncertainty for the NNLL resummed result with the
  default scale choice~\eqref{eq:sz-scale} (NNLL$_\text{fixSZ}$, solid blue)
  compared to a running and a fixed soft scale determined as discussed
  in~\cite{Falgari:2012hx} (NNLL$_\text{fixBN}$, short dashed green, and
  NNLL$_\text{run}$, dashed red) for squark-antisquark (top-left),
  squark-squark (top-right), squark-gluino (bottom-left) and gluino-gluino
  (bottom-right) at LHC with $\sqrt{s} = 13\,$ TeV.  The central lines
  represent the $K$-factors for the default scale choice, while the bands give
  the resummation uncertainties associated with the results.  See text for
  explanation.}
\label{fig:fix_vs_run}  
\end{figure}  
%%%%%%  

In Figure~\ref{fig:fix_vs_run} we compare the NNLL K-factors, defined in~\eqref{eq:Kfact} below, and the  
resummation uncertainty for our default implementation with the fixed  
soft scale~\eqref{eq:sz-scale} to an implementation using a running  
scale, as in the previous NLL results in~\cite{Falgari:2012hx}, and a  
fixed scale determined using the method of Becher and  
Neubert~\cite{Becher:2006nr}. In the running-scale results the  
variation of $\mu_s$ in the resummation uncertainty is replaced by a  
variation of the parameters $\beta_\text{cut}$ and $k_s$ according to  
the procedure given in \cite{Beneke:2011mq}.  The curves do not  
include the $C^{(2)}$ error estimate, which is common to all  
implementations.  The central predictions obtained with the different  
scale-setting methods agree well for all production processes, while  
the estimate of the resummation uncertainty is  
larger for the running scale prescription.  
However, we find that the total theoretical uncertainty including  
factorization scale variation is similar for the three methods.
Note that the  
agreement of the different scale choices is significantly  
improved compared to the NLL results (see Figure~6  
in~\cite{Falgari:2012hx}). Therefore contrary to resummation at NLL level, we find that at NNLL the   
ambiguity in the choice of the soft scale prescription is negligible with respect to the total theoretical uncertainty.

\subsection{Results}  
\label{sec:results}  
  
We present results for  squark and gluino production at the LHC  
for five different higher-order approximations:  
\begin{itemize}  
\item {\bf NNLL}: the default implementation. Contains the full combined soft and Coulomb resummation, Eq.~\eqref{eq:resum-sigma}  
including bound-state contributions~\eqref{eq:J_bound} below threshold, matched to NNLO$_{\text{app}}$ according to~\eqref{eq:match2}.   
For the soft scale we adopt the fixed scale given in~\eqref{eq:sz-scale}.   
\item {\bf NNLL$_{\text{fixed-C}}$}:  as above but using the  
  fixed-order NNLO Coulomb terms~\eqref{eq:fixed-C} without bound-state
  effects  interfering with  
  resummed soft radiation, and for $\mu_h=\mu_f$.    
\item {\bf NNLO$_{\text{app}}$}: The approximate NNLO
  corrections~\cite{Beneke:2009ye} including the spin-dependent
  non-Coulomb~\eqref{eq:non-C} and annihilation terms.
\item {\bf NLL}: The NLL corrections from~\cite{Falgari:2012hx} with combined
  soft and Coulomb resummation and bound-state effects. Note that the scale
  choice~\eqref{eq:sz-scale} is used for the NLL results as well, whereas a
  running scale was used as the default in~\cite{Falgari:2012hx}.   
\item {\bf NLL$_{s}$:} Soft NLL resummation without Coulomb resummation and
  for $\mu_h=\mu_f$.
\end{itemize}

\begin{figure}[t!]  
  \centering  
\includegraphics[width=0.48 \linewidth]{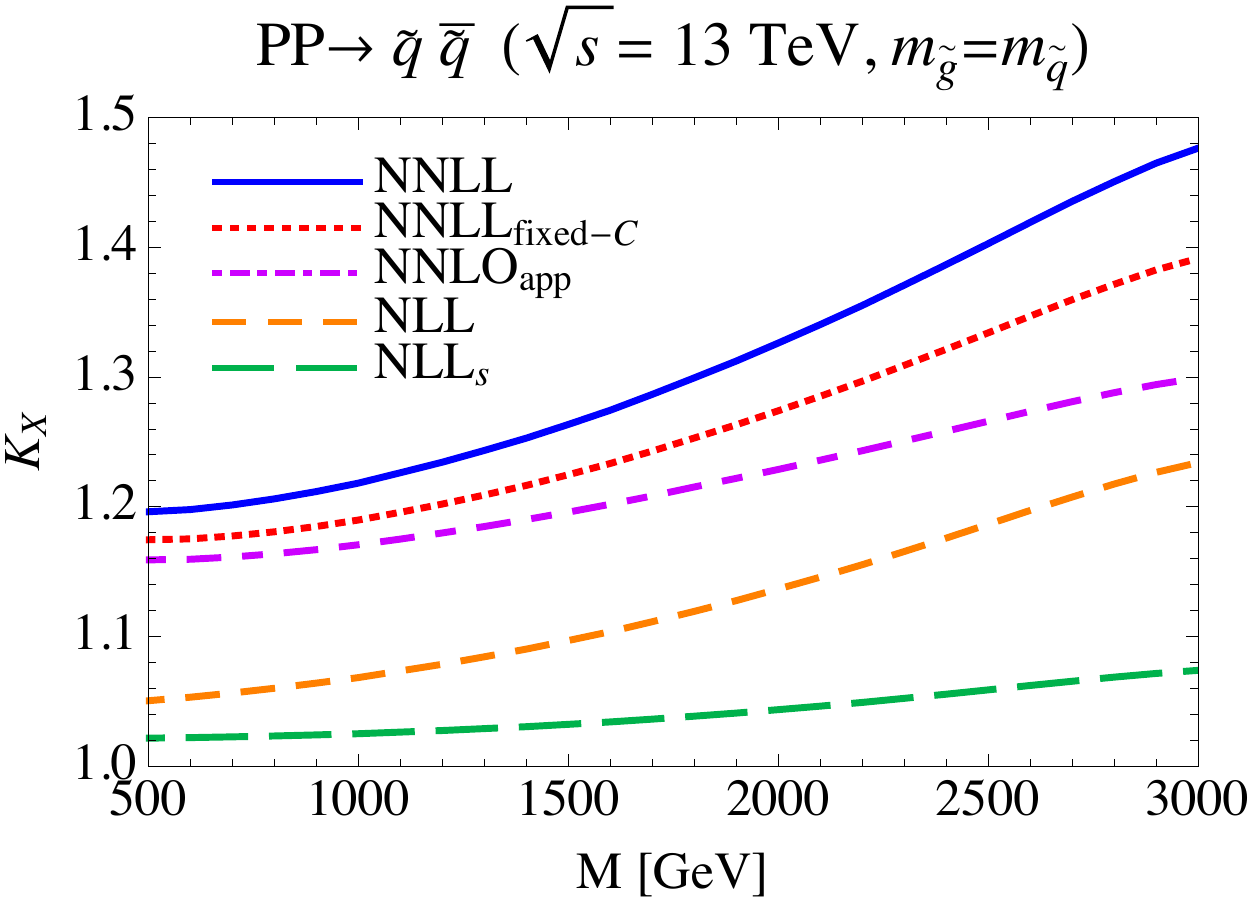}  
\includegraphics[width=0.48 \linewidth]{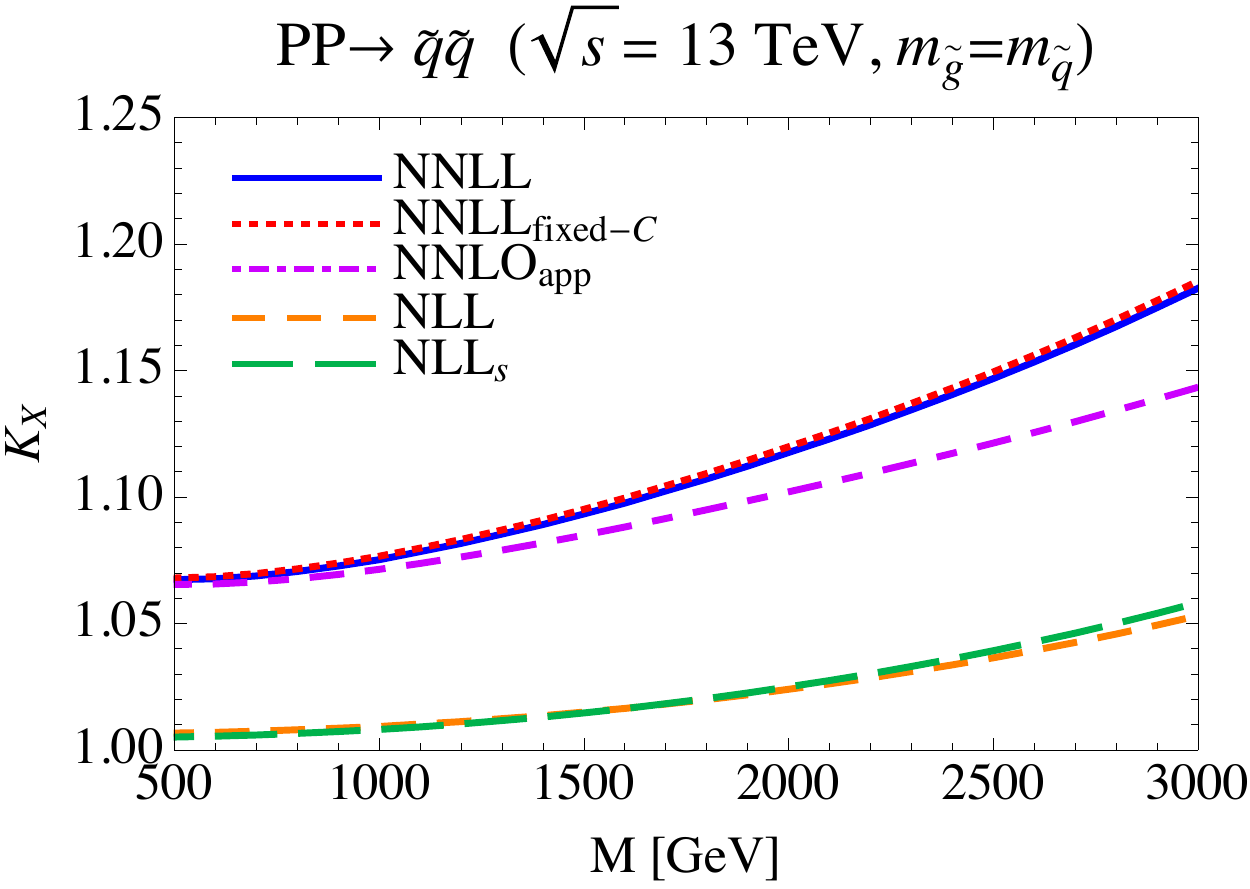}\\[.5cm]  
\includegraphics[width=0.48 \linewidth]{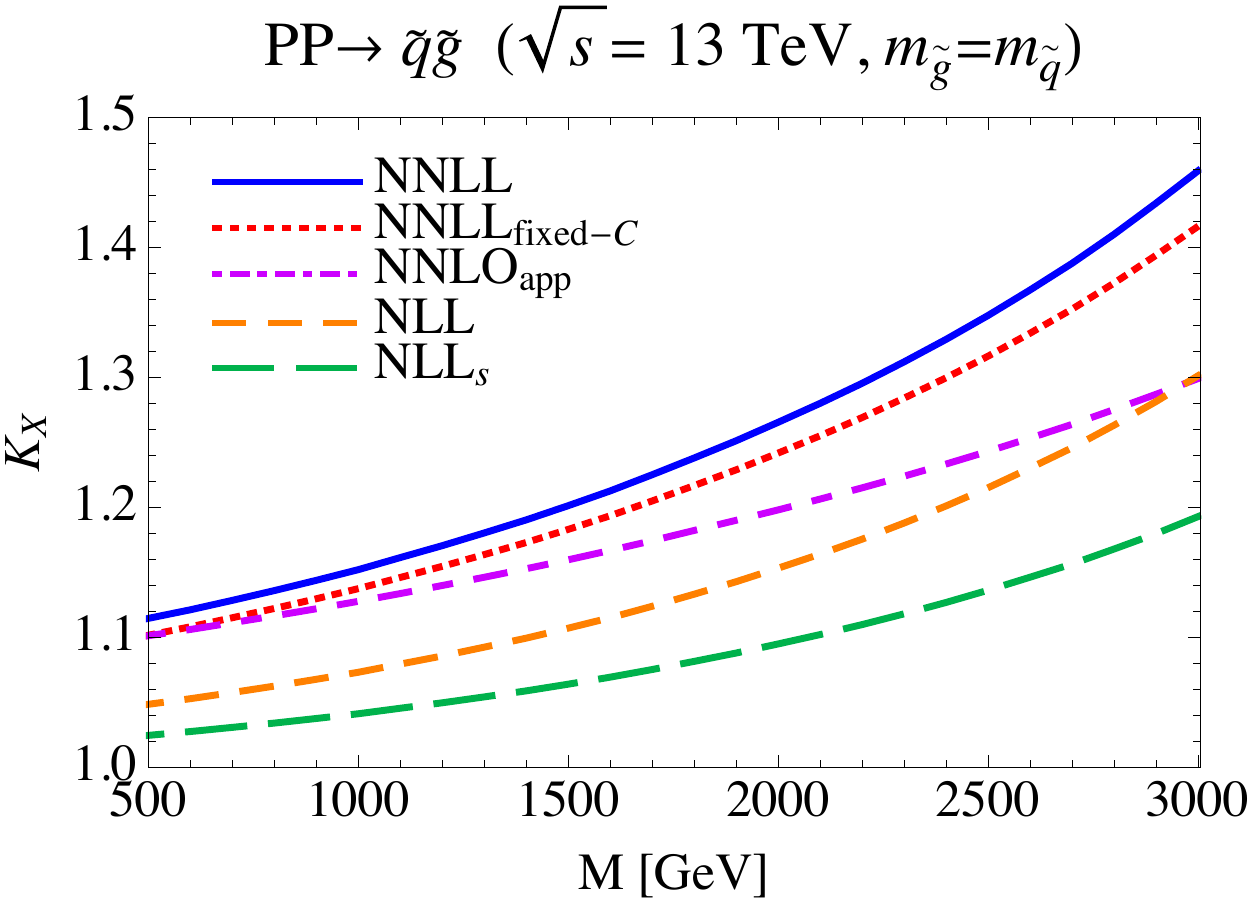}  
\includegraphics[width=0.48 \linewidth]{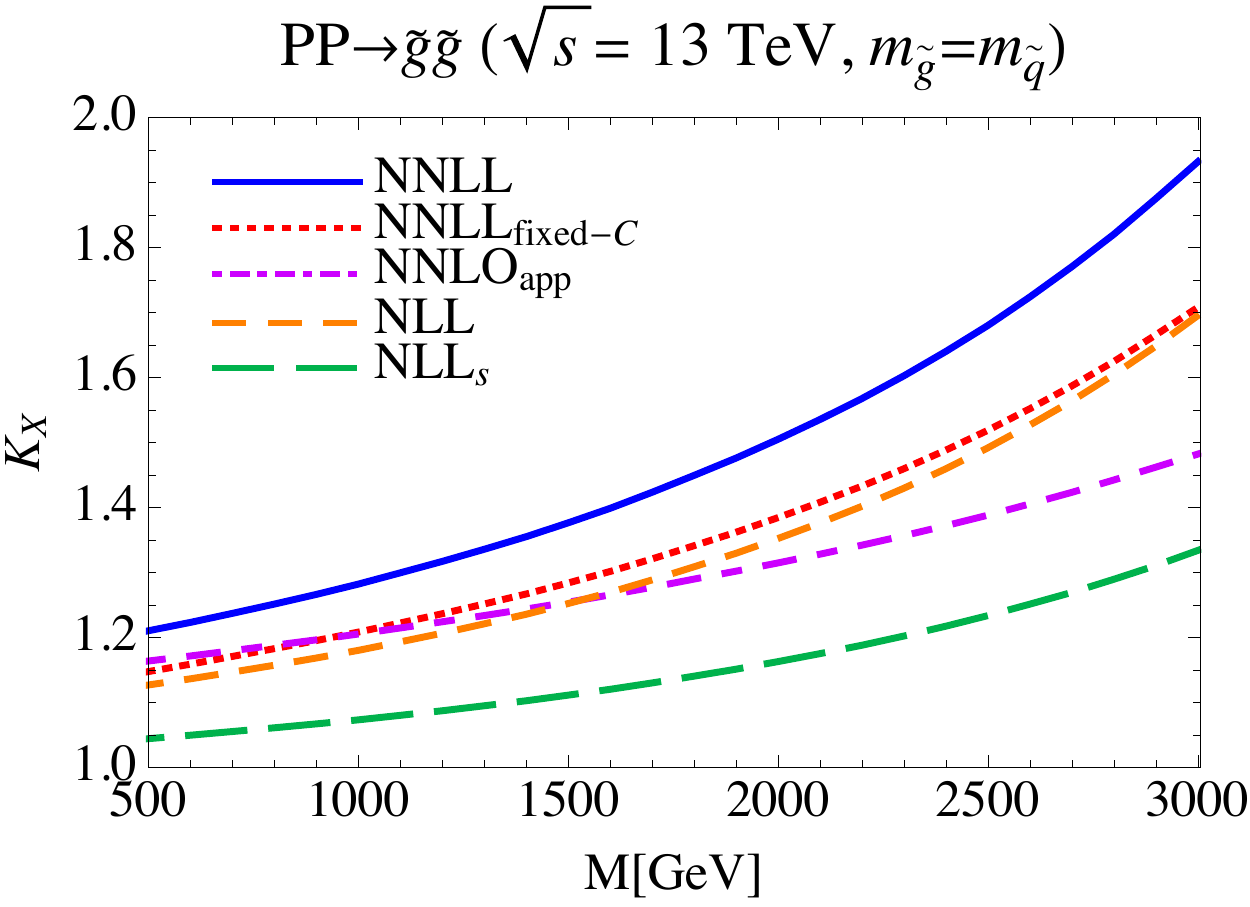} 
\caption{ Higher-order corrections relative to the NLO cross section for
  squark and gluino production at the LHC with $\sqrt{s}=13$~TeV for full
  NNLL resummation (solid blue), NNLL with fixed-order Coulomb corrections
  (dotted red), approximate NNLO (dot-dashed pink), NLL (dashed orange) and
  NLL soft (long-dashed green).  
The \texttt{PDF4LHC15\_nnlo\_30} PDFs are used throughout.}
\label{fig:KNNLL13}  
\end{figure}

Our NNLL predictions for the LHC with $\sqrt{s}=13$ ($14$~TeV) are
provided  as grids for $m_{\tilde q},
m_{\tilde g}=200$--$3000$~GeV ($200$--$3500$~GeV)~\cite{SUSYNNLL}.
 We also provide a
\texttt{Mathematica} file with interpolations of the cross sections.  As
an illustration, our results for the four squark and gluino production
processes at the LHC with $\sqrt s=13\,\mathrm{TeV}$\footnote{%
  Results at $\sqrt s=8\,\mathrm{TeV}$ have been presented
  in~\cite{Beneke:2013opa} for a slightly different setup using the running
  soft scale prescription, the MSTW08 PDFs and omitting the annihilation
  contribution.  } %
for equal squark and gluino masses are shown in Figure~\ref{fig:KNNLL13} in
the form of K-factors beyond NLO,
\begin{equation} \label{eq:Kfact}  
K_{\text{X}}=\frac{\sigma_{\text{X}}}{\sigma_{\text{NLO}}} .  
\end{equation}  
Here $X$ denotes one of the approximations NLL, NNLL$_{\text{fixed-C}}$,
NNLO$_{\text{app}}$ and NNLL defined above and $\sigma_\text{NLO}$ is the
fixed-order NLO result obtained using \texttt{PROSPINO}.  We use the
\texttt{PDF4LHC15\_nnlo} PDFs for all results, including the
NLO normalization in~\eqref{eq:Kfact}, in order to isolate the effects of the
higher-order corrections to the partonic cross sections.
The corrections relative to NLO can become large for the full NNLL
resummation, ranging from up to $18\%$ for squark-squark production to $90\%$
for gluino pair production.
Compared to the NLL results, the NNLL corrections provide a shift of the cross
section by $10$--$20\%$ normalized to the NLO prediction.  This shows that a
stabilization of the perturbative behaviour is achieved by the resummation, in
particular for the processes with large corrections such as gluino-pair
production.  Note that this is only the case for the joint soft-Coulomb
resummation performed here and in~\cite{Beneke:2010da,Falgari:2012hx}, whereas
a large NNLL correction is observed relative to the NLL$_s$ prediction that
does not include Coulomb resummation.  This observation is consistent with the
results of Mellin-space resummation~\cite{Beenakker:2014sma}.  The effect
of Coulomb resummation beyond NNLO and the bound-state effects can be seen by
comparing the NNLL and $\text{NNLL}_{\text{fixed-C}}$ results and is important
in particular for squark-antisquark and gluino-pair production. For
squark-squark production, the effect of Coulomb resummation beyond NNLO is
small. A similar behaviour was also observed for the NLL corrections beyond
NLO in~\cite{Falgari:2012hx} and originates from cancellations between the
negative corrections arising from the repulsive colour-sextet channel and the
positive corrections arising from the attractive colour-triplet channel.  
The comparison to the approximate NNLO results shows that corrections beyond
NNLO become sizeable beyond sparticle masses of about $1.5$~TeV.  The
  $\text{NNLL}_{\text{fixed-C}}$ correction factors in Figure~\ref{fig:KNNLL13} cannot be directly compared to
the corresponding results using the Mellin-space
formalism~\cite{Beenakker:2014sma} that are given using
the MSTW2008 set of PDFs.
However, a comparison to our own earlier results for the
$\text{NNLL}_{\text{fixed-C}}$ approximation at $8\,\mathrm{TeV}$ using the
same PDFs~\cite{Beneke:2013opa} shows overall good agreement, which is
reassuring given the  different methods used for the resummation.  As
already noted in~\cite{Beenakker:2014sma}, the largest difference of order
$10\%$ appears for gluino pair-production at large masses, which is larger
than the estimated resummation uncertainty of approximately $5\%$ of our
result. Therefore, a more detailed comparison of the two approaches will be
useful.

\begin{figure}[t!]  
  \centering  
  \includegraphics[width=0.6 \linewidth]{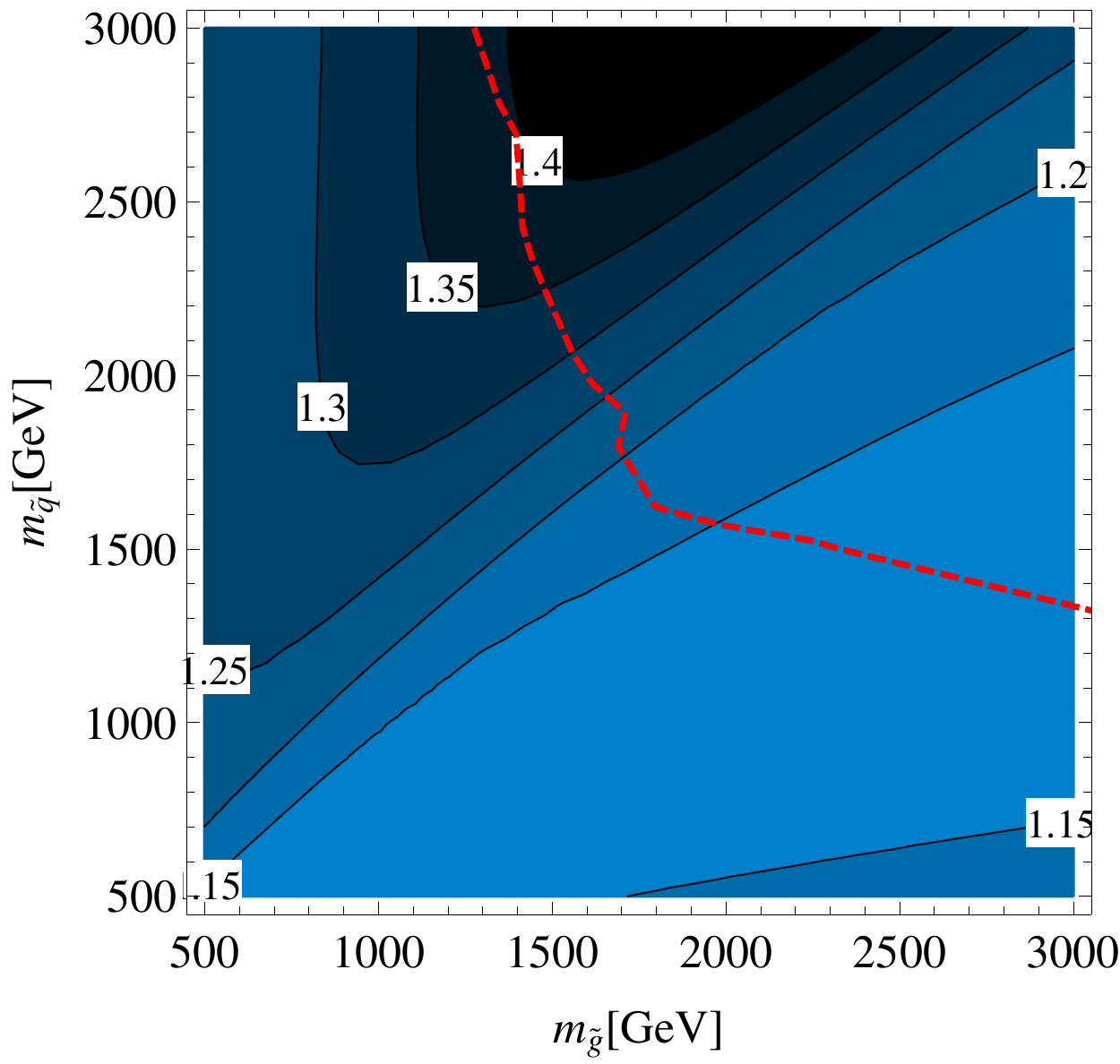}  
 \caption{NNLL $K$-factor for the total SUSY pair production rate at the LHC with $\sqrt{ s}=13$~TeV, as a function of the squark mass $m_{\tilde{q}}$ and gluino
  mass $m_{\tilde{g}}$. The red, dashed line is the ATLAS exclusion bound from
  searches at $\sqrt{s}=8$~TeV in a simplified model with a massless neutralino~\cite{Aad:2014wea}.}  
\label{fig:contour}  
\end{figure}

Figure \ref{fig:contour} shows the NNLL $K$-factor for the total SUSY
production rate, i.e. the sum of all squark and gluino pair production
processes, at the $13$ TeV LHC as a contour plot in the
$(m_{\tilde{g}},m_{\tilde{q}})$-plane. The NNLL corrections are larger in the
region with $m_{\tilde g}< m_{\tilde q}$ where squark-gluino and gluino-pair
production with the corresponding larger $K$-factors dominate the total SUSY
production rate, see e.g.~\cite{Falgari:2012hx}.  In contrast, for $m_{\tilde
  q}\leq m_{\tilde g}$ the total rate is dominated by squark-squark
production with a resulting smaller $K$-factor.  Since squark-antisquark
production is suppressed compared to squark-squark production at the LHC unless
$m_{\tilde g}\gg m_{\tilde q}$, the large $K$-factor in this process has
little impact on the total SUSY production rate for the mass range considered
in Fig.~\ref{fig:contour}.
 For
illustration, the plot also shows a Run 1 ATLAS exclusion bound in a
simplified model with a massless neutralino~\cite{Aad:2014wea} that shows that
corrections larger than $40\%$ arise in a region relevant for current
searches.

\begin{figure}[t!]  
  \centering  
  \includegraphics[width=0.48 \linewidth]{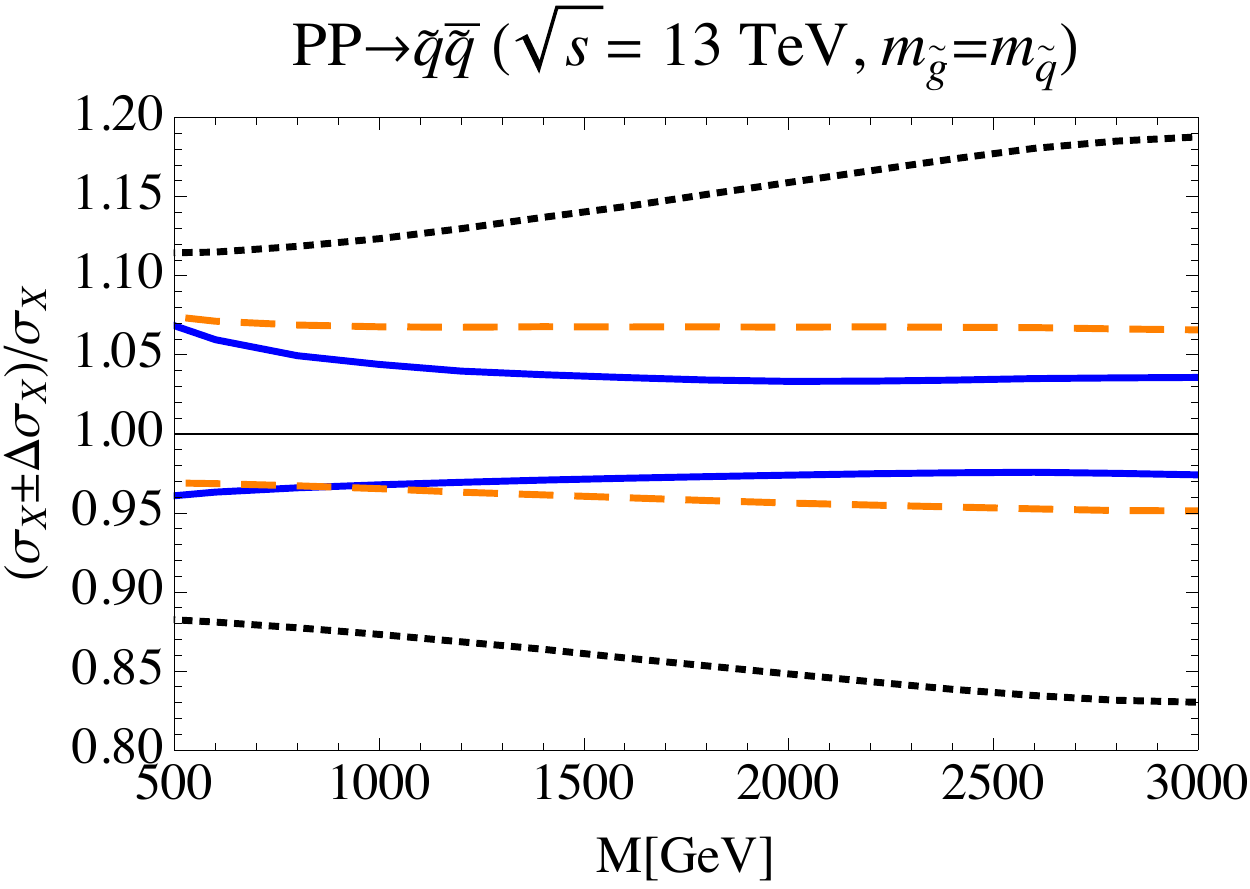}  
  \includegraphics[width=0.48 \linewidth]{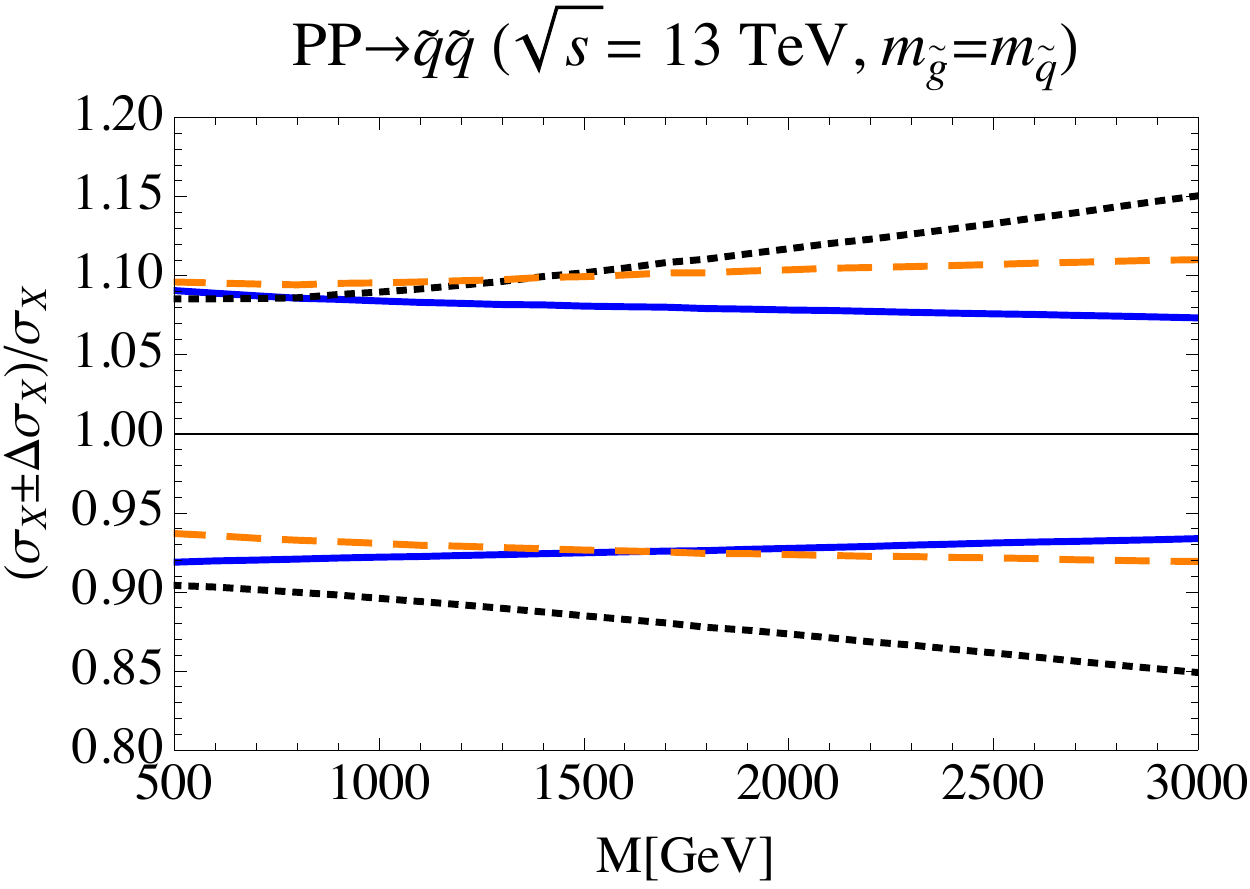}\\[0.5cm]  
 \includegraphics[width=0.48 \linewidth]{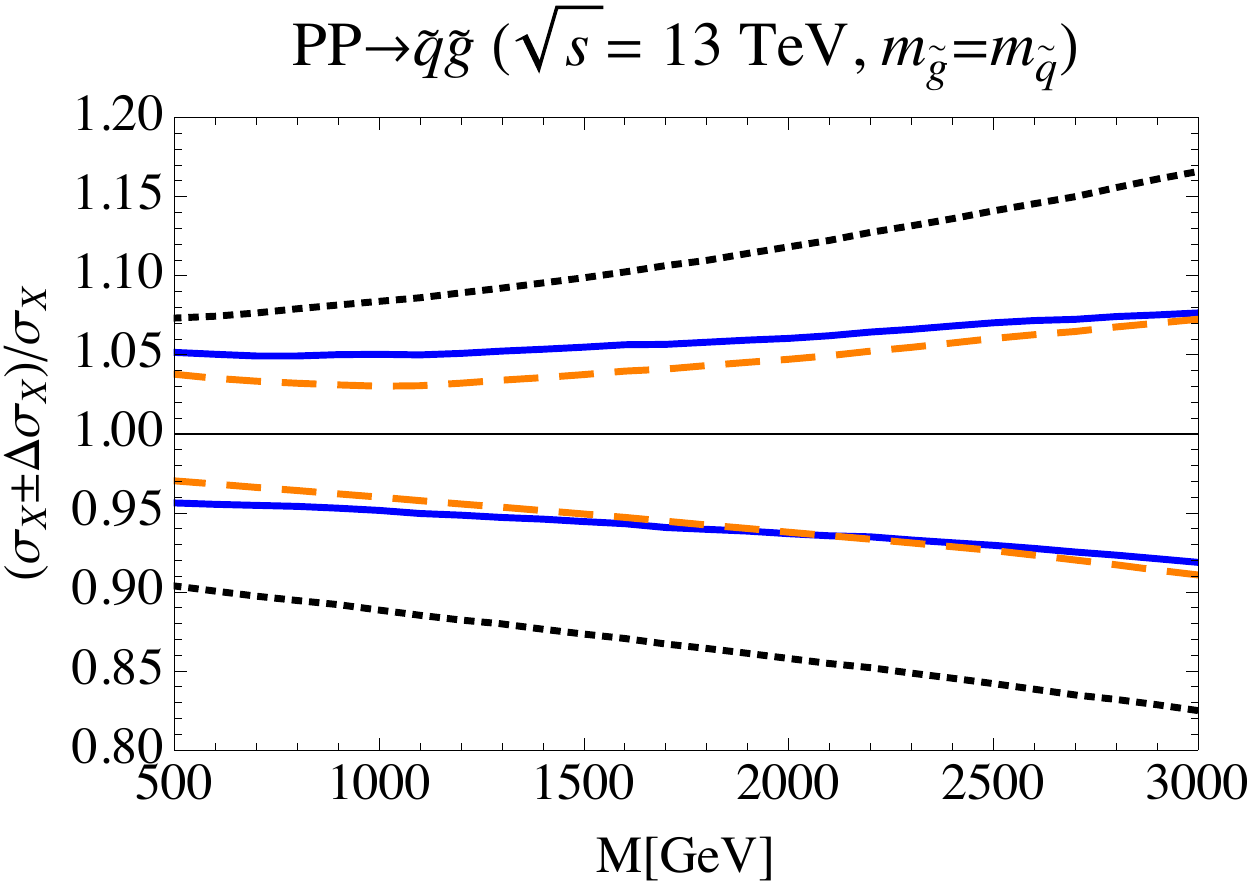}  
 \includegraphics[width=0.48 \linewidth]{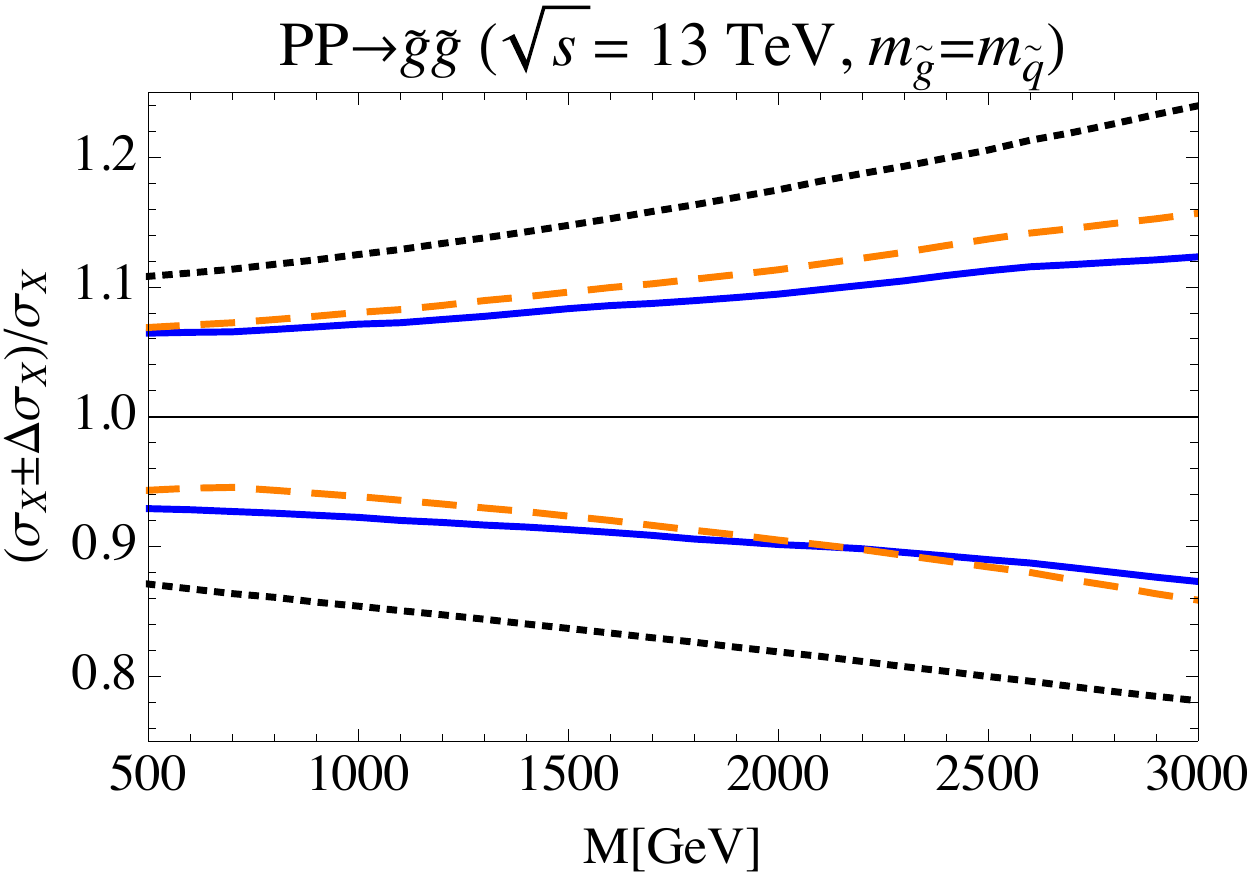}
 \caption{Total theoretical uncertainty (excluding PDF uncertainty) of  
   the NLO approximation (dotted black), NLL (dashed orange) and NNLL  
   (solid blue) resummed results at the LHC with $\sqrt{s}=\,13$~TeV.  
   The uncertainty estimate is given by the scale uncertainty, the  
   resummation uncertainty (for NLL and NNLL) and the estimate of  
   missing higher-order corrections (for NNLL).  All cross sections are  
   normalized to the one at the central value of the scales.}  
\label{fig:uncertainty}  
\end{figure}  
  
The estimate of the theory uncertainty of the NLO, NLL and NNLL approximations
is shown in Figure~\ref{fig:uncertainty}, normalized to the central value of
the respective prediction. Following the discussion in
Section~\ref{sec:setup}, the NLO uncertainty is estimated by the factorization
scale variation, the NLL uncertainty includes in addition the estimate of the
resummation ambiguities and the NNLL uncertainty further includes the estimate
of the two-loop constant.  The PDF$+\alpha_s$~uncertainty is not included in
the results shown in Figure~\ref{fig:uncertainty}.  One observes that the
uncertainty is reduced from up to $20\%$ at NLO to the $10\%$ level at NLL.
While a significant reduction of the uncertainty compared to the NLL results
is observed for the squark-antisquark production process at NNLL, only a
slight reduction or even an increase is observed for the other processes. We
further investigate the different sources of uncertainties in
Figure~\ref{fig:mufvar} and Table~\ref{tab:lhc13-resum} below, where it is
seen that this behaviour is due to the scale uncertainty, while the
resummation uncertainty is generally reduced at NNLL.

%%%%%%%%%%%%%%%%%%%%%%%%%%%%%%%%%%%%%%%%%%%%%%%%%%%%%%%%%%%%%%%%%%%%%%%%%%%%  
\begin{table}[p]  
\begin{center}  
 \resizebox{\textwidth}{!}{ 
\begin{tabular}{|c|c|l|l|l|c|}  
\hline  
$m_{\tilde q}$, $m_{\tilde g}\,(\mathrm{TeV})$&$\tilde s\tilde s'$ &  
$\sigma_{\mathrm{NLO}}(\mathrm{pb})$   &
$\sigma_{\mathrm{NLL}}(\mathrm{pb})$  &    
$\sigma_{\mathrm{NNLL}}(\mathrm{pb})$  & $ K_{\text{NNLL}}$  
\\\hline  
%%%%%%%%%%%%%%%%%  
$1.3$, $1.5$ &$ \tilde q\bar{\tilde q}$&
$2.81_ {-0.37}^{+0.36}{}_{-0.18}^{+0.18}\times 10^{-2}$ &   
$ 3.06_ {-0.11}^{+0.19}{}_{-0.19}^{+0.19}\times 10^{-2} $   &
$  3.49_ {-0.09}^{+0.13}{}_{-0.20}^{+0.20}\times 10^{-2} $ & $1.24$  \\  
%%%%
& $ \tilde q\tilde q$&$8.64_{-0.90}^{+0.74}{}_{-0.25}^{+0.25}\times 10^{-2}$
&$8.76_ {-0.55}^{+0.76}{}_{-0.25}^{+0.25}\times 10^{-2}$ &
$9.37_{-0.53}^{+0.56}{}_{-0.26}^{+0.26}\times 10^{-2}$&$1.08$ \\
%%%%
& $ \tilde q\tilde g$& $7.25_ {-0.86}^{+0.62}{}_{-0.50}^{+0.50}\times
10^{-2}$& $7.93_ {-0.37}^{+0.28}{}_{-0.51}^{+0.51}\times 10^{-2}$ &
$ 8.49_ {-0.46}^{+0.43}{}_ {-0.52}^{+0.52}\times 10^{-2}$& $1.17$ \\
%%%%
& $ \tilde g\tilde g$& $6.39_ {-1.01}^{+0.88}{}_{-1.34}^{+1.34}\times 10^{-3}$
&$8.03_{-0.66}^{+0.81}{}_{-1.37}^{+1.37}\times 10^{-3}$ &$8.71_ {-0.89}^{+0.82}{}_{-1.40}^{+1.40}\times 10^{-3}$
  &  $1.36$
\\\hline
%%%%%%%%%%%%%%%%%  
$1.5$, $1.3$ &$ \tilde q\bar{\tilde q}$&
 $9.33_ {-1.26}^{+1.23}{}_{-0.70}^{+0.70}\times 10^{-3} $  &
$ 1.02_{-0.04}^{+0.06}{}_{-0.07}^{+0.07}\times 10^{-2}$  
& $1.17_ {-0.04}^{+0.05}{}_{-0.08}^{+0.08}\times 10^{-2}$& $1.25$ \\
%%%%
& $ \tilde q\tilde q$&$4.20_ {-0.50}^{+0.47}{}_{-0.13}^{+0.13}\times 10^{-2}$
& $4.25_{-0.35}^{+0.46}{}_{-0.13}^{+0.13}\times 10^{-2}$& 
$4.58_ {-0.44}^{+0.47}{}_{-0.14}^{+0.14}\times 10^{-2}$&$1.09$  \\
%%%%
& $ \tilde q\tilde g$& $8.70_ {-1.15}^{+0.97}{}_ {-0.60}^{0.60}\times 10^{-2}$&  
$9.59_ {-0.47}^{+0.35}{}_{-0.61}^{+0.61}\times 10^{-2}$& 
$1.06_ {-0.05}^{+0.06}{}_{-0.06}^{+0.06}\times 10^{-1}$& $1.22$ \\
%%%%
& $ \tilde g\tilde g$&$2.45_{-0.40}^{+0.38}{}_{-0.42}^{+0.42}\times 10^{-2}$
&$2.97_ {-0.19}^{+0.24}{}_{-0.43}^{+0.43}\times 10^{-2}$ & 
 $3.36_ {-0.22}^{+0.22}{}_{-0.45}^{+0.45}\times 10^{-2}$& $1.37$  
\\\hline\hline
%%%%%%%%%%%%%%%%%  
$1.8$, $2$ &
$ \tilde q\bar{\tilde q}$&$1.66_ {-0.24}^{+0.24}{}_{-0.17}^{+0.17}\times 10^{-3} $ &   
$1.87_ {-0.08}^{+0.12}{}_{-0.18}^{+0.18}\times 10^{-3}$ 
& $2.16_ {-0.05}^{+0.07}{}_{-0.19}^{+0.19}\times 10^{-3} $& $1.30$ \\
%%%%
& $ \tilde q\tilde q$&$ 9.61_{-1.13}^{+0.99}{}_{-0.34}^{+0.34}\times 10^{-3}$&
$9.82_{-0.68}^{+0.92}{}_{-0.34}^{+0.34}\times 10^{-3}$&
$ 1.06_ {-0.06}^{+0.07}{}_{-0.04}^{+0.04}\times 10^{-2}$& $1.11$ \\
%%%%
& $ \tilde q\tilde g$& $5.15_{-0.69}^{+0.54}{}_{-0.59}^{+0.59}\times 10^{-3}$&
 $5.86_{-0.35}^{+0.27}{}_{-0.60}^{+0.60}\times 10^{-3}$&
$ 6.34_ {-0.39}^{+0.36}{}_{-0.60}^{+0.60}\times 10^{-3}$& $1.23$ \\
%%%%
& $ \tilde g\tilde g$&$3.18_{-0.56}^{+0.51}{}_{-1.05}^{+1.05}\times 10^{-4} $
& $4.33_{-0.45}^{+0.52}{}_{-1.08}^{+1.08}\times 10^{-4}$& 
$4.71_ {-0.54}^{+0.50}{}_{-1.12}^{+1.12}\times 10^{-4}$ & $1.48$ 
\\\hline  
%%%%%%%%%%%%%%%%%  
$2$, $1.8$ &$ \tilde q\bar{\tilde q}$&
$5.84_ {-0.87}^{+0.88}{}_{-0.79}^{+0.79}\times 10^{-4} $ &   
$6.62_ {-0.28}^{+0.42}{}_{-0.80}^{+0.80}\times 10^{-4}  $  
 & $7.69_ {-0.23}^{+0.27}{}_{-0.85}^{+0.85}\times 10^{-4}$& $1.32$  \\  
%%%%
& $ \tilde q\tilde q$& $4.68_{-0.61}^{+0.58}{}_{-0.19}^{+0.19}\times 10^{-3}$&
$4.78_ {-0.40}^{+0.53}{}_{-0.19}^{+0.19}\times 10^{-3}$
 & $5.23_ {-0.45}^{+0.49}{}_{-0.19}^{+0.19}\times 10^{-3}$& $1.12$ \\  
%%%%
& $ \tilde q\tilde g$& $5.96_{-0.86}^{+0.76}{}_{-0.68}^{+0.68}\times 10^{-3}$&
$6.82_ {-0.41}^{+0.28}{}_{-0.69}^{+0.69}\times 10^{-3}$&
$ 7.62_ {-0.44}^{+0.46}{}_{-0.70}^{+0.70}\times 10^{-3}$&$1.28$  \\  
%%%%
& $ \tilde g\tilde g$&$1.07_ {-0.19}^{+0.19}{}_{-0.30}^{+0.30}\times 10^{-3}$
& $1.40_ {-0.11}^{+0.14}{}_{-0.30}^{+0.30}\times 10^{-3}$ & 
$ 1.60_ {-0.13}^{+0.13}{}_{-0.32}^{+0.32}\times 10^{-3}$&  $1.49$
\\\hline\hline
%%%%%%%%%%%%%%%%%  
$2.3$, $2.5$ &$ \tilde q\bar{\tilde q}$&
$1.13_ {-0.18}^{+0.19}{}_{-0.24}^{+0.24}\times 10^{-4}$ &   
$1.32_ {-0.06}^{+0.09}{}_{-0.25}^{+0.25}\times 10^{-4} $  
 & $1.55_ {-0.04}^{+0.05}{}_{-0.26}^{+0.26}\times 10^{-4}$&$1.37$  \\  
%%%%
& $ \tilde q\tilde q$&$1.18_{-0.15}^{+0.14}{}_{-0.06}^{+0.06}\times 10^{-3}$ &
$1.21_ {-0.09}^{+0.12}{}_{-0.06}^{+0.06}\times 10^{-3}$
& $1.33_ {-0.08}^{+0.09}{}_{-0.06}^{+0.06}\times 10^{-3}$ 
& $1.13$ \\  
%%%%
& $ \tilde q\tilde g$& $4.29_ {-0.65}^{+0.55}{}_{-0.77}^{+0.77}\times
10^{-4}$& 
$5.13_ {-0.37}^{+0.30}{}_{-0.78}^{+0.78}\times 10^{-4}$
 & 
$5.60_{-0.39}^{+0.37}{}_{-0.79}^{+0.79}\times 10^{-4}$& $1.31$  \\  
& $ \tilde g\tilde g$&$1.88_{-0.36}^{+0.36}{}_{-0.93}^{+0.93}\times 10^{-5}$
&$2.84_ {-0.36}^{+0.41}{}_{-0.98}^{+0.98}\times 10^{-5}$ &
$ 3.11_ {-0.39}^{+0.38}{}_{-1.04}^{+1.04}\times 10^{-5}$ & $1.65$  
\\\hline
%%%%%%%%%%%%%%%%%  
$2.5$, $2.3$ &$ \tilde q\bar{\tilde q}$&
$ 3.98_{-0.64}^{+0.68}{}_{-1.27}^{+1.27}\times 10^{-5}$ &   
$4.72_ {-0.22}^{+0.30}{}_{-1.29}^{+1.29}\times 10^{-5} $  
 & $5.56_{-0.15}^{+0.20}{}_{-1.36}^{+1.36}\times 10^{-5}$& $1.40$ \\
%%%%  
& $ \tilde q\tilde q$&$5.55_{-0.79}^{+0.77}{}_{-0.33}^{+0.33}\times 10^{-4}$ &
$5.74_{-0.48}^{+0.65}{}_{-0.33}^{+0.33}\times 10^{-4}$ &
$6.36_ {-0.51}^{+0.56}{}_{-0.33}^{+0.33}\times 10^{-4}$ & $1.15$ \\ 
%%%% 
& $ \tilde q\tilde g$& $4.84_{-0.77}^{+0.71}{}_{-0.86}^{+0.86}\times 10^{-4}$&
$5.83_ {-0.41}^{+0.31}{}_{-0.88}^{+0.88}\times 10^{-4}$& 
$6.56_ {-0.43}^{+0.45}{}_{-0.90}^{+0.90}\times 10^{-4}$& $1.36$ \\  
%%%%
& $ \tilde g\tilde g$&$6.07_{-1.20}^{+1.25}{}_{-2.56}^{+2.56}\times 10^{-5}$
& $8.62_ {-0.85}^{+1.03}{}_{-2.66}^{+2.66}\times 10^{-5}$& 
$1.00_ {-0.09}^{+0.10}{}_{-0.28}^{+0.28}\times 10^{-4}$
 &  $1.65$
\\\hline\hline  
%%%%%%%%%%%%%%%%%  
  $2.8$, $3$ &$ \tilde q\bar{\tilde q}$&
  $7.94_ {-1.33}^{+1.45}{}_{-4.56}^{+4.56}\times 10^{-6}$ &   
$ 9.69_{-0.48}^{+0.63}{}_{-4.65}^{+4.65}\times 10^{-6} $  
 &$1.15_{-0.03}^{+0.04}{}_{-0.48}^{+0.48}\times 10^{-5}$ &$1.45$  \\  
%%%%
& $ \tilde q\tilde q$&$1.41_ {-0.20}^{+0.19}{}_{-0.11}^{+0.11}\times 10^{-4}$
&$1.47_ {-0.11}^{+0.15}{}_{-0.11}^{+0.11}\times 10^{-4} $& 
$1.64_ {-0.10}^{+0.11}{}_{-0.11}^{+0.11}\times 10^{-4}$& $1.17$ \\  
%%%%
& $ \tilde q\tilde g$& $3.62_{-0.61}^{+0.56}{}_{-0.97}^{+0.97}\times 10^{-5}$&
$ 4.62_ {-0.40}^{+0.32}{}_{-0.99}^{+0.99}\times 10^{-5}$& 
$5.10_{-0.41}^{+0.37}{}_{-1.01}^{+1.01}\times 10^{-5}$& $1.41$ \\  
%%%%
& $ \tilde g\tilde g$&$1.15_ {-0.24}^{+0.25}{}_{-0.82}^{+0.82}\times 10^{-6}$
&$1.97_ {-0.31}^{+0.33}{}_ {-0.92}^{+0.92}\times 10^{-6} $& 
$2.18_ {-0.31}^{+0.29}{}_{-1.01}^{+1.01}\times 10^{-6}$ &$1.90$  
\\\hline  
%%%%%%%%%%%%%%%%%%%%
  $3$, $2.8$ &$ \tilde q\bar{\tilde q}$&
  $2.83_ {-0.47}^{+0.51}{}_{-2.53}^{+2.53}\times 10^{-6}$ &   
$3.50_ {-0.17}^{+0.22}{}_{-2.56}^{+2.56}\times 10^{-6}$  &
$4.17_{-0.12}^{+0.15}{}_{-2.63}^{+2.63}\times 10^{-6}$ &$1.47$ \\
%%%%
 & $ \tilde q\tilde q$&$6.28_ {-0.96}^{+0.98}{}_{-0.60}^{+0.60}\times 10^{-5}$
 &$6.61_ {-0.56}^{+0.77}{}_{-0.59}^{+0.59}\times 10^{-5}$&
$7.43_ {-0.56}^{+0.61}{}_{-0.59}^{+0.59}\times 10^{-5}$
 & $1.18$\\
%%%%
& $ \tilde q\tilde g$& $4.02_ {-0.71}^{+0.69}{}_{-1.08}^{+1.08}\times
10^{-5}$& 
$5.16_ {-0.43}^{+0.34}{}_{-1.10}^{+1.10}\times 10^{-5}$ &
 $5.87_ {-0.44}^{+0.44}{}_{-1.13}^{+1.13}\times 10^{-5}$
 &$1.46$\\
%%%%  
& $ \tilde g\tilde g$&$3.70_{-0.80}^{+0.88}{}_{-2.27}^{+2.27}\times 10^{-6}$
&$5.90_ {-0.73}^{+0.85}{}_{-2.45}^{+2.45}\times 10^{-6} $& 
$6.94_ {-0.76}^{+0.78}{}_{-2.70}^{+2.70}\times 10^{-6}$
 &$1.87$  
\\\hline  
%%%%%%%%%%%%%%%%%  
\end{tabular}}  
\end{center}\caption{Predictions for the LHC with $\sqrt{s}=13$~TeV using the \texttt{PDF4LHC15\_nnlo\_30} PDFs.    
  The first and second error refer to the theoretical uncertainty, defined as in
  Figure~\ref{fig:uncertainty}, and the
  PDF$+\alpha_s$ uncertainty, respectively.}  
   \label{tab:lhc13-nnll}  
\end{table}

%%%%%%%%%%%%%%%%%%%%%%%%%%%%%%%%%%%%%%%%%%%%%%%%%%%%%%%%%%%%%%%%%%%%%%%%%%%%  
\begin{table}[p]  
\begin{center}  
 \resizebox{\textwidth}{!}{ 
\begin{tabular}{|c|c|l|l|l|c|}  
\hline  
$m_{\tilde q}$, $m_{\tilde g}\,(\mathrm{TeV})$&$\tilde s\tilde s'$ &  
$\sigma_{\mathrm{NLO}}(\mathrm{pb})$   &
$\sigma_{\mathrm{NLL}}(\mathrm{pb})$  &    
$\sigma_{\mathrm{NNLL}}(\mathrm{pb})$  & $ K_{\text{NNLL}}$ \\\hline 
%%%%%%%%%%%%%%%%%%%%%%%%%%%%%%%%%%%%%%%%%%%%%%%%
 $1.8,2$&
 $\tilde q \bar{\tilde q}$ & $2.82_{-0.40}^{+0.40}{}_{-0.26}^{+0.26}\times
   10^{-3}$ &
   $3.14_{-0.13}^{+0.20}{}_{-0.26}^{+0.26}\times 10^{-3}$
   & $3.61_{-0.09}^{+0.12}{}_{-0.28}^{+0.28}\times
   10^{-3}$ & $1.28$ \\
%%%%
& $\tilde q\tilde q$ & $1.40_{-0.16}^{+0.14}{}_{-0.05}^{+0.05}\times
   10^{-2}$ &
   $1.43_{-0.10}^{+0.13}{}_{-0.05}^{+0.05}\times 10^{-2}$
   & $1.54_{-0.09}^{+0.10}{}_{-0.05}^{+0.05}\times
   10^{-2}$ & $1.10$ \\
%%%%
& $\tilde q\tilde g$ & $8.61_{-1.11}^{+0.85}{}_{-0.86}^{+0.86}\times
   10^{-3}$ &
   $9.68_{-0.53}^{+0.41}{}_{-0.87}^{+0.87}\times 10^{-3}$
   & $1.04_{-0.06}^{+0.06}{}_{-0.09}^{+0.09}\times
   10^{-2}$ & $1.21$ \\
%%%%
& $\tilde g\tilde g$ & $6.02_{-1.01}^{+0.91}{}_{-1.75}^{+1.75}\times
   10^{-4}$ &
   $7.97_{-0.78}^{+0.91}{}_{-1.80}^{+1.80}\times 10^{-4}$
   & $8.63_{-0.96}^{+0.89}{}_{-1.84}^{+1.84}\times
   10^{-4}$ & $1.44$ \\\hline
%%%%%%%%%%%%%%
  $2.,1.8$ &
 $\tilde q \bar{\tilde q}$ & $1.06_{-0.15}^{+0.15}{}_{-0.12}^{+0.12}\times
   10^{-3}$ &
   $1.19_{-0.05}^{+0.07}{}_{-0.12}^{+0.12}\times 10^{-3}$
   & $1.37_{-0.04}^{+0.05}{}_{-0.13}^{+0.13}\times
   10^{-3}$ & $1.29$ \\
%%%%
& $\tilde q\tilde q$ & $7.20_{-0.90}^{+0.85}{}_{-0.26}^{+0.26}\times
   10^{-3}$ &
   $7.34_{-0.59}^{+0.80}{}_{-0.26}^{+0.26}\times 10^{-3}$
   & $7.98_{-0.70}^{+0.75}{}_{-0.27}^{+0.27}\times
   10^{-3}$ & $1.11$ \\
%%%%
& $\tilde q\tilde g$ & $9.94_{-1.39}^{+1.20}{}_{-0.99}^{+0.99}\times
   10^{-3}$ &
   $1.12_{-0.06}^{+0.04}{}_{-0.10}^{+0.10}\times 10^{-2}$
   & $1.25_{-0.07}^{+0.07}{}_{-0.10}^{+0.10}\times
   10^{-2}$ & $1.26$ \\
%%%%
& $\tilde g\tilde g$ & $1.91_{-0.33}^{+0.32}{}_{-0.47}^{+0.47}\times
   10^{-3}$ &
   $2.43_{-0.18}^{+0.22}{}_{-0.48}^{+0.48}\times 10^{-3}$
   & $2.77_{-0.21}^{+0.21}{}_{-0.50}^{+0.50}\times
   10^{-3}$ & $1.45$ \\\hline
%%%%%%%%%%%%%%%%%%%%%%%%
 $2.3,2.5$ & 
 $\tilde q \bar{\tilde q}$ & $2.28_{-0.35}^{+0.36}{}_{-0.38}^{+0.38}\times
   10^{-4}$ &
   $2.62_{-0.12}^{+0.17}{}_{-0.38}^{+0.38}\times 10^{-4}$
   & $3.05_{-0.07}^{+0.10}{}_{-0.41}^{+0.41}\times
   10^{-4}$ & 1.34 \\
& $\tilde q\tilde q$ & $1.96_{-0.25}^{+0.22}{}_{-0.09}^{+0.09}\times
   10^{-3}$ &
   $2.02_{-0.14}^{+0.20}{}_{-0.09}^{+0.09}\times 10^{-3}$
   & $2.20_{-0.13}^{+0.14}{}_{-0.09}^{+0.09}\times
   10^{-3}$ & $1.12$ \\
%%%%
& $\tilde q\tilde g$ & $8.31_{-1.20}^{+0.99}{}_{-1.27}^{+1.27}\times
   10^{-4}$ &
   $9.75_{-0.65}^{+0.52}{}_{-1.29}^{+1.29}\times 10^{-4}$
   & $1.06_{-0.07}^{+0.07}{}_{-0.13}^{+0.13}\times
   10^{-3}$ & $1.28$ \\
%%%%
& $\tilde g\tilde g$ & $4.21_{-0.78}^{+0.74}{}_{-1.81}^{+1.81}\times
   10^{-5}$ &
   $6.08_{-0.72}^{+0.83}{}_{-1.88}^{+1.88}\times 10^{-5}$
   & $6.64_{-0.79}^{+0.77}{}_{-1.97}^{+1.97}\times
   10^{-5}$ &$1.58$ \\\hline
%%%%%%%%%%%%%%%%
  $2.5,2.3$ &  $\tilde q \bar{\tilde q}$ & $8.62_{-1.34}^{+1.40}{}_{-1.96}^{+1.96}\times
   10^{-5}$ &
   $1.00_{-0.04}^{+0.06}{}_{-0.20}^{+0.20}\times 10^{-4}$
   & $1.18_{-0.03}^{+0.04}{}_{-0.21}^{+0.21}\times
   10^{-4}$ & $1.36$ \\
%%%%
& $\tilde q\tilde q$ & $9.89_{-1.34}^{+1.30}{}_{-0.50}^{+0.50}\times
   10^{-4}$ &
   $1.02_{-0.08}^{+0.11}{}_{-0.05}^{+0.05}\times 10^{-3}$
   & $1.12_{-0.09}^{+0.10}{}_{-0.05}^{+0.05}\times
   10^{-3}$ & $1.13$ \\
%%%% 
& $\tilde q\tilde g$ & $9.38_{-1.43}^{+1.29}{}_{-1.43}^{+1.43}\times
   10^{-4}$ &
   $1.11_{-0.07}^{+0.05}{}_{-0.15}^{+0.15}\times 10^{-3}$
   & $1.24_{-0.08}^{+0.08}{}_{-0.15}^{+0.15}\times
   10^{-3}$ & $1.32$ \\
%%%%
& $\tilde g\tilde g$ & $1.27_{-0.24}^{+0.24}{}_{-0.46}^{+0.46}\times
   10^{-4}$ &
   $1.74_{-0.16}^{+0.19}{}_{-0.48}^{+0.48}\times 10^{-4}$
   & $2.00_{-0.17}^{+0.18}{}_{-0.50}^{+0.50}\times
   10^{-4}$ & $1.58$ \\\hline
%%%%%%%%%%%%%%%%%%%%%%%%
 $2.8,3$ & 
 $\tilde q \bar{\tilde q}$ & $1.91_{-0.31}^{+0.34}{}_{-0.73}^{+0.73}\times
   10^{-5}$ &
   $2.29_{-0.11}^{+0.15}{}_{-0.74}^{+0.74}\times 10^{-5}$
   & $2.70_{-0.06}^{+0.09}{}_{-0.78}^{+0.78}\times
   10^{-5}$ & $1.41$ \\
%%%%
& $\tilde q\tilde q$ & $2.77_{-0.38}^{+0.36}{}_{-0.18}^{+0.18}\times
   10^{-4}$ &
   $2.88_{-0.22}^{+0.29}{}_{-0.18}^{+0.18}\times 10^{-4}$
   & $3.19_{-0.19}^{+0.21}{}_{-0.18}^{+0.18}\times
   10^{-4}$ & $1.15$ \\
%%%%
& $\tilde q\tilde g$ & $8.41_{-1.35}^{+1.19}{}_{-1.91}^{+1.91}\times
   10^{-5}$ &
   $1.04_{-0.08}^{+0.07}{}_{-0.19}^{+0.19}\times 10^{-4}$
   & $1.14_{-0.09}^{+0.08}{}_{-0.20}^{+0.20}\times
   10^{-4}$ & $1.36$ \\
%%%%
& $\tilde g\tilde g$ & $3.13_{-0.63}^{+0.64}{}_{-1.91}^{+1.91}\times
   10^{-6}$ &
   $5.05_{-0.72}^{+0.79}{}_{-2.07}^{+2.07}\times 10^{-6}$
   & $5.56_{-0.75}^{+0.70}{}_{-2.23}^{+2.23}\times
   10^{-6}$ & $1.78$ \\\hline
%%%%%%%%%%%%%%%%%%%%
$3.,2.8$ & $\tilde q \bar{\tilde q}$ & $7.24_{-1.18}^{+1.27}{}_{-4.15}^{+4.15}\times
   10^{-6}$ &
   $8.78_{-0.42}^{+0.55}{}_{-4.22}^{+4.22}\times 10^{-6}$
   & $1.04_{-0.03}^{+0.04}{}_{-0.44}^{+0.44}\times
   10^{-5}$ & $1.44$ \\
%%%%
& $\tilde q\tilde q$ & $1.34_{-0.20}^{+0.20}{}_{-0.10}^{+0.10}\times
   10^{-4}$ &
   $1.40_{-0.12}^{+0.16}{}_{-0.10}^{+0.10}\times 10^{-4}$
   & $1.56_{-0.12}^{+0.13}{}_{-0.10}^{+0.10}\times
   10^{-4}$ &$ 1.16$ \\
%%%%
& $\tilde q\tilde g$ & $9.34_{-1.56}^{+1.48}{}_{-2.12}^{+2.12}\times
   10^{-5}$ &
   $1.16_{-0.09}^{+0.07}{}_{-0.21}^{+0.21}\times 10^{-4}$
   & $1.31_{-0.10}^{+0.09}{}_{-0.22}^{+0.22}\times
   10^{-4}$ & $1.41$ \\
%%%%
& $\tilde g\tilde g$ & $9.30_{-1.92}^{+2.05}{}_{-4.90}^{+4.90}\times
   10^{-6}$ &
   $1.40_{-0.16}^{+0.19}{}_{-0.52}^{+0.52}\times 10^{-5}$
   & $1.64_{-0.17}^{+0.17}{}_{-0.56}^{+0.56}\times
   10^{-5}$ & $1.76$ \\\hline
%%%%%%%%%%%%%%%%
$ 3.3,3.5$ &
 $\tilde q \bar{\tilde q}$ & $1.68_{-0.28}^{+0.31}{}_{-1.66}^{+1.66}\times
   10^{-6}$ &
   $2.07_{-0.10}^{+0.13}{}_{-1.68}^{+1.68}\times 10^{-6}$
   & $2.46_{-0.06}^{+0.08}{}_{-1.73}^{+1.73}\times
   10^{-6}$ & $1.47$ \\
%%%%
& $\tilde q\tilde q$ & $3.68_{-0.55}^{+0.54}{}_{-0.38}^{+0.38}\times
   10^{-5}$ &
   $3.89_{-0.30}^{+0.41}{}_{-0.37}^{+0.37}\times 10^{-5}$
   & $4.35_{-0.26}^{+0.28}{}_{-0.37}^{+0.37}\times
   10^{-5}$ & $1.18$ \\
%%%% 
& $\tilde q\tilde g$ & $8.18_{-1.44}^{+1.37}{}_{-2.68}^{+2.68}\times
   10^{-6}$ &
   $1.08_{-0.10}^{+0.08}{}_{-0.27}^{+0.27}\times 10^{-5}$
   & $1.20_{-0.10}^{+0.09}{}_{-0.28}^{+0.28}\times
   10^{-5}$ & $1.47$ \\
%%%%
& $\tilde g\tilde g$ & $2.28_{-0.50}^{+0.53}{}_{-1.90}^{+1.90}\times
   10^{-7}$ &
   $4.19_{-0.69}^{+0.74}{}_{-2.26}^{+2.26}\times 10^{-7}$
   & $4.65_{-0.69}^{+0.66}{}_{-2.55}^{+2.55}\times
   10^{-7}$ & $2.04$ \\\hline
%%%%%%%%%%%%%%
 $3.5,3.3$ &
 $\tilde q \bar{\tilde q}$ & $6.72_{-1.11}^{+1.20}{}_{-9.60}^{+9.60}\times
   10^{-7}$ &
   $8.32_{-0.39}^{+0.52}{}_{-9.65}^{+9.65}\times 10^{-7}$
   & $9.93_{-0.28}^{+0.35}{}_{-9.82}^{+9.82}\times
   10^{-7}$ & $1.48$ \\
%%%%
& $\tilde q\tilde q$ & $1.68_{-0.27}^{+0.27}{}_{-0.21}^{+0.21}\times
   10^{-5}$ &
   $1.78_{-0.15}^{+0.21}{}_{-0.20}^{+0.20}\times 10^{-5}$
   & $2.02_{-0.14}^{+0.16}{}_{-0.20}^{+0.20}\times
   10^{-5}$ & $1.20$ \\
%%%%
& $\tilde q\tilde g$ & $8.98_{-1.63}^{+1.63}{}_{-2.93}^{+2.93}\times
   10^{-6}$ &
   $1.19_{-0.11}^{+0.09}{}_{-0.30}^{+0.30}\times 10^{-5}$
   & $1.36_{-0.11}^{+0.11}{}_{-0.31}^{+0.31}\times
   10^{-5}$ & $1.52$ \\
%%%%
& $\tilde g\tilde g$ & $6.89_{-1.53}^{+1.72}{}_{-5.02}^{+5.02}\times
   10^{-7}$ &
   $1.17_{-0.16}^{+0.18}{}_{-0.57}^{+0.57}\times 10^{-6}$
   & $1.38_{-0.16}^{+0.16}{}_{-0.64}^{+0.64}\times
   10^{-6}$ & $2.01$ \\\hline
%%%%%%%%%%%%%%%%%
\end{tabular} } 
\end{center}\caption{Predictions for the LHC with $\sqrt{s}=14$~TeV.    
  The errors are defined as in Table~\ref{tab:lhc13-nnll}.}  
   \label{tab:lhc14-nnll}  
\end{table}  
%%%%%%%%%%%%%%%%%%%%%%%%%%%%%%%%%%%%%%%%%%%%%%%%%%%%%%%%%%%%%%%%%%%%%%%%%%%%%
  
Numerical predictions for the cross sections of squark and gluino production
at the LHC at $\sqrt{s}=13\,\mathrm{TeV}$ are presented in
Table~\ref{tab:lhc13-nnll} for a sample of squark and gluino masses from
$1$--$3\,\mathrm{TeV}$.  The corresponding results for $\sqrt
s=14\,\mathrm{TeV}$ are shown in Table~\ref{tab:lhc14-nnll} for the mass range
from $1.5$--$3.5\,\mathrm{TeV}$.  In these results, we included the
theoretical uncertainty (scale, resummation and higher-order uncertainty as in
Figure~\ref{fig:uncertainty}) and the PDF$+\alpha_s$ uncertainty determined as
discussed in Section~\ref{sec:setup}.  For the processes involving squarks,
the PDF$+\alpha_s$ uncertainty is of the order of $\pm 5$--$10\%$ for lighter
sparticle masses. For heavier sparticles, the larger uncertainty in the gluon
PDF becomes visible, which leads to a growth of the PDF+$\alpha_s$ uncertainty
to $\pm 30\%$ for squark-gluino and over $\pm 100\%$ for squark-antisquark
production at $\sqrt{s}=14\,\mathrm{TeV}$. For the gluino-pair production
process the PDF uncertainty grows from $\pm 20\%$ to over $\pm 80\%$ at the
highest considered masses at $\sqrt{s}=14\,\mathrm{TeV}$. For the
squark-squark production process, where the gluon PDF does not enter at tree
level, the relative PDF+$\alpha_s$ uncertainty is smaller and remains below
$\pm 10\%$ throughout the mass range.  Therefore the PDF uncertainty is
smaller than the NLO scale uncertainty or comparable at smaller masses,
whereas the uncertainty due to the poorly determined gluon PDF becomes very
large at high masses, in particular for squark-antisquark and gluino-pair
production.  It should be taken into account, however, that the largest PDF
uncertainties appear for cross sections of the order of $10^{-7}\,\mathrm{pb}$
that are beyond the reach of even the high-luminosity phase of the LHC.  In
general, the NLL resummation reduces the theory uncertainty below the PDF
uncertainty, apart from squark-squark production where the PDF uncertainties
are very small and typically below the theory uncertainties.  Consistent with
Figure~\ref{fig:uncertainty}, the NNLL resummation further reduces the theory
uncertainty strongly for squark-antisquark production, whereas the effect for
the other processes is moderate.  The size of the NNLL corrections is
consistent with Figure~\ref{fig:KNNLL13}, with corrections relative to NLO of
up to a factor of two for gluino pair production at the highest considered
masses.  Although the PDF4LHC15 set combines the results of several PDF fits,
it should be taken into account, however, that different PDF sets can lead to
results that are not covered by the PDF4LHC15 error estimate, or have a much
larger estimate of the PDF uncertainty, in particular for processes involving
the gluon PDF~\cite{Langenfeld:2012ti,Borschensky:2014cia,Ball:2014uwa}.  We
refer to~\cite{Beenakker:2015rna} for a recent discussion of the effect of
different sets of PDF fits on predictions with NLL soft-gluon resummation.
This includes a PDF set obtained using threshold-resummed cross sections in
the PDF fit~\cite{Bonvini:2015ira} that is in principle appropriate for
resummed calculations, but currently has large uncertainties due to a reduced
data set used in the fit.  For squark and gluino production, non-trivial
changes on the central values were found for the resummed PDFs that, however,
lie inside of the uncertainty band of the standard PDF sets.

\begin{figure}[t!]  
  \centering  
   \includegraphics[width=0.48 \linewidth]{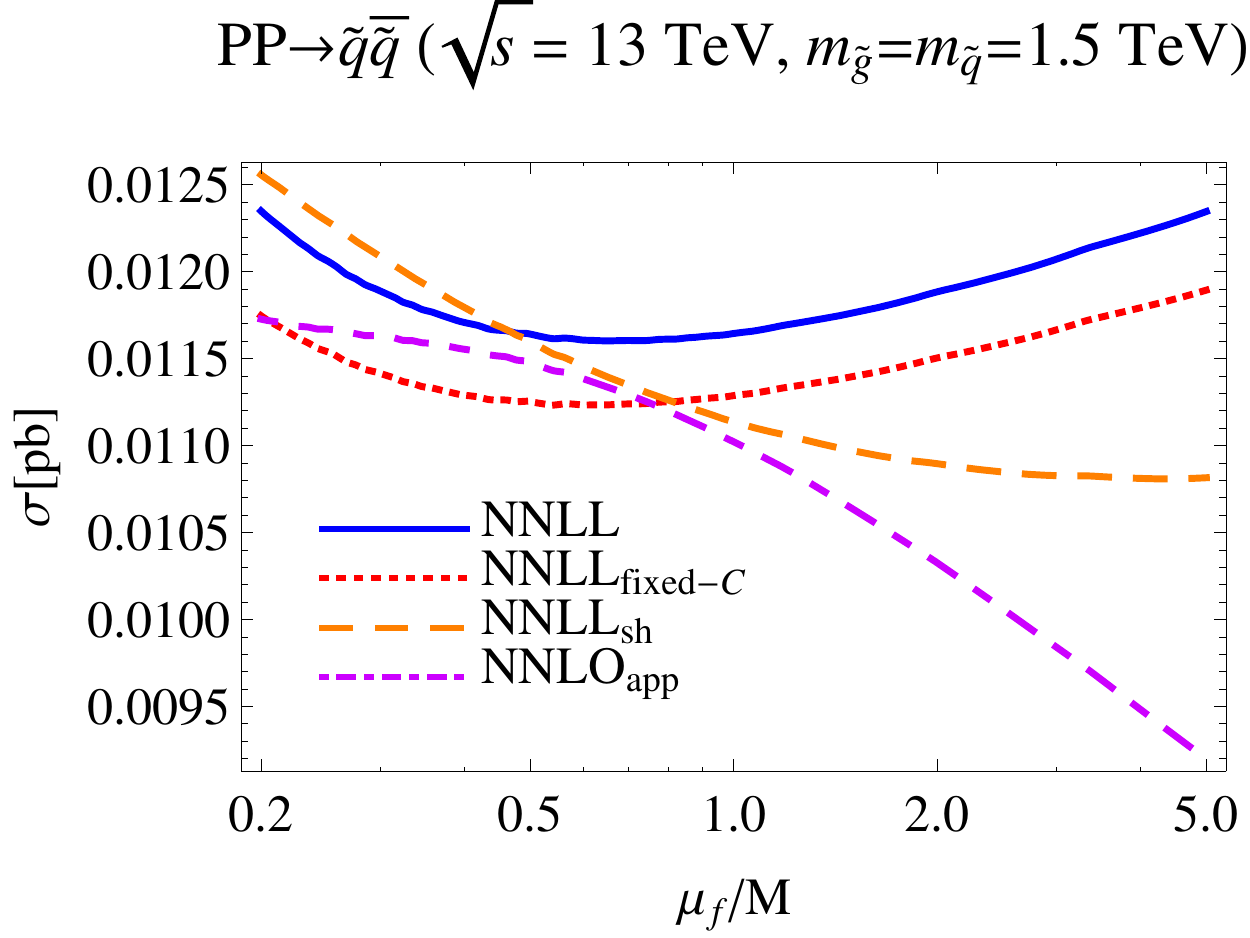}  
  \includegraphics[width=0.48 \linewidth]{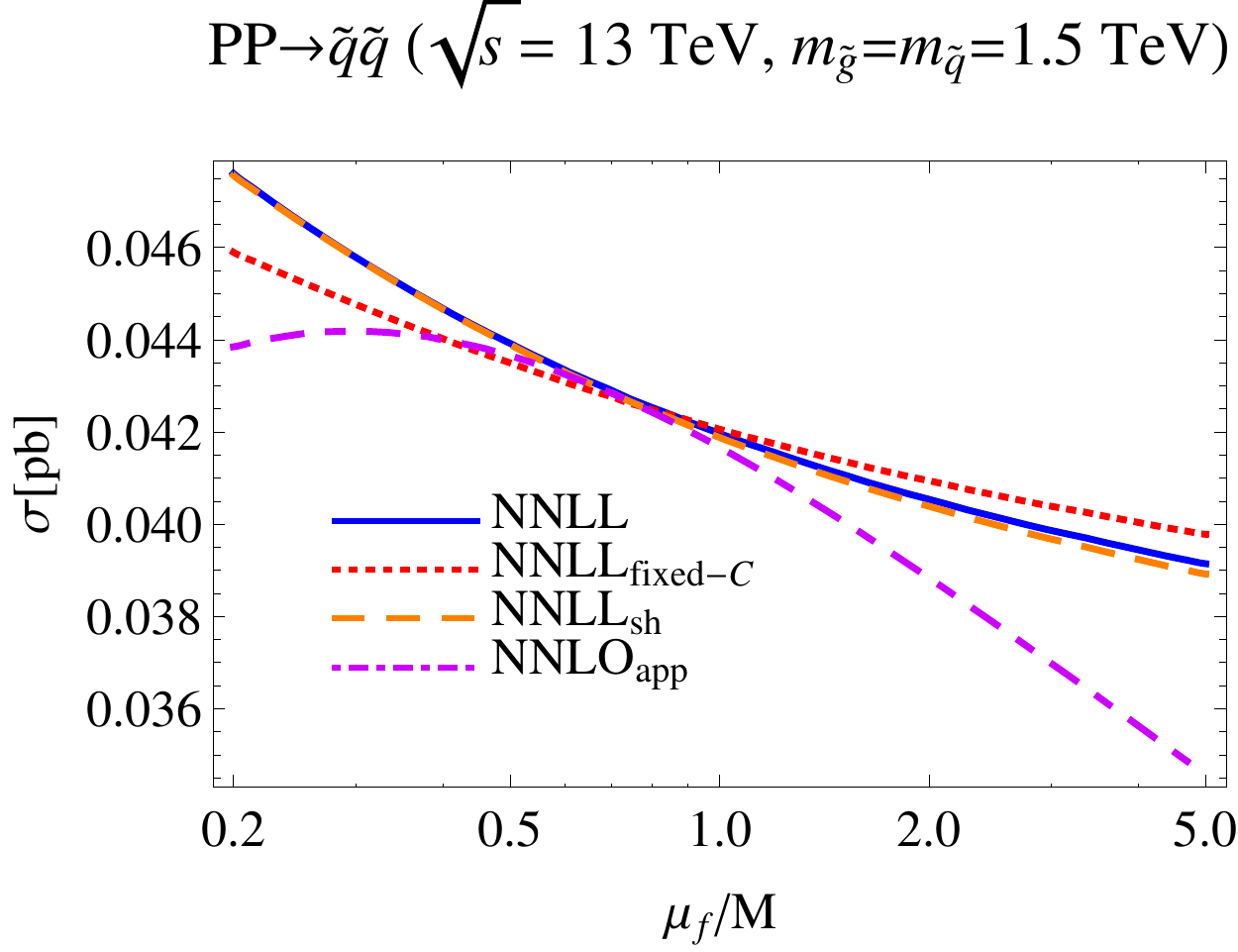}\\[.5cm]    
  \includegraphics[width=0.48 \linewidth]{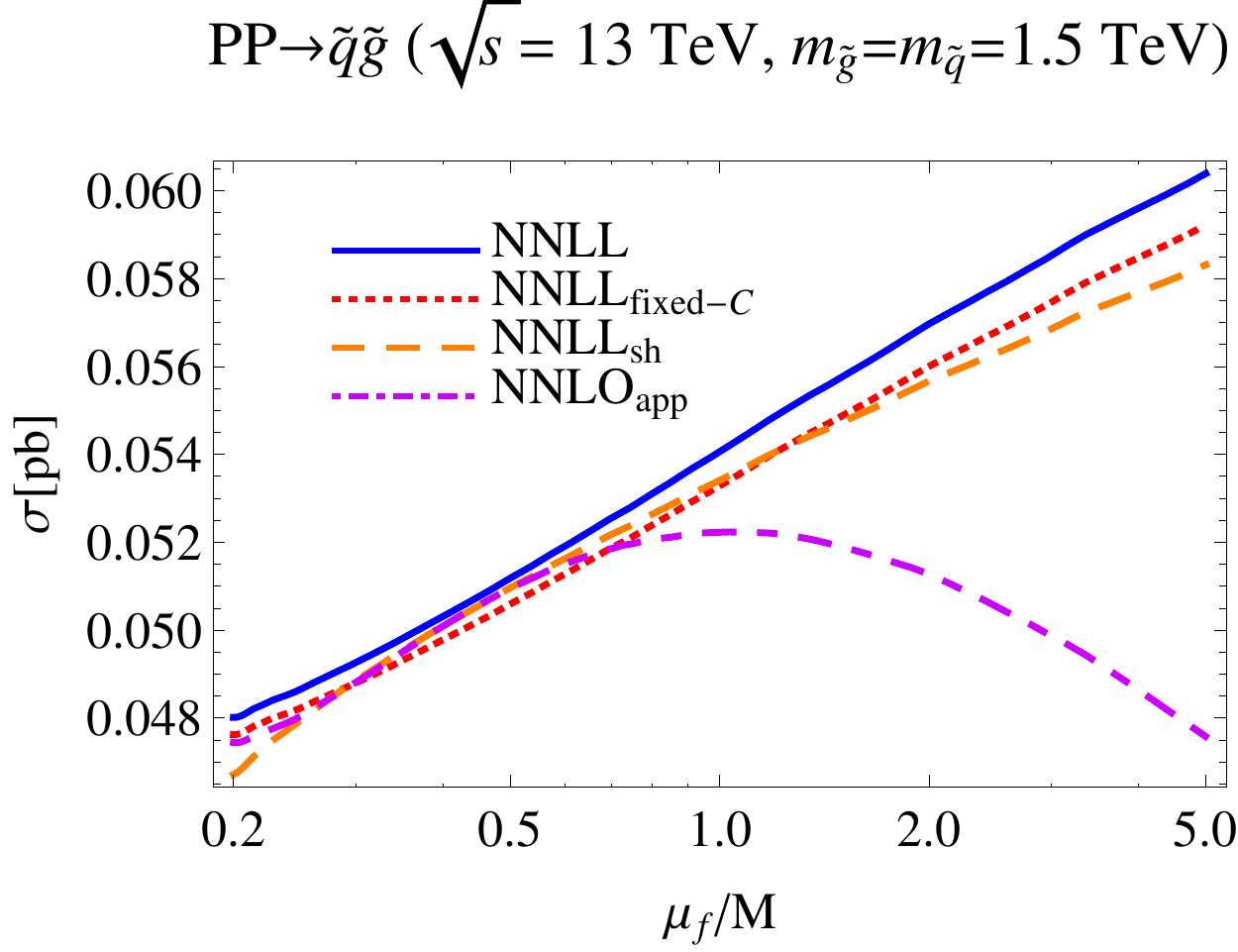}  
  \includegraphics[width=0.48 \linewidth]{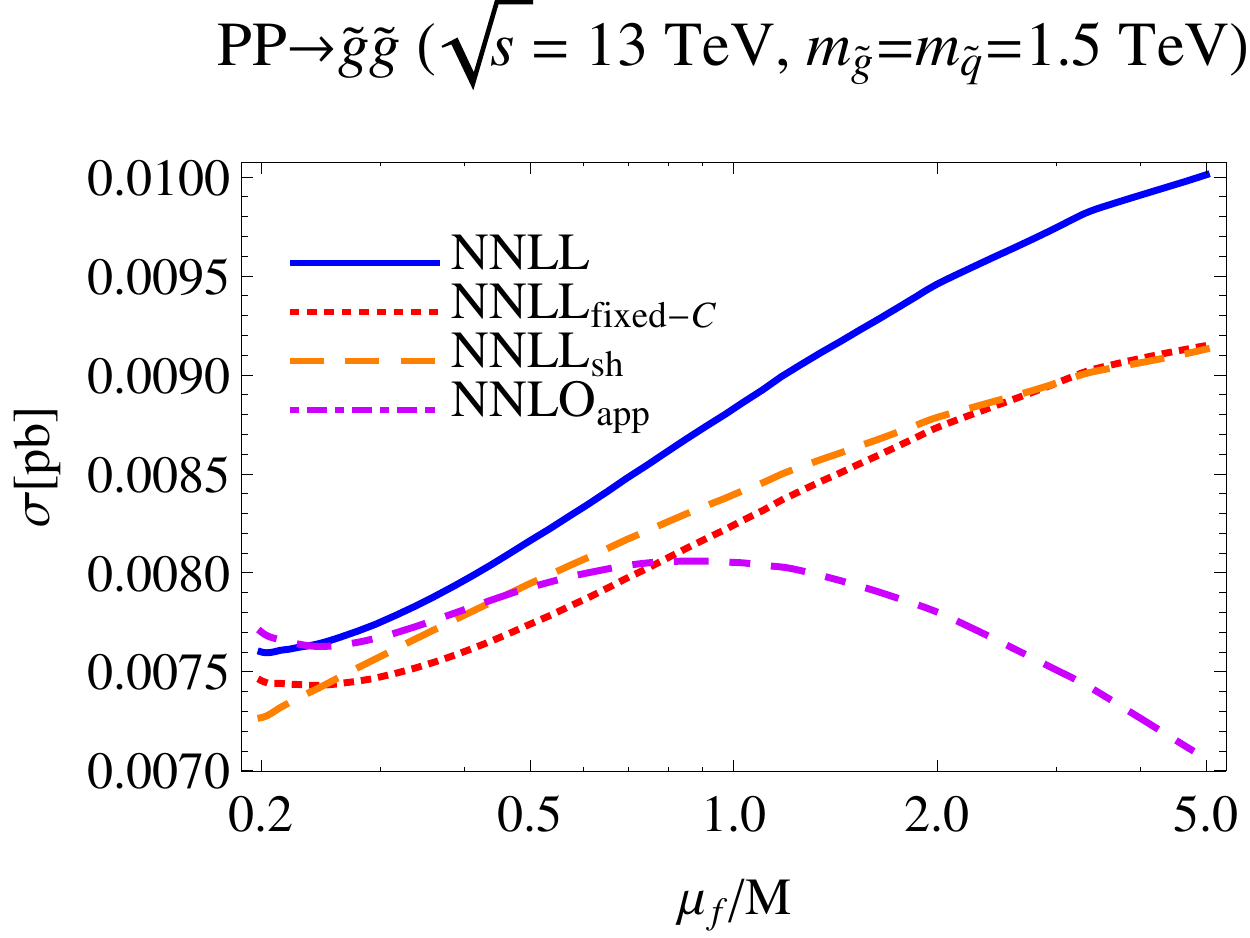}    
 \caption{Cross section for squark and gluino production at the LHC with $\sqrt{ s}=13$~TeV for full NNLL resummation (solid blue), NNLL with  
  fixed-order Coulomb corrections (dotted red), $\text{NNLL}_{\mathrm{sh}}$ (dashed orange) and approximate NNLO  
  (dot-dashed pink), as a function of the ratio of the factorization  
  scale and average produced mass. 
}  
\label{fig:mufvar}  
\end{figure}  
  
The scale dependence of various higher-order approximations is shown  
in Figure~\ref{fig:mufvar}. In addition to the approximate NNLO  
corrections, the full NNLL result and the NNLL soft resummation with  
fixed-order Coulomb corrections $\text{NNLL}_{\text{fixed-C}}$, we  
also consider an approximation $\text{NNLL}_{\text{sh}}$ where Coulomb  
corrections are set to zero in the resummation formula and only  
included through the matching to $\text{NNLO}_{\text{app}}$. It is  
seen that soft resummation in $\text{NNLL}_{\text{sh}}$ significantly affects
the shape of the scale dependence and reduces the  
scale uncertainty for squark-antisquark and squark-squark production.
For the mass values considered here, the approximate NNLO cross section has a
maximum near the
default scale $\mu_f=M$, so the variation of the factorization scale alone
leads to an asymmetrical error estimate, whereas the soft resummation leads to
a more symmetrical behaviour.  
 The  
inclusion of soft-Coulomb interference in the  
$\text{NNLL}_{\text{fixed-C}}$ prediction further reduces the  
uncertainty for the case of squark-antisquark and squark-squark  
production, while it is  increased for gluino-pair and squark-gluino  
production. This observation is consistent with the results obtained  
in the Mellin-space approach for gluino-pair production~\cite{Beenakker:2014sma}.  
The further resummation of Coulomb corrections in the NNLL prediction provides an overall  
shift of the cross section while the scale dependence  is  
qualitatively similar to the $\text{NNLL}_{\text{fixed-C}}$ approximation.  

%%%%%%%%%%%%%%%%%%%%%%%%%%%%%%%%%%%%%%%%%%%%%%%%%%%%%%%%%%%%%%%%%%%%%%%  
\begin{table}[t]  
\begin{center}  
 \resizebox{\textwidth}{!}{ 
%%%%%%%%%%%%%%%%%%  
\begin{tabular}{|@{}c@{}|@{}c@{}|l|l|l|l|}  
\hline  
\,$m_{\tilde q}$,$m_{\tilde g}(\mathrm{TeV})$&\,$\tilde s\tilde s'$ &  
$\sigma_{\mathrm{NLL}}(\mathrm{pb})$   &
$\sigma_{\mathrm{NNLO}_{\mathrm{app}}}(\mathrm{pb})$  &    
$\sigma_{\mathrm{NNLL}_{\mathrm{fixed-C}}}(\mathrm{pb})$ &
$\sigma_{\mathrm{NNLL}}(\mathrm{pb})$ 
\\\hline  
%%%%%%%%%%%%%%%%%
 1.3,1.5 &
 $\tilde q\overline{\tilde q}$ & $3.06_{-0.02}^{+0.14}{}_{-0.11}^{+0.12}\times 10^{-2}$ &
   $3.32_{-0.18}^{+0.09}{}_{-0.07}^{+0.07}\times 10^{-2}$ &
   $3.39_{-0.04}^{+0.08}{}_{-0.09}^{+0.12}\times 10^{-2}$ &
   $3.49_{-0.03}^{+0.09}{}_{-0.08}^{+0.09}\times 10^{-2}$ \\
%%%%
& $\tilde q\tilde q$ & $8.76_{-0.55}^{+0.75}{}_{-0.07}^{+0.09}\times 10^{-2}$ &
   $9.31_{-0.53}^{+0.31}{}_{-0.46}^{+0.46}\times 10^{-2}$ &
   $9.38_{-0.19}^{+0.21}{}_{-0.46}^{+0.48}\times 10^{-2}$ &
   $9.37_{-0.25}^{+0.32}{}_{-0.47}^{+0.46}\times 10^{-2}$ \\
%%%% 
&$\tilde q\tilde g$ & $7.93_{-0.02}^{+0.20}{}_{-0.37}^{+0.21}\times 10^{-2}$ &
   $8.27^{+0.00}_{-0.26}{}_{-0.06}^{+0.06}\times 10^{-2}$ &
   $8.38_{-0.42}^{+0.39}{}_{-0.09}^{+0.07}\times 10^{-2}$ &
   $8.49_{-0.46}^{+0.42}{}_{-0.08}^{+0.06}\times 10^{-2}$ \\
%%%% 
&$\tilde g\tilde g$ & $8.03_{-0.32}^{+0.55}{}_{-0.58}^{+0.59}\times 10^{-3}$ &
   $7.97^{+0.00}_{-0.25}{}_{-0.41}^{+0.41}\times 10^{-3}$ &
   $8.14_{-0.61}^{+0.55}{}_{-0.42}^{+0.41}\times 10^{-3}$ &
   $8.71_{-0.77}^{+0.69}{}_{-0.44}^{+0.44}\times 10^{-3}$ \\\hline
%%%%%%%%%%%%%%%%%%
 1.5,1.3 & 
 $\tilde q\overline{\tilde q}$ & $1.02_{-0.01}^{+0.05}{}_{-0.04}^{+0.04}\times 10^{-2}$ &
   $1.11_{-0.07}^{+0.04}{}_{-0.04}^{+0.04}\times 10^{-2}$ &
   $1.13_{-0.01}^{+0.02}{}_{-0.04}^{+0.05}\times 10^{-2}$ &
   $1.17_{-0.01}^{+0.02}{}_{-0.04}^{+0.04}\times 10^{-2}$ \\
%%%%
& $\tilde q\tilde q$ & $4.25_{-0.34}^{+0.46}{}_{-0.04}^{+0.04}\times 10^{-2}$ &
   $4.55_{-0.34}^{+0.26}{}_{-0.39}^{+0.39}\times 10^{-2}$ &
   $4.59_{-0.15}^{+0.20}{}_{-0.39}^{+0.40}\times 10^{-2}$ &
   $4.58_{-0.19}^{+0.26}{}_{-0.39}^{+0.39}\times 10^{-2}$ \\
%%%%
 &$\tilde q\tilde g$ & $9.59^{+0.23}_{-0.00}{}_{-0.47}^{+0.27}\times 10^{-2}$ &
   $1.02^{+0.00}_{-0.03}{}_{-0.02}^{+0.02}\times 10^{-1}$ &
   $1.05_{-0.05}^{+0.05}{}_{-0.03}^{+0.03}\times 10^{-1}$ &
   $1.06_{-0.05}^{+0.06}{}_{-0.03}^{+0.02}\times 10^{-1}$ \\
%%%% 
&$\tilde g\tilde g$ & $2.97_{-0.03}^{+0.12}{}_{-0.19}^{+0.20}\times 10^{-2}$ &
   $3.08_{-0.12}^{+0.02}{}_{-0.08}^{+0.08}\times 10^{-2}$ &
   $3.14_{-0.13}^{+0.15}{}_{-0.09}^{+0.10}\times 10^{-2}$ &
   $3.36_{-0.19}^{+0.19}{}_{-0.10}^{+0.10}\times 10^{-2}$ \\\hline
%%%%%%%%%%%%%%%%%  
 1.8,2  & 
 $\tilde q\overline{\tilde q}$ & $1.87_{-0.02}^{+0.08}{}_{-0.08}^{+0.08}\times 10^{-3}$ &
   $2.02_{-0.14}^{+0.09}{}_{-0.04}^{+0.04}\times 10^{-3}$ &
   $2.08_{-0.01}^{+0.04}{}_{-0.06}^{+0.08}\times 10^{-3}$ &
   $2.16_{-0.01}^{+0.04}{}_{-0.05}^{+0.05}\times    10^{-3}$ \\
%%%%
 &$\tilde q\tilde q$ & $9.82_{-0.67}^{+0.91}{}_{-0.13}^{+0.12}\times 10^{-3}$ &
   $1.05_{-0.07}^{+0.05}{}_{-0.05}^{+0.05}\times 10^{-2}$ &
   $1.06_{-0.02}^{+0.03}{}_{-0.06}^{+0.06}\times 10^{-2}$ &
   $1.06_{-0.03}^{+0.04}{}_{-0.06}^{+0.05}\times 10^{-2}$ \\
%%%%
&$\tilde q\tilde g$ & $5.86_{-0.05}^{+0.19} {}_{-0.34}^{+0.19}\times 10^{-3}$ &
   $6.07^{+0.00}_{-0.15}{}_{-0.04}^{+0.04}\times 10^{-3}$ &
   $6.23_{-0.35}^{+0.33}{}_{-0.08}^{+0.05}\times 10^{-3}$ &
   $6.34_{-0.38}^{+0.36}{}_{-0.08}^{+0.04}\times 10^{-3}$ \\
%%%%
& $\tilde g\tilde g$ & $4.33_{-0.27}^{+0.38}{}_{-0.37}^{+0.36}\times 10^{-4}$ &
   $4.14^{+0.00}_{-0.16} {}_{-0.23}^{+0.23}\times 10^{-4}$ &
   $4.35_{-0.38}^{+0.35}{}_{-0.23}^{+0.23}\times 10^{-4}$ &
   $4.71_{-0.48}^{+0.43}{}_{-0.24}^{+0.25}\times 10^{-4}$ \\\hline
%%%%%%%%%%%%%%%% 
2.,1.8 & 
 $\tilde q\overline{\tilde q}$ & $6.62_{-0.07}^{+0.28}{}_{-0.27}^{+0.31}\times 10^{-4}$ &
   $7.15_{-0.53}^{+0.35}{}_{-0.20}^{+0.20}\times 10^{-4}$ &
   $7.39_{-0.04}^{+0.11}{}_{-0.24}^{+0.31}\times 10^{-4}$ &
   $7.69_{-0.04}^{+0.14}{}_{-0.22}^{+0.23}\times 10^{-4}$ \\
%%%%
& $\tilde q\tilde q$ & $4.78_{-0.39}^{+0.53}{}_{-0.07}^{+0.05}\times 10^{-3}$ &
   $5.16_{-0.42}^{+0.32}{}_{-0.40}^{+0.40}\times 10^{-3}$ &
   $5.24_{-0.17}^{+0.22}{}_{-0.40}^{+0.41}\times 10^{-3}$ &
   $5.23_{-0.21}^{+0.28}{}_{-0.40}^{+0.40}\times 10^{-3}$ \\
%%%% 
&$\tilde q\tilde g$ & $6.82_{-0.01}^{+0.17}{}_{-0.41}^{+0.22}\times 10^{-3}$ &
   $7.20_{-0.26}^{+0.02}{}_{-0.13}^{+0.13}\times 10^{-3}$ &
   $7.48_{-0.39}^{+0.42}{}_{-0.17}^{+0.15}\times 10^{-3}$ &
   $7.62_{-0.41}^{+0.44}{}_{-0.17}^{+0.13}\times 10^{-3}$ \\
%%%% 
&$\tilde g\tilde g$ & $1.40_{-0.03}^{+0.08}{}_{-0.11}^{+0.11}\times 10^{-3}$ &
   $1.41_{-0.08}^{+0.01}{}_{-0.04}^{+0.04}\times 10^{-3}$ &
   $1.47_{-0.08}^{+0.09}{}_{-0.05}^{+0.05}\times 10^{-3}$ &
   $1.60_{-0.12}^{+0.12}{}_{-0.05}^{+0.05}\times
                                                                             10^{-3}$ \\\hline
%%%%%%%%%%%%%%%%%
2.3,2.5 & 
 $\tilde q\overline{\tilde q}$ & $1.32_{-0.01}^{+0.06}{}_{-0.06}^{+0.07}\times 10^{-4}$ &
   $1.41_{-0.12}^{+0.09}{}_{-0.03}^{+0.03}\times 10^{-4}$ &
   $1.48_{-0.01}^{+0.02}{}_{-0.04}^{+0.06}\times 10^{-4}$ &
   $1.55_{-0.01}^{+0.03}{}_{-0.03}^{+0.04}\times 10^{-4}$ \\
%%%%
& $\tilde q\tilde q$ & $1.21_{-0.09}^{+0.12}{}_{-0.02}^{+0.02}\times 10^{-3}$ &
   $1.31_{-0.10}^{+0.07}{}_{-0.07}^{+0.07}\times 10^{-3}$ &
   $1.33_{-0.03}^{+0.04}{}_{-0.07}^{+0.07}\times 10^{-3}$ &
   $1.33_{-0.04}^{+0.05}{}_{-0.07}^{+0.07}\times 10^{-3}$ \\
%%%% 
&$\tilde q\tilde g$ & $5.13_{-0.08}^{+0.23}{}_{-0.36}^{+0.19}\times 10^{-4}$ &
   $5.23^{+0.00}_{-0.19} {}_{-0.03}^{+0.03}\times 10^{-4}$ &
   $5.49_{-0.35}^{+0.34}{}_{-0.09}^{+0.04}\times 10^{-4}$ &
   $5.60_{-0.38}^{+0.37}{}_{-0.09}^{+0.04}\times 10^{-4}$ \\
%%%% 
&$\tilde g\tilde g$ & $2.84_{-0.24}^{+0.32}{}_{-0.27}^{+0.26}\times 10^{-5}$ &
   $2.59_{-0.14}^{+0.01}{}_{-0.12}^{+0.12}\times 10^{-5}$ &
   $2.82_{-0.29}^{+0.28}{}_{-0.13}^{+0.13}\times 10^{-5}$ &
   $3.11_{-0.36}^{+0.35}{}_{-0.14}^{+0.14}\times 10^{-5}$ \\\hline
 2.5,2.3 & 
 $\tilde q\overline{\tilde q}$ & $4.72_{-0.03}^{+0.17}{}_{-0.22}^{+0.25}\times 10^{-5}$ &
   $5.03_{-0.42}^{+0.31}{}_{-0.12}^{+0.12}\times 10^{-5}$ &
   $5.29_{-0.04}^{+0.09}{}_{-0.17}^{+0.22}\times 10^{-5}$ &
   $5.56_{-0.05}^{+0.13}{}_{-0.14}^{+0.16}\times 10^{-5}$ \\
%%%%
& $\tilde q\tilde q$ & $5.74_{-0.47}^{+0.65}{}_{-0.11}^{+0.08}\times 10^{-4}$ &
   $6.22_{-0.55}^{+0.44}{}_{-0.44}^{+0.44}\times 10^{-4}$ &
   $6.38_{-0.20}^{+0.27}{}_{-0.45}^{+0.46}\times 10^{-4}$ &
   $6.36_{-0.24}^{+0.34}{}_{-0.45}^{+0.44}\times 10^{-4}$ \\
%%%% 
&$\tilde q\tilde g$ & $5.83_{-0.04}^{+0.22}{}_{-0.41}^{+0.22}\times 10^{-4}$ &
   $6.05_{-0.28}^{+0.05}{}_{-0.09}^{+0.09}\times 10^{-4}$ &
   $6.42_{-0.38}^{+0.41}{}_{-0.15}^{+0.12}\times 10^{-4}$ &
   $6.56_{-0.40}^{+0.44}{}_{-0.15}^{+0.09}\times 10^{-4}$ \\
%%%%
& $\tilde g\tilde g$ & $8.62_{-0.39}^{+0.72}{}_{-0.76}^{+0.74}\times 10^{-5}$ &
   $8.38_{-0.56}^{+0.19}{}_{-0.24}^{+0.24}\times 10^{-5}$ &
   $9.07_{-0.65}^{+0.71}{}_{-0.27}^{+0.30}\times 10^{-5}$ &
   $1.00_{-0.09}^{+0.09}{}_{-0.03}^{+0.03}\times 10^{-4}$ \\\hline
%%%%%%%%%%%%%%%%%%%% 
2.8,3 & 
 $\tilde q\overline{\tilde q}$ & $9.69_{-0.03}^{+0.32}{}_{-0.47}^{+0.54}\times 10^{-6}$ &
   $1.02_{-0.09}^{+0.07}{}_{-0.02}^{+0.02}\times 10^{-5}$ &
   $1.09_{-0.01}^{+0.02}{}_{-0.03}^{+0.04}\times 10^{-5}$ &
   $1.15_{-0.01}^{+0.03}{}_{-0.02}^{+0.03}\times 10^{-5}$ \\
%%%%
& $\tilde q\tilde q$ & $1.47_{-0.11}^{+0.15}{}_{-0.04}^{+0.02}\times 10^{-4}$ &
   $1.59_{-0.14}^{+0.10}{}_{-0.08}^{+0.08}\times 10^{-4}$ &
   $1.64_{-0.04}^{+0.05}{}_{-0.08}^{+0.09}\times 10^{-4}$ &
   $1.64_{-0.05}^{+0.07}{}_{-0.08}^{+0.08}\times 10^{-4}$ \\
%%%% 
&$\tilde q\tilde g$ & $4.62_{-0.14}^{+0.25}{}_{-0.38}^{+0.20}\times 10^{-5}$ &
   $4.61_{-0.24}^{+0.03}{}_{-0.03}^{+0.03}\times 10^{-5}$ &
   $4.97_{-0.36}^{+0.34}{}_{-0.10}^{+0.04}\times 10^{-5}$ &
   $5.10_{-0.39}^{+0.37}{}_{-0.10}^{+0.03}\times 10^{-5}$ \\
%%%% 
&$\tilde g\tilde g$ & $1.97_{-0.22}^{+0.26}{}_{-0.22}^{+0.20}\times 10^{-6}$ &
   $1.68_{-0.13}^{+0.02}{}_{-0.08}^{+0.08}\times 10^{-6}$ &
   $1.93_{-0.23}^{+0.22}{}_{-0.09}^{+0.08}\times 10^{-6}$ &
   $2.18_{-0.30}^{+0.27}{}_{-0.09}^{+0.10}\times 10^{-6}$ \\\hline
%%%%%%%%%%%%%%%%%%%%%%%
 3.,2.8& 
 $\tilde q\overline{\tilde q}$ & $3.50_{-0.01}^{+0.09}{}_{-0.17}^{+0.20}\times 10^{-6}$ &
   $3.68_{-0.34}^{+0.25}{}_{-0.08}^{+0.08}\times 10^{-6}$ &
   $3.93_{-0.05}^{+0.08}{}_{-0.12}^{+0.17}\times 10^{-6}$ &
   $4.17_{-0.07}^{+0.10}{}_{-0.10}^{+0.11}\times
                                                                             10^{-6}$ \\
%%%%%%
& $\tilde q\tilde q$ & $6.61_{-0.54}^{+0.76}{}_{-0.16}^{+0.11}\times 10^{-5}$ &
   $7.19_{-0.70}^{+0.58}{}_{-0.47}^{+0.47}\times 10^{-5}$ &
   $7.45_{-0.23}^{+0.31}{}_{-0.48}^{+0.50}\times 10^{-5}$ &
   $7.43_{-0.28}^{+0.39}{}_{-0.48}^{+0.48}\times 10^{-5}$ \\
%%%% 
&$\tilde q\tilde g$ & $5.16_{-0.09}^{+0.25}{}_{-0.42}^{+0.23}\times 10^{-5}$ &
   $5.24_{-0.32}^{+0.07}{}_{-0.07}^{+0.07}\times 10^{-5}$ &
   $5.71_{-0.39}^{+0.40}{}_{-0.14}^{+0.10}\times 10^{-5}$ &
   $5.87_{-0.42}^{+0.43}{}_{-0.14}^{+0.07}\times
                                                                             10^{-5}$ \\
%%%%
& $\tilde g\tilde g$ & $5.90_{-0.42}^{+0.63}{}_{-0.60}^{+0.57}\times 10^{-6}$ &
   $5.43_{-0.46}^{+0.20}{}_{-0.16}^{+0.16}\times 10^{-6}$ &
   $6.16_{-0.56}^{+0.58}{}_{-0.19}^{+0.21}\times 10^{-6}$ &
   $6.94_{-0.73}^{+0.75}{}_{-0.20}^{+0.21}\times 10^{-6}$ \\\hline
\end{tabular}}  
\end{center}\caption{
Different higher-order approximations for the LHC with $\sqrt{s}=13$~TeV.    
The first error denotes the scale variation  
  while the second error refers to the estimate of the remaining  
  theoretical uncertainty. The latter is given by the resummation  
  uncertainty (NLL), the variation of the two-loop  
  constant~(NNLO$_{\text{app}}$),  while for the two NNLL results the two errors are added in quadrature.  
}  
\label{tab:lhc13-resum}  
\end{table}

In Table~\ref{tab:lhc13-resum} we provide numerical results for the
higher-order approximations NLL, NLLO$_{\text{app}}$,
NNLL$_{\mathrm{fixed-C}}$ and NNLL defined above.  In order to study the
contributions of the different sources of uncertainties, the scale uncertainty
is shown separately from the remaining theoretical uncertainties.  The results
are shown for the same mass values as in Table~\ref{tab:lhc13-nnll}.  The
magnitude of the corrections from the successive improvement in accuracy is
consistent with that seen in Figure~\ref{fig:KNNLL13}. The difference between
the approximate NNLO results and the resummed predictions is moderate at
smaller masses but grows more sizeable for heavy sparticles.  The higher-order
Coulomb corrections and bound-state effects only included in the full NNLL
results become important in particular for squark-antisquark and gluino-pair
production at high masses, whereas the corrections are moderate for
squark-gluino production and small for squark-squark production. The
uncertainty from resummation ambiguities and missing higher-order corrections
is strongly reduced from NLL to NNLL for all processes with the exception of
squark-squark production. It is seen that this uncertainty is dominated by the
two-loop constant variation, which is identical for NNLO$_{\text{app}}$ and
the two NNLL implementations.  The scale uncertainty alone is usually reduced
for the NNLO$_{\text{app}}$ approximation, but can be very asymmetric as seen
already in Figure~\ref{fig:mufvar}. For a more realistic uncertainty
estimate at this order, the renormalization scale should be varied
independently.  The scale uncertainty is further reduced at NNLL  for
squark-antisquark and squark-squark production but
increased for squark-gluino and gluino-pair production, consistent with
Figure~\ref{fig:uncertainty}.
  
%%%%%%%%%%%%%%%%%%%%%%%%%%%%%%%%%%%%%%%%%%%%%%%%%%%%%%%%%%%%%%%%%%%%%%%%%%  
\section{Conclusions}  
\label{sec:conclusion}  
  
We performed a combined NNLL resummation of soft-gluon and Coulomb
corrections for all squark- and gluino-pair production channels at the LHC
based on the method developed for top-quark pair
production~\cite{Beneke:2011mq}, extending an earlier NLL
study~\cite{Falgari:2012hx}.
Grids with our NNLL predictions for the LHC with $\sqrt{s}=13$ and
$14$~TeV for $m_{\tilde q}, m_{\tilde g}=200$--$3000$~GeV and $200$--$3500$~GeV,  
respectively, are publicly available~\cite{SUSYNNLL}.
We furthermore completed the result for the
NNLO threshold expansion of the total cross section~\cite{Beneke:2009ye} by
deriving the spin-dependent non-Coulomb corrections and the process-specific
annihilation contributions, which both give rise to a single-logarithmic NNLO
correction.

Our NNLL results show generally moderate corrections to the NLL predictions
with combined soft-Coulomb corrections~\cite{Falgari:2012hx}, which shows that
the combined resummation is the adequate method to control the QCD corrections
in the region of large sparticle masses, where both the NLO SQCD and the NLL
soft-gluon and Coulomb corrections can become very large, especially for
gluino-pair production.  Corrections beyond NNLO included in the resummed
results become sizeable for sparticle masses above $1.5\, \mathrm{TeV}$.  We
carefully estimated uncertainties due to scale choices and ambiguities of
the resummation formalism and found that the total theoretical uncertainty of
the squark and gluino pair production processes due to missing higher-order
corrections is reduced to the $10\%$ level.  We also compared different
scale-setting procedures for the soft scale in the momentum-space formalism
for soft-gluon resummation and found a better agreement compared to the NLL
calculation. The NNLL
calculation leaves the PDF uncertainties as the dominant source of
uncertainties, which can hopefully be reduced in the future using constraints
from measurements at the LHC.

%%%%%%%%%%%%%%%%%%%%
\paragraph{Note added:} 
 In the final stages of this work we became aware of
related work on the combined soft-Coulomb resummation in the Mellin-space formalism~\cite{NNLLfast}, where a detailed comparison to our results from~\cite{Beneke:2013opa} is performed.
%%%%%%%%%%%%%%%%%%%%%%%%%%%%%%%%%%%%%%%%%%%%%%%%%%%%%%%%%%%%%%%%%%%%%%%%%%  
\section*{Acknowledgements}  
We would like to thank Pietro Falgari for collaboration in the
early stages of this work, Anna Kulesza for correspondence on the hard
matching coefficients computed in~\cite{Beenakker:2013mva} and Michael
Kr\"amer for sharing a draft of~\cite{NNLLfast}.
The work of MB is supported by the BMBF grant 
05H15WOCAA.   
The work of CW was  
partially funded by Research Funding Program ARISTEIA, HOCTools  
(co-financed by the European Union (European Social Fund ESF) and  
Greek national funds through the Operational Program "Education and  
Lifelong Learning" of the National Strategic Reference Framework  
(NSRF)).    
CS is supported by the Heisenberg Programme of the DFG.  
JP and CS  
acknowledge support by the Munich Institute for Astro- and Particle  
Physics (MIAPP) of the DFG cluster of excellence "Origin and Structure  
of the Universe" and the Mainz Institute for Theoretical Physics  
(MITP) during parts of this work.  

%%%%%%%%%%%%%%%%%%%%%%%%%%%%%%%%%%%%%%%%%%%%%%%%%%%%%%%%%%%%%%%%%%%%%%%%%%%%%  
\appendix  
\section{Explicit formulae}  
\subsection{Expansion of the NNLL  cross section}  
\label{app:expansion}  
  
In this appendix we collect the expansions of the NNLL correction factors to $\mathcal{O}(\alpha_s)$ and $\mathcal{O}(\alpha_s^2)$, respectively.  
The expansion to NLO accuracy yields all threshold-enhanced NLO terms and the constant term,  
\begin{eqnarray}  
\label{eq:NLOapprox}  
f^{\mathrm{NNLL} (1)}_{p p',i}  
&=& -\frac{2 \pi^2 D_{R_\alpha}}{\beta} \sqrt{\frac{2 m_r}{\mbar}}  
+4 C_{rr'} \left[\logE[2]{}+6\ln 2 \, \logE{}\right]\nonumber\\ 
&&-4 (C_{R_\alpha}+4 C_{rr'})\logE{}+C^{(1)}_{pp',i}(\mu) +{\cal O}(\beta).  
\end{eqnarray}  
Here and in the following we use the notation  
\begin{equation}  
  \logPow{1}{x}=\ln\left(\frac{x}{\mu_f}\right)\;,  
\end{equation}  
and the sum of the two quadratic Casimir operators of the colour representation of the incoming partons has been defined as  
\begin{equation}  
   C_{rr'}= C_{r}+ C_{r'}  .
\end{equation}  
The constant term can be expressed in terms of the one-loop hard coefficient in~\eqref{eq:hard-def}  
 by the relation  
\begin{equation}  
\label{eq:C1}  
C^{(1)}_{pp',i}(\mu)= h^{(1)}_{i}(\mu)+   
4\,C_{rr'} \bigg[9\ln^22-12\ln 2+8 -\frac{11 \pi^2}{24}\bigg] -  
 12  C_{R} \left[\ln 2-1\right]\,.  
\end{equation}  
The expansion to NNLO accuracy reads\footnote{In the corresponding formula in
  the journal version of this paper we have set $C_A=3$.}   
\begin{align}  
f^{\mathrm{NNLL} (2)}_{p p',i}=&  
\frac{4 \pi ^4 D_{R_\alpha}^2 }{3\beta^2} \frac{2\mred}{\mbar} +  
\frac{2\pi ^2 D_{R_\alpha}}{\beta} \sqrt{\frac{2\mred}{\mbar}}   
\Biggl\{-4  
   C_{rr'} \left(\logE[2]{2}+\logh(\logh-2  
   \logm)-\frac{\pi ^2}{8}\right)\nonumber\\  
 &+ \logE{2} (\beta_0+4 C_{R_\alpha})  
-2 \logh \left(2C_{R_\alpha}+\gamma_{rr'}^{(0)}-2\beta_0\right)  
   +\beta_0 \logmr \nonumber\\  
   &- a_1-4 C_{R_\alpha}- h_1^{(i)}(\mu_h)\Biggr\}  \nonumber\\  
   &+8 \pi ^2 D_{R_\alpha} \left(C_A-2 D_{R_\alpha}  
   (1+\nus)-\frac{1}{2}\frac{4 \mred^2}{\mbar^2}\nua\right) \ln\frac{E}{M} 
\nonumber\\  
%%%%%%%%%%%%%%%%%%%%%%%%%%%%&  
&+8 C_{rr'}^2 \left(\logE[4]{8}-\logEs[4]{8}  
+\logh[4] \right)  
-16   C_{rr'}  
   \left(\frac{\beta_0}{6}+4 C_{rr'}+C_{R_\alpha}\right)  
 \left(\logE[3]{8}-\logEs[3]{8}\right)  
\nonumber\\  
& -8 C_{rr'}  
\Bigl[  
4 C_{rr'}  
   \logm+  
\frac{8 \beta_0}{3}-\gamma_{rr'}^{(0)}-2 C_{R_\alpha}  
\Bigr]\logh[3]  
\nonumber\\  
%%%%%%%%%%%%%%%%%%%%%%%%%%%%%%%%%%%%%%%%%%%  
&+\Biggl\{C_{rr'}^2\left(384-\frac{70 \pi ^2}{3}\right) + 4C_{rr'}  
   \left[4 \beta_0+28 C_{R_\alpha}+  
       C_A\left(\frac{67}{9}-\frac{\pi^2}{3}\right)-\frac{20 n_f T_F}{9}\right]  
\nonumber\\  
&+4 C_{R_\alpha}  (\beta_0+2 C_{R_\alpha}) \Biggr\}   
  \left(\logE[2]{8}-\logEs[2]{8}\right)  
\nonumber\\  
%%%%%%%%%%%%%%%%%%%%%%%%%%%%%%%%%%%%%%  
&+\Biggl\{  
16C_{rr'}^2 \left(  
   \logE[2]{8}-4 \logE{8}+2 \logm[2]-\frac{11 \pi  
   ^2}{24}+8\right)  
\nonumber\\  
&+C_{rr'}\Bigl[  
  -16 C_{R_\alpha} \logE{8}+48 C_{R_\alpha}-\frac{80n_f T_F}{9}  
  -4 C_A\left(\frac{\pi^2}{3}-\frac{67}{9}\right)\nonumber\\  
&  
+\left(40 \beta_0-16 \gamma_{rr'}^{(0)}-32  
   C_{R_\alpha}\right)\logm  
\Bigr]  
\nonumber\\  
&  
+12 \beta_0^2-10 \beta_0 (\gamma_{rr'}^{(0)}+2 C_{R_\alpha})  
+2 (\gamma_{rr'}^{(0)}+2 C_{R_\alpha})^2 
\Biggr\}\,\logh[2]  
\nonumber\\  
%%%%%%%%%%%%%%%%%%%%%%%%%%%%%%%%%%%%%%%%  
&+\Biggl\{C_{rr'}^2 \left(448  
   \zeta (3)-1536+\frac{280 \pi ^2}{3}\right)  
 + C_{rr'} \Biggl[\left(\frac{11  
   \pi ^2}{3}-64\right) \beta_0  
\nonumber\\  
&+\left(\frac{70 \pi  
   ^2}{3}-448\right) C_{R_\alpha}+C_A\left(28 \zeta (3)+\frac{59 \pi  
   ^2}{9}-\frac{4024}{27}\right) + n_f T_F \left(\frac{1184}{27}-\frac{4 \pi  
   ^2}{9} \right) \Biggr]  
\nonumber\\  
&+C_{R_\alpha} \left[-24 \beta_0+C_A\left(-8  
   \zeta (3)+\frac{4 \pi ^2}{3}-\frac{196}{9}\right)
 +\frac{80 n_f T_F}{9}\right]-48 C_{R_\alpha}^2\Biggr\}\,  
\left(\logE{8}-\logEs{8}\right)  
 \nonumber\\  
%%%%%%%%%%%%%%%%%%%%%%%%%%%%%%%%%%%%%%%%%%%%%%%%%%%%%%%%%%  
&+ \Biggl\{4C_{rr'}  
  \left(\logE[2]{8}-4 \logE{8}+\logh[2]-\frac{11  
   \pi ^2}{24}+8\right)\nonumber\\  
&  
-4C_{R_\alpha} \left(\logE{8}-3\right)  
-2\logh \Bigl[4  C_{rr'} \logm  
- 2C_{R_\alpha}+3 \beta_0-\gamma_{rr'}^{(0)}\Bigr]  
\Biggr\}\,h_1^{(i)}(\mu_h)   
\nonumber\\  
%%%%%%%%%%%%%%%%%%%%%%%%%%%%%%%%%%%%%%%%  
&+\Biggl\{  
\left(4 C_{rr'}\logm +2\beta_0-\gamma_{rr'}^{(0)}  
-2 C_{R_\alpha}\right) \Biggl[-8 C_{rr'}  
   \logE[2]{8}  
+8 (4 C_{rr'}+C_{R_\alpha})\logE{8}  
\nonumber\\  
& +C_{rr'}\left(-64 +\frac{11 \pi ^2}{3}\right)-24  
   C_{R_\alpha}\Biggr]  
-C_{rr'} \left[C_A\left(\frac{536}{9}-\frac{8\pi^2}{3}\right)-\frac{160 n_f T_F}{9}\right]  
\logm
\nonumber\\  
&+C_{rr'} \left[C_A\left(\frac{808}{27}-\frac{11 \pi^2}{9}-28 \zeta (3)\right)
  -\left(\frac{224}{27}-\frac{4\pi^2}{9}\right) n_f T_F
\right]  
\nonumber\\  
&+2 \gamma_{rr'}^{(1)}+C_{R_\alpha} \left[C_A\left(8 \zeta  
   (3)+\frac{196}{9} -\frac{4\pi ^2}{3}\right)-\frac{-80 n_f T_F}{9}\right]-4\beta_1  
\Biggr\}\,\logh
+C_{pp',i}^{(2)} . 
%%%%%%%%%%%%%%%%%%%%%%%%%%%%%%%%%%%%%%%%%  
\end{align}  
Here we further defined $\logEs{8}=\ln\left(\frac{8 E}{\mu_s}\right)$.  
The two-loop beta-function coefficient is given by $\beta_1=\frac{34}{3}C_A^2
-\frac{20}{3} C_A T_F n_f -4 C_F T_F n_f$.
The anomalous-dimension coefficients appearing in this formula are  
related to the incoming partons and are defined as  
\begin{equation}  
\gamma_{rr'}^{(n)}= \gamma^{\phi,r(n)}+\gamma^{\phi,r'(n)} ,   
\end{equation}  
with  
\begin{eqnarray}  
  \gamma^{\phi,3(0)}&=&3C_F ,\\  
 \gamma^{\phi,3(1)}&=&C_F^2\left(\frac{3}{2}-2\pi^2+24\zeta_3\right)  
   +C_AC_F\left(\frac{17}{6}+\frac{22\pi^2}{9}-12\zeta_3\right)\nonumber\\   
   && -\,C_F T_F n_f\left(\frac{2}{3}+\frac{8\pi^2}{9}\right),\\  
  \gamma^{\phi,8(0)}&=&\beta_0 =\frac{11}{3} C_A- \frac{4}{3} T_F n_f ,\\  
 \gamma^{\phi,8(1)}&=&4C_A^2\left(\frac{8}{3}+3\zeta_3\right)  
    -\frac{16}{3}C_AT_F n_f- 4C_FT_F n_f.  
\end{eqnarray}  
  
\subsection{Analytic NNLL result for fixed-order Coulomb corrections}  
\label{app:fixC}  
If the Coulomb corrections are treated at fixed NNLO accuracy through  
the factor~\eqref{eq:fixed-C}, the  
$\omega$-convolution in the resummed cross  
section~\eqref{eq:resum-sigma} can be performed explicitly, resulting in  
an analytic expression:  
\begin{align}  
\hat\sigma^{\text{res}}_{pp'}(\hat s,\mu)=&  
\sum_{i}  
\sigma^{(0)}_{pp',i}(\mu)  
 U_{R_\alpha}(\mu_h,\mu_s,\mu_f)  
 \left(\frac{2\mbar}{\mu_s}\right)^{-2\eta}   
\tilde{s}_i^{R_\alpha}(\partial_\eta,\mu_s)\,  
\mathcal{C}_{\text{hC}}^{\text{NNLO}}(E,\mu_h,\mu_s,\mu_f) , 
\label{eq:resum-fixC}  
\end{align}  
with the Laplace transform of the NLO soft function~\eqref{eq:nlo-soft} and
where the function $\mathcal{C}_{\text{hC}}^{\text{NNLO}}$ is given by  
\begin{align}  
\mathcal{C}_{\text{hC}}^{\text{NNLO}}(E,\mu_h,\mu_s,\mu_f)&=  
\left(\frac{2E e^{-\gamma_E}}{\mu_s}\right)^{2\eta}  
\sum_{n=0}^2 \left(\frac{\alpha_s}{4\pi}\right)^n  
\mathcal{C}_{\text{hC}}^{(n)}(E,\mu_h,\mu_s,\mu_f), \label{eq:chc}\\  
%%%%%%%%%%%%%%%%%%%%%%%%%%%%%%%%%%%%%%%%%%%%%%%%%  
  \mathcal{C}_{\text{hC}}^{(0)}(E,\mu_h,\mu_s,\mu_f)&=    
  \frac{\sqrt\pi}{2\Gamma(2\eta+\frac{3}{2})}, \label{eq:chc-lo}\\  
%%%%%%%%%%%%%%%%%%%%%%%%%%%%%%%%%%%%%%%%%%%%%%%%%  
\mathcal{C}_{\text{hC}}^{(1)}(E,\mu_h,\mu_s,\mu_f)=&    
 -\frac{ (2\pi^2 D_{R_\alpha})}{\Gamma(2\eta+1)}  
 \sqrt{ \frac{2\mred}{E}}  
 +  
  \frac{\sqrt\pi}{2\Gamma(2\eta+\frac{3}{2})} \, h^{(1)}_i(\mu_h)  ,
\\  
%%%%%%%%%%%%%%%%%%%%%%%%%%%%%%%%%%%%%%%%%%%%%%%%%  
  \mathcal{C}_{\text{hC}}^{(2)}(E,\mu_h,\mu_s,\mu_f)=&    
 \frac{\sqrt\pi (2\pi^2D_{R_\alpha})^2}{3\Gamma(2\eta+\frac{1}{2})}  
 \left(\frac{2\mred}{E}\right)   
-\frac{ (2\pi^2 D_{R_\alpha})}{\Gamma(2\eta+1)}  
 \sqrt{ \frac{2\mred}{E}} \nonumber\\  
&\times  \left[h_1^{(i)}(\mu_h)+a_1  
           - \beta_0\left(\ln\left(\frac{8E \mred}{\mu^2}\right)  
               -\psi^{(0)}(2\eta+1)-\gamma_E  
           \right)\right]  
\nonumber\\  
&+   \frac{2\sqrt\pi (2\pi^2   D_{R_\alpha})}{\Gamma(2\eta+\frac{3}{2})}   
      \left(C_A-2 D_{R_\alpha}(v_{\text{spin}}+1)
        -\frac{\nua}{2}\frac{4 \mred^2}{\mbar^2}\right)\nonumber\\  
&\times         \left(\ln\left(\frac{E}{\mbar}\right)  
          -\psi^{(0)}\left(2\eta+\frac{3}{2}\right)  
          -\gamma_E+2-2\ln 2\right)\nonumber\\
&+   \frac{\sqrt\pi}{2\Gamma(2\eta+\frac{3}{2})} \, h^{(1)}_i(\mu_h)\, (-2\beta_0)
\ln\left(\frac{\mu_h}{\mu_f}\right)  ,  \label{eq:chc-nnlo}
\end{align}  
with the digamma function $\psi^{(0)}(x)=\frac{d\ln\Gamma(x)}{dx}$.  It is
straightforward to evaluate the action of the derivative with respect to
$\eta$  in~\eqref{eq:resum-fixC} on the
factors~\eqref{eq:chc-lo}--\eqref{eq:chc-nnlo}.
  
%%%%%%%%%%%

\providecommand{\href}[2]{#2}\begingroup\raggedright\endgroup

\end{document}